\documentclass[twocolumn,traditabstract,longauth]{aa}

\usepackage[english]{babel} 
\usepackage{graphicx,amsmath}
\usepackage{epstopdf}
\usepackage{epsf,color}
\usepackage{natbib}
\bibpunct{(}{)}{;}{a}{}{,} 
\usepackage{txfonts}

\def\setsymbol#1#2{\expandafter\def\csname #1\endcsname{#2}}
\def\getsymbol#1{\csname #1\endcsname}

\def\Planck{{\it Planck\/}}

\def\allearlypapers{\nocite{planck2011-1.1, planck2011-1.3, planck2011-1.4, planck2011-1.5, planck2011-1.6, planck2011-1.7, planck2011-1.10, planck2011-1.10sup, planck2011-5.1a, planck2011-5.1b, planck2011-5.2a, planck2011-5.2b, planck2011-5.2c, planck2011-6.1, planck2011-6.2, planck2011-6.3a, planck2011-6.4a, planck2011-6.4b, planck2011-6.6, planck2011-7.0, planck2011-7.2, planck2011-7.3, planck2011-7.7a, planck2011-7.7b, planck2011-7.12, planck2011-7.13}}

\newbox\tablebox    \newdimen\tablewidth
\def\leaderfil{\leaders\hbox to 5pt{\hss.\hss}\hfil}
\def\endPlancktablewide{\tablewidth=\textwidth 
    $$\hss\copy\tablebox\hss$$
    \vskip-\lastskip\vskip -2pt}
\def\tablenote#1 #2\par{\begingroup \parindent=0.8em
    \abovedisplayshortskip=0pt\belowdisplayshortskip=0pt
    \noindent
    $$\hss\vbox{\hsize\tablewidth \hangindent=\parindent \hangafter=1 \noindent
    \hbox to \parindent{\sup{\rm #1}\hss}\strut#2\strut\par}\hss$$
    \endgroup}

\def\L2{\ifmmode L_2\else $L_2$\fi}

\def\DeltaT{\ifmmode \Delta T\else $\Delta T$\fi}
\def\deltat{\ifmmode \Delta t\else $\Delta t$\fi}
\def\fknee{\ifmmode f_{\rm knee}\else $f_{\rm knee}$\fi}
\def\Fmax{\ifmmode F_{\rm max}\else $F_{\rm max}$\fi}
\def\solar{\ifmmode{\rm M}_{\mathord\odot}\else${\rm M}_{\mathord\odot}$\fi}

\def\inv{\ifmmode^{-1}\else$^{-1}$\fi}
\def\mo{\ifmmode^{-1}\else$^{-1}$\fi}
\def\sup#1{\ifmmode ^{\rm #1}\else $^{\rm #1}$\fi}
\def\expo#1{\ifmmode \times 10^{#1}\else $\times 10^{#1}$\fi}
\def\,{\thinspace}
\def\lsim{\mathrel{\raise .4ex\hbox{\rlap{$<$}\lower 1.2ex\hbox{$\sim$}}}}
\def\gsim{\mathrel{\raise .4ex\hbox{\rlap{$>$}\lower 1.2ex\hbox{$\sim$}}}}

\def\simprop{\mathrel{\raise .4ex\hbox{\rlap{$\propto$}\lower 1.2ex\hbox{$\sim$}}}}
\def\deg{\ifmmode^\circ\else$^\circ$\fi}
\def\pdeg{\ifmmode $\setbox0=\hbox{$^{\circ}$}\rlap{\hskip.11\wd0 .}$^{\circ}
          \else \setbox0=\hbox{$^{\circ}$}\rlap{\hskip.11\wd0 .}$^{\circ}$\fi}
\def\arcs{\ifmmode {^{\scriptstyle\prime\prime}}
          \else $^{\scriptstyle\prime\prime}$\fi}
\def\arcm{\ifmmode {^{\scriptstyle\prime}}
          \else $^{\scriptstyle\prime}$\fi}
\newdimen\sa  \newdimen\sb
\def\parcs{\sa=.07em \sb=.03em
     \ifmmode \hbox{\rlap{.}}^{\scriptstyle\prime\kern -\sb\prime}\hbox{\kern -\sa}
     \else \rlap{.}$^{\scriptstyle\prime\kern -\sb\prime}$\kern -\sa\fi}
\def\parcm{\sa=.08em \sb=.03em
     \ifmmode \hbox{\rlap{.}\kern\sa}^{\scriptstyle\prime}\hbox{\kern-\sb}
     \else \rlap{.}\kern\sa$^{\scriptstyle\prime}$\kern-\sb\fi}
\def\ra[#1 #2 #3.#4]{#1\sup{h}#2\sup{m}#3\sup{s}\llap.#4}
\def\dec[#1 #2 #3.#4]{#1\deg#2\arcm#3\arcs\llap.#4}
\def\deco[#1 #2 #3]{#1\deg#2\arcm#3\arcs}
\def\rra[#1 #2]{#1\sup{h}#2\sup{m}}

\def\dots{\relax\ifmmode \ldots\else $\ldots$\fi}
\def\WHzsr{\ifmmode $W\,Hz\mo\,sr\mo$\else W\,Hz\mo\,sr\mo\fi}
\def\mHz{\ifmmode $\,mHz$\else \,mHz\fi}
\def\GHz{\ifmmode $\,GHz$\else \,GHz\fi}
\def\mKs{\ifmmode $\,mK\,s$^{1/2}\else \,mK\,s$^{1/2}$\fi}
\def\muKs{\ifmmode \,\mu$K\,s$^{1/2}\else \,$\mu$K\,s$^{1/2}$\fi}
\def\muKRJs{\ifmmode \,\mu$K$_{\rm RJ}$\,s$^{1/2}\else \,$\mu$K$_{\rm RJ}$\,s$^{1/2}$\fi}
\def\muKHz{\ifmmode \,\mu$K\,Hz$^{-1/2}\else \,$\mu$K\,Hz$^{-1/2}$\fi}
\def\MJysr{\ifmmode \,$MJy\,sr\mo$\else \,MJy\,sr\mo\fi}
\def\MJysrmK{\ifmmode \,$MJy\,sr\mo$\,mK$_{\rm CMB}\mo\else \,MJy\,sr\mo\,mK$_{\rm CMB}\mo$\fi}
\def\microns{\ifmmode \,\mu$m$\else \,$\mu$m\fi}

\def\muK{\ifmmode \,\mu$K$\else \,$\mu$\hbox{K}\fi}
\def\microK{\ifmmode \,\mu$K$\else \,$\mu$\hbox{K}\fi}
\def\muW{\ifmmode \,\mu$W$\else \,$\mu$\hbox{W}\fi}
\def\kms{\ifmmode $\,km\,s$^{-1}\else \,km\,s$^{-1}$\fi}
\def\kmsMpc{\ifmmode $\,\kms\,Mpc\mo$\else \,\kms\,Mpc\mo\fi}
\setsymbol{LFI:center:frequency:70GHz:units}{70.3\,GHz}
\setsymbol{LFI:center:frequency:44GHz:units}{44.1\,GHz}
\setsymbol{LFI:center:frequency:30GHz:units}{28.5\,GHz}

\setsymbol{LFI:center:frequency:70GHz}{70.3}
\setsymbol{LFI:center:frequency:44GHz}{44.1}
\setsymbol{LFI:center:frequency:30GHz}{28.5}

\setsymbol{LFI:center:frequency:LFI18:Rad:M:units}{71.7\GHz}
\setsymbol{LFI:center:frequency:LFI19:Rad:M:units}{67.5\GHz}
\setsymbol{LFI:center:frequency:LFI20:Rad:M:units}{69.2\GHz}
\setsymbol{LFI:center:frequency:LFI21:Rad:M:units}{70.4\GHz}
\setsymbol{LFI:center:frequency:LFI22:Rad:M:units}{71.5\GHz}
\setsymbol{LFI:center:frequency:LFI23:Rad:M:units}{70.8\GHz}
\setsymbol{LFI:center:frequency:LFI24:Rad:M:units}{44.4\GHz}
\setsymbol{LFI:center:frequency:LFI25:Rad:M:units}{44.0\GHz}
\setsymbol{LFI:center:frequency:LFI26:Rad:M:units}{43.9\GHz}
\setsymbol{LFI:center:frequency:LFI27:Rad:M:units}{28.3\GHz}
\setsymbol{LFI:center:frequency:LFI28:Rad:M:units}{28.8\GHz}
\setsymbol{LFI:center:frequency:LFI18:Rad:S:units}{70.1\GHz}
\setsymbol{LFI:center:frequency:LFI19:Rad:S:units}{69.6\GHz}
\setsymbol{LFI:center:frequency:LFI20:Rad:S:units}{69.5\GHz}
\setsymbol{LFI:center:frequency:LFI21:Rad:S:units}{69.5\GHz}
\setsymbol{LFI:center:frequency:LFI22:Rad:S:units}{72.8\GHz}
\setsymbol{LFI:center:frequency:LFI23:Rad:S:units}{71.3\GHz}
\setsymbol{LFI:center:frequency:LFI24:Rad:S:units}{44.1\GHz}
\setsymbol{LFI:center:frequency:LFI25:Rad:S:units}{44.1\GHz}
\setsymbol{LFI:center:frequency:LFI26:Rad:S:units}{44.1\GHz}
\setsymbol{LFI:center:frequency:LFI27:Rad:S:units}{28.5\GHz}
\setsymbol{LFI:center:frequency:LFI28:Rad:S:units}{28.2\GHz}

\setsymbol{LFI:center:frequency:LFI18:Rad:M}{71.7}
\setsymbol{LFI:center:frequency:LFI19:Rad:M}{67.5}
\setsymbol{LFI:center:frequency:LFI20:Rad:M}{69.2}
\setsymbol{LFI:center:frequency:LFI21:Rad:M}{70.4}
\setsymbol{LFI:center:frequency:LFI22:Rad:M}{71.5}
\setsymbol{LFI:center:frequency:LFI23:Rad:M}{70.8}
\setsymbol{LFI:center:frequency:LFI24:Rad:M}{44.4}
\setsymbol{LFI:center:frequency:LFI25:Rad:M}{44.0}
\setsymbol{LFI:center:frequency:LFI26:Rad:M}{43.9}
\setsymbol{LFI:center:frequency:LFI27:Rad:M}{28.3}
\setsymbol{LFI:center:frequency:LFI28:Rad:M}{28.8}
\setsymbol{LFI:center:frequency:LFI18:Rad:S}{70.1}
\setsymbol{LFI:center:frequency:LFI19:Rad:S}{69.6}
\setsymbol{LFI:center:frequency:LFI20:Rad:S}{69.5}
\setsymbol{LFI:center:frequency:LFI21:Rad:S}{69.5}
\setsymbol{LFI:center:frequency:LFI22:Rad:S}{72.8}
\setsymbol{LFI:center:frequency:LFI23:Rad:S}{71.3}
\setsymbol{LFI:center:frequency:LFI24:Rad:S}{44.1}
\setsymbol{LFI:center:frequency:LFI25:Rad:S}{44.1}
\setsymbol{LFI:center:frequency:LFI26:Rad:S}{44.1}
\setsymbol{LFI:center:frequency:LFI27:Rad:S}{28.5}
\setsymbol{LFI:center:frequency:LFI28:Rad:S}{28.2}

\setsymbol{LFI:white:noise:sensitivity:70GHz:units}{152.6\muKs}
\setsymbol{LFI:white:noise:sensitivity:44GHz:units}{173.1\muKs}
\setsymbol{LFI:white:noise:sensitivity:30GHz:units}{146.8\muKs}

\setsymbol{LFI:white:noise:sensitivity:70GHz}{152.6}
\setsymbol{LFI:white:noise:sensitivity:44GHz}{173.1}
\setsymbol{LFI:white:noise:sensitivity:30GHz}{146.8}

\setsymbol{LFI:white:noise:sensitivity:LFI18:Rad:M:units}{512.0\muKs}
\setsymbol{LFI:white:noise:sensitivity:LFI19:Rad:M:units}{581.4\muKs}
\setsymbol{LFI:white:noise:sensitivity:LFI20:Rad:M:units}{590.8\muKs}
\setsymbol{LFI:white:noise:sensitivity:LFI21:Rad:M:units}{455.2\muKs}
\setsymbol{LFI:white:noise:sensitivity:LFI22:Rad:M:units}{492.0\muKs}
\setsymbol{LFI:white:noise:sensitivity:LFI23:Rad:M:units}{507.7\muKs}
\setsymbol{LFI:white:noise:sensitivity:LFI24:Rad:M:units}{462.2\muKs}
\setsymbol{LFI:white:noise:sensitivity:LFI25:Rad:M:units}{413.6\muKs}
\setsymbol{LFI:white:noise:sensitivity:LFI26:Rad:M:units}{478.6\muKs}
\setsymbol{LFI:white:noise:sensitivity:LFI27:Rad:M:units}{277.7\muKs}
\setsymbol{LFI:white:noise:sensitivity:LFI28:Rad:M:units}{312.3\muKs}
\setsymbol{LFI:white:noise:sensitivity:LFI18:Rad:S:units}{465.7\muKs}
\setsymbol{LFI:white:noise:sensitivity:LFI19:Rad:S:units}{555.6\muKs}
\setsymbol{LFI:white:noise:sensitivity:LFI20:Rad:S:units}{623.2\muKs}
\setsymbol{LFI:white:noise:sensitivity:LFI21:Rad:S:units}{564.1\muKs}
\setsymbol{LFI:white:noise:sensitivity:LFI22:Rad:S:units}{534.4\muKs}
\setsymbol{LFI:white:noise:sensitivity:LFI23:Rad:S:units}{542.4\muKs}
\setsymbol{LFI:white:noise:sensitivity:LFI24:Rad:S:units}{399.2\muKs}
\setsymbol{LFI:white:noise:sensitivity:LFI25:Rad:S:units}{392.6\muKs}
\setsymbol{LFI:white:noise:sensitivity:LFI26:Rad:S:units}{418.6\muKs}
\setsymbol{LFI:white:noise:sensitivity:LFI27:Rad:S:units}{302.9\muKs}
\setsymbol{LFI:white:noise:sensitivity:LFI28:Rad:S:units}{285.3\muKs}

\setsymbol{LFI:white:noise:sensitivity:LFI18:Rad:M}{512.0}
\setsymbol{LFI:white:noise:sensitivity:LFI19:Rad:M}{581.4}
\setsymbol{LFI:white:noise:sensitivity:LFI20:Rad:M}{590.8}
\setsymbol{LFI:white:noise:sensitivity:LFI21:Rad:M}{455.2}
\setsymbol{LFI:white:noise:sensitivity:LFI22:Rad:M}{492.0}
\setsymbol{LFI:white:noise:sensitivity:LFI23:Rad:M}{507.7}
\setsymbol{LFI:white:noise:sensitivity:LFI24:Rad:M}{462.2}
\setsymbol{LFI:white:noise:sensitivity:LFI25:Rad:M}{413.6}
\setsymbol{LFI:white:noise:sensitivity:LFI26:Rad:M}{478.6}
\setsymbol{LFI:white:noise:sensitivity:LFI27:Rad:M}{277.7}
\setsymbol{LFI:white:noise:sensitivity:LFI28:Rad:M}{312.3}
\setsymbol{LFI:white:noise:sensitivity:LFI18:Rad:S}{465.7}
\setsymbol{LFI:white:noise:sensitivity:LFI19:Rad:S}{555.6}
\setsymbol{LFI:white:noise:sensitivity:LFI20:Rad:S}{623.2}
\setsymbol{LFI:white:noise:sensitivity:LFI21:Rad:S}{564.1}
\setsymbol{LFI:white:noise:sensitivity:LFI22:Rad:S}{534.4}
\setsymbol{LFI:white:noise:sensitivity:LFI23:Rad:S}{542.4}
\setsymbol{LFI:white:noise:sensitivity:LFI24:Rad:S}{399.2}
\setsymbol{LFI:white:noise:sensitivity:LFI25:Rad:S}{392.6}
\setsymbol{LFI:white:noise:sensitivity:LFI26:Rad:S}{418.6}
\setsymbol{LFI:white:noise:sensitivity:LFI27:Rad:S}{302.9}
\setsymbol{LFI:white:noise:sensitivity:LFI28:Rad:S}{285.3}

\setsymbol{LFI:knee:frequency:70GHz:units}{29.5\mHz}
\setsymbol{LFI:knee:frequency:44GHz:units}{56.2\mHz}
\setsymbol{LFI:knee:frequency:30GHz:units}{113.7\mHz}

\setsymbol{LFI:knee:frequency:70GHz}{29.5}
\setsymbol{LFI:knee:frequency:44GHz}{56.2}
\setsymbol{LFI:knee:frequency:30GHz}{113.7}

\setsymbol{LFI:knee:frequency:LFI18:Rad:M:units}{16.3\mHz}
\setsymbol{LFI:knee:frequency:LFI19:Rad:M:units}{15.1\mHz}
\setsymbol{LFI:knee:frequency:LFI20:Rad:M:units}{18.7\mHz}
\setsymbol{LFI:knee:frequency:LFI21:Rad:M:units}{37.2\mHz}
\setsymbol{LFI:knee:frequency:LFI22:Rad:M:units}{12.7\mHz}
\setsymbol{LFI:knee:frequency:LFI23:Rad:M:units}{34.6\mHz}
\setsymbol{LFI:knee:frequency:LFI24:Rad:M:units}{46.2\mHz}
\setsymbol{LFI:knee:frequency:LFI25:Rad:M:units}{24.9\mHz}
\setsymbol{LFI:knee:frequency:LFI26:Rad:M:units}{67.6\mHz}
\setsymbol{LFI:knee:frequency:LFI27:Rad:M:units}{187.4\mHz}
\setsymbol{LFI:knee:frequency:LFI28:Rad:M:units}{122.2\mHz}
\setsymbol{LFI:knee:frequency:LFI18:Rad:S:units}{17.7\mHz}
\setsymbol{LFI:knee:frequency:LFI19:Rad:S:units}{22.0\mHz}
\setsymbol{LFI:knee:frequency:LFI20:Rad:S:units}{8.7\mHz}
\setsymbol{LFI:knee:frequency:LFI21:Rad:S:units}{25.9\mHz}
\setsymbol{LFI:knee:frequency:LFI22:Rad:S:units}{15.8\mHz}
\setsymbol{LFI:knee:frequency:LFI23:Rad:S:units}{129.8\mHz}
\setsymbol{LFI:knee:frequency:LFI24:Rad:S:units}{100.9\mHz}
\setsymbol{LFI:knee:frequency:LFI25:Rad:S:units}{38.9\mHz}
\setsymbol{LFI:knee:frequency:LFI26:Rad:S:units}{58.9\mHz}
\setsymbol{LFI:knee:frequency:LFI27:Rad:S:units}{104.4\mHz}
\setsymbol{LFI:knee:frequency:LFI28:Rad:S:units}{40.7\mHz}

\setsymbol{LFI:knee:frequency:LFI18:Rad:M}{16.3}
\setsymbol{LFI:knee:frequency:LFI19:Rad:M}{15.1}
\setsymbol{LFI:knee:frequency:LFI20:Rad:M}{18.7}
\setsymbol{LFI:knee:frequency:LFI21:Rad:M}{37.2}
\setsymbol{LFI:knee:frequency:LFI22:Rad:M}{12.7}
\setsymbol{LFI:knee:frequency:LFI23:Rad:M}{34.6}
\setsymbol{LFI:knee:frequency:LFI24:Rad:M}{46.2}
\setsymbol{LFI:knee:frequency:LFI25:Rad:M}{24.9}
\setsymbol{LFI:knee:frequency:LFI26:Rad:M}{67.6}
\setsymbol{LFI:knee:frequency:LFI27:Rad:M}{187.4}
\setsymbol{LFI:knee:frequency:LFI28:Rad:M}{122.2}
\setsymbol{LFI:knee:frequency:LFI18:Rad:S}{17.7}
\setsymbol{LFI:knee:frequency:LFI19:Rad:S}{22.0}
\setsymbol{LFI:knee:frequency:LFI20:Rad:S}{8.7}
\setsymbol{LFI:knee:frequency:LFI21:Rad:S}{25.9}
\setsymbol{LFI:knee:frequency:LFI22:Rad:S}{15.8}
\setsymbol{LFI:knee:frequency:LFI23:Rad:S}{129.8}
\setsymbol{LFI:knee:frequency:LFI24:Rad:S}{100.9}
\setsymbol{LFI:knee:frequency:LFI25:Rad:S}{38.9}
\setsymbol{LFI:knee:frequency:LFI26:Rad:S}{58.9}
\setsymbol{LFI:knee:frequency:LFI27:Rad:S}{104.4}
\setsymbol{LFI:knee:frequency:LFI28:Rad:S}{40.7}

\setsymbol{LFI:slope:70GHz:units}{$-1.03$\mHz}
\setsymbol{LFI:slope:44GHz:units}{$-0.89$\mHz}
\setsymbol{LFI:slope:30GHz:units}{$-0.87$\mHz}

\setsymbol{LFI:slope:70GHz}{$-1.03$}
\setsymbol{LFI:slope:44GHz}{$-0.89$}
\setsymbol{LFI:slope:30GHz}{$-0.87$}

\setsymbol{LFI:slope:LFI18:Rad:M:units}{$-1.04$\mHz}
\setsymbol{LFI:slope:LFI19:Rad:M:units}{$-1.09$\mHz}
\setsymbol{LFI:slope:LFI20:Rad:M:units}{$-0.69$\mHz}
\setsymbol{LFI:slope:LFI21:Rad:M:units}{$-1.56$\mHz}
\setsymbol{LFI:slope:LFI22:Rad:M:units}{$-1.01$\mHz}
\setsymbol{LFI:slope:LFI23:Rad:M:units}{$-0.96$\mHz}
\setsymbol{LFI:slope:LFI24:Rad:M:units}{$-0.83$\mHz}
\setsymbol{LFI:slope:LFI25:Rad:M:units}{$-0.91$\mHz}
\setsymbol{LFI:slope:LFI26:Rad:M:units}{$-0.95$\mHz}
\setsymbol{LFI:slope:LFI27:Rad:M:units}{$-0.87$\mHz}
\setsymbol{LFI:slope:LFI28:Rad:M:units}{$-0.88$\mHz}
\setsymbol{LFI:slope:LFI18:Rad:S:units}{$-1.15$\mHz}
\setsymbol{LFI:slope:LFI19:Rad:S:units}{$-1.00$\mHz}
\setsymbol{LFI:slope:LFI20:Rad:S:units}{$-0.95$\mHz}
\setsymbol{LFI:slope:LFI21:Rad:S:units}{$-0.92$\mHz}
\setsymbol{LFI:slope:LFI22:Rad:S:units}{$-1.01$\mHz}
\setsymbol{LFI:slope:LFI23:Rad:S:units}{$-0.95$\mHz}
\setsymbol{LFI:slope:LFI24:Rad:S:units}{$-0.73$\mHz}
\setsymbol{LFI:slope:LFI25:Rad:S:units}{$-1.16$\mHz}
\setsymbol{LFI:slope:LFI26:Rad:S:units}{$-0.79$\mHz}
\setsymbol{LFI:slope:LFI27:Rad:S:units}{$-0.82$\mHz}
\setsymbol{LFI:slope:LFI28:Rad:S:units}{$-0.91$\mHz}

\setsymbol{LFI:slope:LFI18:Rad:M}{$-1.04$}
\setsymbol{LFI:slope:LFI19:Rad:M}{$-1.09$}
\setsymbol{LFI:slope:LFI20:Rad:M}{$-0.69$}
\setsymbol{LFI:slope:LFI21:Rad:M}{$-1.56$}
\setsymbol{LFI:slope:LFI22:Rad:M}{$-1.01$}
\setsymbol{LFI:slope:LFI23:Rad:M}{$-0.96$}
\setsymbol{LFI:slope:LFI24:Rad:M}{$-0.83$}
\setsymbol{LFI:slope:LFI25:Rad:M}{$-0.91$}
\setsymbol{LFI:slope:LFI26:Rad:M}{$-0.95$}
\setsymbol{LFI:slope:LFI27:Rad:M}{$-0.87$}
\setsymbol{LFI:slope:LFI28:Rad:M}{$-0.88$}
\setsymbol{LFI:slope:LFI18:Rad:S}{$-1.15$}
\setsymbol{LFI:slope:LFI19:Rad:S}{$-1.00$}
\setsymbol{LFI:slope:LFI20:Rad:S}{$-0.95$}
\setsymbol{LFI:slope:LFI21:Rad:S}{$-0.92$}
\setsymbol{LFI:slope:LFI22:Rad:S}{$-1.01$}
\setsymbol{LFI:slope:LFI23:Rad:S}{$-0.95$}
\setsymbol{LFI:slope:LFI24:Rad:S}{$-0.73$}
\setsymbol{LFI:slope:LFI25:Rad:S}{$-1.16$}
\setsymbol{LFI:slope:LFI26:Rad:S}{$-0.79$}
\setsymbol{LFI:slope:LFI27:Rad:S}{$-0.82$}
\setsymbol{LFI:slope:LFI28:Rad:S}{$-0.91$}

\setsymbol{LFI:FWHM:70GHz:units}{13\parcm01}
\setsymbol{LFI:FWHM:44GHz:units}{27\parcm92}
\setsymbol{LFI:FWHM:30GHz:units}{32\parcm65}

\setsymbol{LFI:FWHM:70GHz}{13.01}
\setsymbol{LFI:FWHM:44GHz}{27.92}
\setsymbol{LFI:FWHM:30GHz}{32.65}

\setsymbol{LFI:FWHM:LFI18:units}{13\parcm39}
\setsymbol{LFI:FWHM:LFI19:units}{13\parcm01}
\setsymbol{LFI:FWHM:LFI20:units}{12\parcm75}
\setsymbol{LFI:FWHM:LFI21:units}{12\parcm74}
\setsymbol{LFI:FWHM:LFI22:units}{12\parcm87}
\setsymbol{LFI:FWHM:LFI23:units}{13\parcm27}
\setsymbol{LFI:FWHM:LFI24:units}{22\parcm98}
\setsymbol{LFI:FWHM:LFI25:units}{30\parcm46}
\setsymbol{LFI:FWHM:LFI26:units}{30\parcm31}
\setsymbol{LFI:FWHM:LFI27:units}{32\parcm65}
\setsymbol{LFI:FWHM:LFI28:units}{32\parcm66}

\setsymbol{LFI:FWHM:LFI18}{13.39}
\setsymbol{LFI:FWHM:LFI19}{13.01}
\setsymbol{LFI:FWHM:LFI20}{12.75}
\setsymbol{LFI:FWHM:LFI21}{12.74}
\setsymbol{LFI:FWHM:LFI22}{12.87}
\setsymbol{LFI:FWHM:LFI23}{13.27}
\setsymbol{LFI:FWHM:LFI24}{22.98}
\setsymbol{LFI:FWHM:LFI25}{30.46}
\setsymbol{LFI:FWHM:LFI26}{30.31}
\setsymbol{LFI:FWHM:LFI27}{32.65}
\setsymbol{LFI:FWHM:LFI28}{32.66}

\setsymbol{LFI:FWHM:uncertainty:LFI18:units}{0.170\arcm}
\setsymbol{LFI:FWHM:uncertainty:LFI19:units}{0.174\arcm}
\setsymbol{LFI:FWHM:uncertainty:LFI20:units}{0.170\arcm}
\setsymbol{LFI:FWHM:uncertainty:LFI21:units}{0.156\arcm}
\setsymbol{LFI:FWHM:uncertainty:LFI22:units}{0.164\arcm}
\setsymbol{LFI:FWHM:uncertainty:LFI23:units}{0.171\arcm}
\setsymbol{LFI:FWHM:uncertainty:LFI24:units}{0.652\arcm}
\setsymbol{LFI:FWHM:uncertainty:LFI25:units}{1.075\arcm}
\setsymbol{LFI:FWHM:uncertainty:LFI26:units}{1.131\arcm}
\setsymbol{LFI:FWHM:uncertainty:LFI27:units}{1.266\arcm}
\setsymbol{LFI:FWHM:uncertainty:LFI28:units}{1.287\arcm}

\setsymbol{LFI:FWHM:uncertainty:LFI18}{0.170}
\setsymbol{LFI:FWHM:uncertainty:LFI19}{0.174}
\setsymbol{LFI:FWHM:uncertainty:LFI20}{0.170}
\setsymbol{LFI:FWHM:uncertainty:LFI21}{0.156}
\setsymbol{LFI:FWHM:uncertainty:LFI22}{0.164}
\setsymbol{LFI:FWHM:uncertainty:LFI23}{0.171}
\setsymbol{LFI:FWHM:uncertainty:LFI24}{0.652}
\setsymbol{LFI:FWHM:uncertainty:LFI25}{1.075}
\setsymbol{LFI:FWHM:uncertainty:LFI26}{1.131}
\setsymbol{LFI:FWHM:uncertainty:LFI27}{1.266}
\setsymbol{LFI:FWHM:uncertainty:LFI28}{1.287}

\setsymbol{HFI:center:frequency:100GHz:units}{100\,GHz}
\setsymbol{HFI:center:frequency:143GHz:units}{143\,GHz}
\setsymbol{HFI:center:frequency:217GHz:units}{217\,GHz}
\setsymbol{HFI:center:frequency:353GHz:units}{353\,GHz}
\setsymbol{HFI:center:frequency:545GHz:units}{545\,GHz}
\setsymbol{HFI:center:frequency:857GHz:units}{857\,GHz}

\setsymbol{HFI:center:frequency:100GHz}{100}
\setsymbol{HFI:center:frequency:143GHz}{143}
\setsymbol{HFI:center:frequency:217GHz}{217}
\setsymbol{HFI:center:frequency:353GHz}{353}
\setsymbol{HFI:center:frequency:545GHz}{545}
\setsymbol{HFI:center:frequency:857GHz}{857}

\setsymbol{HFI:Ndetectors:100GHz}{8}
\setsymbol{HFI:Ndetectors:143GHz}{11}
\setsymbol{HFI:Ndetectors:217GHz}{12}
\setsymbol{HFI:Ndetectors:353GHz}{12}
\setsymbol{HFI:Ndetectors:545GHz}{3}
\setsymbol{HFI:Ndetectors:857GHz}{4}

\setsymbol{HFI:FWHM:Maps:100GHz:units}{9\parcm88}
\setsymbol{HFI:FWHM:Maps:143GHz:units}{7\parcm18}
\setsymbol{HFI:FWHM:Maps:217GHz:units}{4\parcm87}
\setsymbol{HFI:FWHM:Maps:353GHz:units}{4\parcm65}
\setsymbol{HFI:FWHM:Maps:545GHz:units}{4\parcm72}
\setsymbol{HFI:FWHM:Maps:857GHz:units}{4\parcm39}
\setsymbol{HFI:FWHM:Maps:100GHz}{9.88}
\setsymbol{HFI:FWHM:Maps:143GHz}{7.18}
\setsymbol{HFI:FWHM:Maps:217GHz}{4.87}
\setsymbol{HFI:FWHM:Maps:353GHz}{4.65}
\setsymbol{HFI:FWHM:Maps:545GHz}{4.72}
\setsymbol{HFI:FWHM:Maps:857GHz}{4.39}

\setsymbol{HFI:beam:ellipticity:Maps:100GHz}{1.15}
\setsymbol{HFI:beam:ellipticity:Maps:143GHz}{1.01}
\setsymbol{HFI:beam:ellipticity:Maps:217GHz}{1.06}
\setsymbol{HFI:beam:ellipticity:Maps:353GHz}{1.05}
\setsymbol{HFI:beam:ellipticity:Maps:545GHz}{1.14}
\setsymbol{HFI:beam:ellipticity:Maps:857GHz}{1.19}

\setsymbol{HFI:FWHM:Mars:100GHz:units}{9\parcm37}
\setsymbol{HFI:FWHM:Mars:143GHz:units}{7\parcm04}
\setsymbol{HFI:FWHM:Mars:217GHz:units}{4\parcm68}
\setsymbol{HFI:FWHM:Mars:353GHz:units}{4\parcm43}
\setsymbol{HFI:FWHM:Mars:545GHz:units}{3\parcm80}
\setsymbol{HFI:FWHM:Mars:857GHz:units}{3\parcm67}

\setsymbol{HFI:FWHM:Mars:100GHz}{9.37}
\setsymbol{HFI:FWHM:Mars:143GHz}{7.04}
\setsymbol{HFI:FWHM:Mars:217GHz}{4.68}
\setsymbol{HFI:FWHM:Mars:353GHz}{4.43}
\setsymbol{HFI:FWHM:Mars:545GHz}{3.80}
\setsymbol{HFI:FWHM:Mars:857GHz}{3.67}

\setsymbol{HFI:beam:ellipticity:Mars:100GHz}{1.18}
\setsymbol{HFI:beam:ellipticity:Mars:143GHz}{1.03}
\setsymbol{HFI:beam:ellipticity:Mars:217GHz}{1.14}
\setsymbol{HFI:beam:ellipticity:Mars:353GHz}{1.09}
\setsymbol{HFI:beam:ellipticity:Mars:545GHz}{1.25}
\setsymbol{HFI:beam:ellipticity:Mars:857GHz}{1.03}

\setsymbol{HFI:CMB:relative:calibration:100GHz}{$\lsim 1\%$}
\setsymbol{HFI:CMB:relative:calibration:143GHz}{$\lsim 1\%$}
\setsymbol{HFI:CMB:relative:calibration:217GHz}{$\lsim 1\%$}
\setsymbol{HFI:CMB:relative:calibration:353GHz}{$\lsim 1\%$}
\setsymbol{HFI:CMB:relative:calibration:545GHz}{}
\setsymbol{HFI:CMB:relative:calibration:857GHz}{}

\setsymbol{HFI:CMB:absolute:calibration:100GHz}{$\lsim 2\%$}
\setsymbol{HFI:CMB:absolute:calibration:143GHz}{$\lsim 2\%$}
\setsymbol{HFI:CMB:absolute:calibration:217GHz}{$\lsim 2\%$}
\setsymbol{HFI:CMB:absolute:calibration:353GHz}{$\lsim 2\%$}
\setsymbol{HFI:CMB:absolute:calibration:545GHz}{}
\setsymbol{HFI:CMB:absolute:calibration:857GHz}{}

\setsymbol{HFI:FIRAS:gain:calibration:accuracy:statistical:100GHz}{}
\setsymbol{HFI:FIRAS:gain:calibration:accuracy:statistical:143GHz}{}
\setsymbol{HFI:FIRAS:gain:calibration:accuracy:statistical:217GHz}{}
\setsymbol{HFI:FIRAS:gain:calibration:accuracy:statistical:353GHz}{2.5\%}
\setsymbol{HFI:FIRAS:gain:calibration:accuracy:statistical:545GHz}{1\%}
\setsymbol{HFI:FIRAS:gain:calibration:accuracy:statistical:857GHz}{0.5\%}

\setsymbol{HFI:FIRAS:gain:calibration:accuracy:systematic:100GHz}{}
\setsymbol{HFI:FIRAS:gain:calibration:accuracy:systematic:143GHz}{}
\setsymbol{HFI:FIRAS:gain:calibration:accuracy:systematic:217GHz}{}
\setsymbol{HFI:FIRAS:gain:calibration:accuracy:systematic:353GHz}{}
\setsymbol{HFI:FIRAS:gain:calibration:accuracy:systematic:545GHz}{7\%}
\setsymbol{HFI:FIRAS:gain:calibration:accuracy:systematic:857GHz}{7\%}

\setsymbol{HFI:FIRAS:zero:point:accuracy:100GHz:units}{0.8\MJysr}
\setsymbol{HFI:FIRAS:zero:point:accuracy:143GHz:units}{}
\setsymbol{HFI:FIRAS:zero:point:accuracy:217GHz:units}{}
\setsymbol{HFI:FIRAS:zero:point:accuracy:353GHz:units}{1.4\MJysr}
\setsymbol{HFI:FIRAS:zero:point:accuracy:545GHz:units}{2.2\MJysr}
\setsymbol{HFI:FIRAS:zero:point:accuracy:857GHz:units}{1.7\MJysr}

\setsymbol{HFI:FIRAS:zero:point:accuracy:100GHz}{0.8}
\setsymbol{HFI:FIRAS:zero:point:accuracy:143GHz}{}
\setsymbol{HFI:FIRAS:zero:point:accuracy:217GHz}{}
\setsymbol{HFI:FIRAS:zero:point:accuracy:353GHz}{1.4}
\setsymbol{HFI:FIRAS:zero:point:accuracy:545GHz}{2.2}
\setsymbol{HFI:FIRAS:zero:point:accuracy:857GHz}{1.7}

\setsymbol{HFI:unit:conversion:100GHz:units}{0.2415\MJysrmK}
\setsymbol{HFI:unit:conversion:143GHz:units}{0.3694\MJysrmK}
\setsymbol{HFI:unit:conversion:217GHz:units}{0.4811\MJysrmK}
\setsymbol{HFI:unit:conversion:353GHz:units}{0.2883\MJysrmK}
\setsymbol{HFI:unit:conversion:545GHz:units}{0.05826\MJysrmK}
\setsymbol{HFI:unit:conversion:857GHz:units}{0.002238\MJysrmK}

\setsymbol{HFI:unit:conversion:100GHz}{0.2415}
\setsymbol{HFI:unit:conversion:143GHz}{0.3694}
\setsymbol{HFI:unit:conversion:217GHz}{0.4811}
\setsymbol{HFI:unit:conversion:353GHz}{0.2883}
\setsymbol{HFI:unit:conversion:545GHz}{0.05826}
\setsymbol{HFI:unit:conversion:857GHz}{0.002238}

\setsymbol{HFI:colour:correction:alpha=-2:V1.01:100GHz}{0.9893}
\setsymbol{HFI:colour:correction:alpha=-2:V1.01:143GHz}{0.9759}
\setsymbol{HFI:colour:correction:alpha=-2:V1.01:217GHz}{1.0007}
\setsymbol{HFI:colour:correction:alpha=-2:V1.01:353GHz}{1.0028}
\setsymbol{HFI:colour:correction:alpha=-2:V1.01:545GHz}{1.0019}
\setsymbol{HFI:colour:correction:alpha=-2:V1.01:857GHz}{0.9889}

\setsymbol{HFI:colour:correction:alpha=0:V1.01:100GHz}{1.0008}
\setsymbol{HFI:colour:correction:alpha=0:V1.01:143GHz}{1.0148}
\setsymbol{HFI:colour:correction:alpha=0:V1.01:217GHz}{0.9909}
\setsymbol{HFI:colour:correction:alpha=0:V1.01:353GHz}{0.9888}
\setsymbol{HFI:colour:correction:alpha=0:V1.01:545GHz}{0.9878}
\setsymbol{HFI:colour:correction:alpha=0:V1.01:857GHz}{1.0014}

\newcommand{\hi}{{\sc Hi}}
\newcommand{\poker}{{\small P{\tiny OKER}}}
\newcommand{\master}{{\small M{\tiny ASTER}}}

\newcommand{\lmmin}{$\log_{10}M_{\rm min}$ }

\definecolor{purple}{rgb}{0.60,0.3,0.9}

\newcommand{\healpix}{HEALPix}
\def\setsymbol#1#2{\expandafter\def\csname #1\endcsname{#2}} \def\getsymbol#1{\csname #1\endcsname}
\setsymbol{tabHEP}{3}

\begin{document} 
\author{\small
Planck Collaboration:
P.~A.~R.~Ade\inst{74}
\and
N.~Aghanim\inst{48}
\and
M.~Arnaud\inst{60}
\and
M.~Ashdown\inst{58, 4}
\and
J.~Aumont\inst{48}
\and
C.~Baccigalupi\inst{72}
\and
A.~Balbi\inst{30}
\and
A.~J.~Banday\inst{78, 8, 65}
\and
R.~B.~Barreiro\inst{55}
\and
J.~G.~Bartlett\inst{3, 56}
% ...
}
\institute{\small
Aalto University Mets\"{a}hovi Radio Observatory, Mets\"{a}hovintie 114, FIN-02540 Kylm\"{a}l\"{a}, Finland\\
\and
Agenzia Spaziale Italiana Science Data Center, c/o ESRIN, via Galileo Galilei, Frascati, Italy\\
\and
Astroparticule et Cosmologie, CNRS (UMR7164), Universit\'{e} Denis Diderot Paris 7, B\^{a}timent Condorcet, 10 rue A. Domon et L\'{e}onie Duquet, Paris, France\\
\and
Astrophysics Group, Cavendish Laboratory, University of Cambridge, J J Thomson Avenue, Cambridge CB3 0HE, U.K.\\
\and
Atacama Large Millimeter/submillimeter Array, ALMA Santiago Central Offices, Alonso de Cordova 3107, Vitacura, Casilla 763 0355, Santiago, Chile\\
\and
Australia Telescope National Facility, CSIRO, P.O. Box 76, Epping, NSW 1710, Australia\\
\and
CITA, University of Toronto, 60 St. George St., Toronto, ON M5S 3H8, Canada\\
\and
CNRS, IRAP, 9 Av. colonel Roche, BP 44346, F-31028 Toulouse cedex 4, France\\
\and
California Institute of Technology, Pasadena, California, U.S.A.\\
\and
Centre of Mathematics for Applications, University of Oslo, Blindern, Oslo, Norway\\
\and
DAMTP, University of Cambridge, Centre for Mathematical Sciences, Wilberforce Road, Cambridge CB3 0WA, U.K.\\
\and
DSM/Irfu/SPP, CEA-Saclay, F-91191 Gif-sur-Yvette Cedex, France\\
\and
DTU Space, National Space Institute, Juliane Mariesvej 30, Copenhagen, Denmark\\
\and
Departamento de F\'{\i}sica, Universidad de Oviedo, Avda. Calvo Sotelo s/n, Oviedo, Spain\\
\and
Department of Astronomy and Astrophysics, University of Toronto, 50 Saint George Street, Toronto, Ontario, Canada\\
\and
Department of Physics \& Astronomy, University of British Columbia, 6224 Agricultural Road, Vancouver, British Columbia, Canada\\
\and
Department of Physics and Astronomy, University of Sussex, Brighton BN1 9QH, U.K.\\
\and
Department of Physics, Gustaf H\"{a}llstr\"{o}min katu 2a, University of Helsinki, Helsinki, Finland\\
\and
Department of Physics, Princeton University, Princeton, New Jersey, U.S.A.\\
\and
Department of Physics, Purdue University, 525 Northwestern Avenue, West Lafayette, Indiana, U.S.A.\\
\and
Department of Physics, University of California, Berkeley, California, U.S.A.\\
\and
Department of Physics, University of California, One Shields Avenue, Davis, California, U.S.A.\\
\and
Department of Physics, University of California, Santa Barbara, California, U.S.A.\\
\and
Department of Physics, University of Illinois at Urbana-Champaign, 1110 West Green Street, Urbana, Illinois, U.S.A.\\
\and
Dipartimento di Fisica G. Galilei, Universit\`{a} degli Studi di Padova, via Marzolo 8, 35131 Padova, Italy\\
\and
Dipartimento di Fisica, Universit\`{a} La Sapienza, P. le A. Moro 2, Roma, Italy\\
\and
Dipartimento di Fisica, Universit\`{a} degli Studi di Milano, Via Celoria, 16, Milano, Italy\\
\and
Dipartimento di Fisica, Universit\`{a} degli Studi di Trieste, via A. Valerio 2, Trieste, Italy\\
\and
Dipartimento di Fisica, Universit\`{a} di Ferrara, Via Saragat 1, 44122 Ferrara, Italy\\
\and
Dipartimento di Fisica, Universit\`{a} di Roma Tor Vergata, Via della Ricerca Scientifica, 1, Roma, Italy\\
\and
Discovery Center, Niels Bohr Institute, Blegdamsvej 17, Copenhagen, Denmark\\
\and
Dpto. Astrof\'{i}sica, Universidad de La Laguna (ULL), E-38206 La Laguna, Tenerife, Spain\\
\and
European Southern Observatory, ESO Vitacura, Alonso de Cordova 3107, Vitacura, Casilla 19001, Santiago, Chile\\
\and
European Space Agency, ESAC, Planck Science Office, Camino bajo del Castillo, s/n, Urbanizaci\'{o}n Villafranca del Castillo, Villanueva de la Ca\~{n}ada, Madrid, Spain\\
\and
European Space Agency, ESTEC, Keplerlaan 1, 2201 AZ Noordwijk, The Netherlands\\
\and
Helsinki Institute of Physics, Gustaf H\"{a}llstr\"{o}min katu 2, University of Helsinki, Helsinki, Finland\\
\and
INAF - Osservatorio Astrofisico di Catania, Via S. Sofia 78, Catania, Italy\\
\and
INAF - Osservatorio Astronomico di Padova, Vicolo dell'Osservatorio 5, Padova, Italy\\
\and
INAF - Osservatorio Astronomico di Roma, via di Frascati 33, Monte Porzio Catone, Italy\\
\and
INAF - Osservatorio Astronomico di Trieste, Via G.B. Tiepolo 11, Trieste, Italy\\
\and
INAF/IASF Bologna, Via Gobetti 101, Bologna, Italy\\
\and
INAF/IASF Milano, Via E. Bassini 15, Milano, Italy\\
\and
INRIA, Laboratoire de Recherche en Informatique, Universit\'{e} Paris-Sud 11, B\^{a}timent 490, 91405 Orsay Cedex, France\\
\and
IPAG: Institut de Plan\'{e}tologie et d'Astrophysique de Grenoble, Universit\'{e} Joseph Fourier, Grenoble 1 / CNRS-INSU, UMR 5274, Grenoble, F-38041, France\\
\and
Imperial College London, Astrophysics group, Blackett Laboratory, Prince Consort Road, London, SW7 2AZ, U.K.\\
\and
Infrared Processing and Analysis Center, California Institute of Technology, Pasadena, CA 91125, U.S.A.\\
\and
Institut N\'{e}el, CNRS, Universit\'{e} Joseph Fourier Grenoble I, 25 rue des Martyrs, Grenoble, France\\
\and
Institut d'Astrophysique Spatiale, CNRS (UMR8617) Universit\'{e} Paris-Sud 11, B\^{a}timent 121, Orsay, France\\
\and
Institut d'Astrophysique de Paris, CNRS UMR7095, Universit\'{e} Pierre \& Marie Curie, 98 bis boulevard Arago, Paris, France\\
\and
Institut de Ci\`{e}ncies de l'Espai, CSIC/IEEC, Facultat de Ci\`{e}ncies, Campus UAB, Torre C5 par-2, Bellaterra 08193, Spain\\
\and
Institute of Astronomy and Astrophysics, Academia Sinica, Taipei, Taiwan\\
\and
Institute of Astronomy, University of Cambridge, Madingley Road, Cambridge CB3 0HA, U.K.\\
\and
Institute of Theoretical Astrophysics, University of Oslo, Blindern, Oslo, Norway\\
\and
Instituto de Astrof\'{\i}sica de Canarias, C/V\'{\i}a L\'{a}ctea s/n, La Laguna, Tenerife, Spain\\
\and
Instituto de F\'{\i}sica de Cantabria (CSIC-Universidad de Cantabria), Avda. de los Castros s/n, Santander, Spain\\
\and
Jet Propulsion Laboratory, California Institute of Technology, 4800 Oak Grove Drive, Pasadena, California, U.S.A.\\
\and
Jodrell Bank Centre for Astrophysics, Alan Turing Building, School of Physics and Astronomy, The University of Manchester, Oxford Road, Manchester, M13 9PL, U.K.\\
\and
Kavli Institute for Cosmology Cambridge, Madingley Road, Cambridge, CB3 0HA, U.K.\\
\and
LERMA, CNRS, Observatoire de Paris, 61 Avenue de l'Observatoire, Paris, France\\
\and
Laboratoire AIM, IRFU/Service d'Astrophysique - CEA/DSM - CNRS - Universit\'{e} Paris Diderot, B\^{a}t. 709, CEA-Saclay, F-91191 Gif-sur-Yvette Cedex, France\\
\and
Laboratoire Traitement et Communication de l'Information, CNRS (UMR 5141) and T\'{e}l\'{e}com ParisTech, 46 rue Barrault F-75634 Paris Cedex 13, France\\
\and
Laboratoire de Physique Subatomique et de Cosmologie, CNRS/IN2P3, Universit\'{e} Joseph Fourier Grenoble I, Institut National Polytechnique de Grenoble, 53 rue des Martyrs, 38026 Grenoble cedex, France\\
\and
Laboratoire de l'Acc\'{e}l\'{e}rateur Lin\'{e}aire, Universit\'{e} Paris-Sud 11, CNRS/IN2P3, Orsay, France\\
\and
Lawrence Berkeley National Laboratory, Berkeley, California, U.S.A.\\
\and
Max-Planck-Institut f\"{u}r Astrophysik, Karl-Schwarzschild-Str. 1, 85741 Garching, Germany\\
\and
MilliLab, VTT Technical Research Centre of Finland, Tietotie 3, Espoo, Finland\\
\and
NRAO, P.O. Box 2, Rt 28/92, Green Bank, WV 24944-0002, U.S.A.\\
\and
National University of Ireland, Department of Experimental Physics, Maynooth, Co. Kildare, Ireland\\
\and
Niels Bohr Institute, Blegdamsvej 17, Copenhagen, Denmark\\
\and
Observational Cosmology, Mail Stop 367-17, California Institute of Technology, Pasadena, CA, 91125, U.S.A.\\
\and
Optical Science Laboratory, University College London, Gower Street, London, U.K.\\
\and
SISSA, Astrophysics Sector, via Bonomea 265, 34136, Trieste, Italy\\
\and
SUPA, Institute for Astronomy, University of Edinburgh, Royal Observatory, Blackford Hill, Edinburgh EH9 3HJ, U.K.\\
\and
School of Physics and Astronomy, Cardiff University, Queens Buildings, The Parade, Cardiff, CF24 3AA, U.K.\\
\and
Space Research Institute (IKI), Russian Academy of Sciences, Profsoyuznaya Str, 84/32, Moscow, 117997, Russia\\
\and
Space Sciences Laboratory, University of California, Berkeley, California, U.S.A.\\
\and
Stanford University, Dept of Physics, Varian Physics Bldg, 382 Via Pueblo Mall, Stanford, California, U.S.A.\\
\and
Universit\'{e} de Toulouse, UPS-OMP, IRAP, F-31028 Toulouse cedex 4, France\\
% ...
}

\title{\textit{Planck} Early Results XVIII: The power spectrum of cosmic infrared background anisotropies}

\date{Received 10 January 2011/Accepted 17 June 2011}

\abstract{Using \Planck\ maps of six regions of low Galactic dust emission with a total area of about 140 ${\rm deg}^2$, we determine the angular power spectra of cosmic infrared background (CIB) anisotropies from multipole $\ell = 200$ to $\ell = 2000$ at 217, 353, 545 and 857\,GHz.  We use 21-cm observations of \hi\  as a tracer of thermal dust emission to reduce the already low level of Galactic dust emission and use the 143\,GHz \Planck\ maps in these fields to clean out cosmic microwave background anisotropies.  Both of these cleaning processes are necessary to avoid significant contamination of the CIB signal.  We measure correlated CIB structure across frequencies. As expected, the correlation decreases with increasing frequency separation, because the contribution of high-redshift galaxies to CIB anisotropies increases with wavelengths.  We find no significant difference between the frequency spectrum of the CIB anisotropies and the CIB mean, with $\Delta I / I$=15\% from 217 to 857\,GHz. In terms of clustering properties, the \Planck\ data alone rule out the linear scale- and redshift-independent bias model. Non-linear corrections are significant. Consequently, we develop an alternative model that couples a dusty galaxy, parametric evolution model with a simple halo-model approach. It provides an excellent fit to the measured anisotropy angular power spectra and suggests that a different halo occupation distribution is required at each frequency, which is consistent with our expectation that each frequency is dominated by contributions from different redshifts. In our best-fit model, half of the anisotropy power at $\ell$=2000 comes from redshifts $z<0.8$ at 857\,GHz and $z<1.5$ at 545\,GHz, while about  90\% come from redshifts $z>$2 at 353 and 217\,GHz, respectively.
}

\authorrunning{Planck Collaboration}
\titlerunning{CIB anisotropies with \textit{Planck}}

\maketitle

\allearlypapers

\section{Introduction}

In addition to instrument noise, deep cosmological surveys in the far-infrared to millimeter spectral range are limited
in depth by confusion from extragalactic sources \citep[e.g.][]{blain1998,lagache2003,dole2004,fernandez-conde2008,nguyen2010}. 
This limitation arises from the high density of faint, distant galaxies that produce signal fluctuations within the telescope beam. As a consequence, the cosmic infrared background (CIB), which records much of the radiant energy released by processes of structure formation that have occurred since the decoupling of matter and radiation following the Big Bang
\citep{puget1996, hauser2001, dole2006},
is barely resolved into its constituents.  Indeed, less than 10\% of the CIB is resolved by {\it Spitzer} at 160~$\mu$m  \citep{bether2010}, about 10\% by {\it Herschel} at 350~$\mu$m \citep{oliver2010} and a negligible fraction
is resolved by {\it Planck}\footnote{\Planck\ (http://www.esa.int/\Planck ) is a project of the European Space Agency (ESA) with instruments provided by two scientific consortia funded by ESA member states (in particular the lead countries France and Italy), with contributions from NASA (USA) and telescope reflectors provided by a collaboration between ESA and a scientific consortium led and funded by Denmark.}
\citep{fernandez-conde2008}.
Thus, in the absence of foreground (Galactic dust) and background (cosmic microwave background, CMB) emissions, and when the instrument noise is subdominant, maps of the diffuse emission at the angular resolution probed by the current surveys reveal a web of structures, characteristic of CIB anisotropies. 
With the advent of large area far-infrared to millimeter surveys ({\it Herschel}, {\it Planck}, SPT and ACT), CIB anisotropies constitute a new tool for structure formation and evolution study.

Cosmic infrared background anisotropies are expected to trace large-scale structures and probe the clustering properties of galaxies, which
in turn are linked to those of their hosting dark matter halos. Because the clustering of dark matter is well understood, observations of anisotropies in the CIB constrain the relationship between dusty, star-forming galaxies and the dark matter distribution.
The connection between a population of galaxies and dark matter halos can be described by its halo occupation distribution
\citep[HOD;][]{PeaSmi00,Sel00,Ben00,WHS01,BerWei02,Cooray:2002dia},
which specifies the probability distribution of the number of objects with a given property (e.g., luminosity, stellar mass, or star-formation rate) within a dark matter halo of a given mass and their radial distribution within the halo.
The HOD and the halo model provide a powerful theoretical framework for describing the connection between galaxies and dark matter halos.  Once decisions are made about which properties of the halos and their environment the HOD
depends upon, what the moments of the HOD are and what the radial profile of objects within halos is, the halo model
can be used to predict any clustering-related observable.  In particular, the halo model predicts that the bias,
describing the clustering of galaxies in relation to the dark matter, becomes scale-independent at large scales.
This assumption of a scale-independent bias is often made in modelling the CIB.

The way galaxies populate dark matter halos is not the only ingredient that enters into the CIB anisotropy modelling.
Correlated anisotropies also depend on the mean emissivity per comoving unit volume of dusty, star-forming galaxies, that results from dusty galaxies evolution models. Such models are more and more constrained thanks to the increasing number of observations (mainly galaxies number counts and luminosity functions), but remain largely empirical.  So far, CIB anisotropy models have combined (i) a scale-independent bias clustering with a very simple emissivity model based on the CIB mean \citep{knox2001,hall2010} or an empirical model of dusty galaxy evolution \citep{lagache2007} or the predictions of the physical model by \cite{granato2004} for the formation and evolution of spheroidal galaxies \citep{negrello2007}; (ii) a HOD with the \cite{lagache2003} dusty galaxies evolution model \citep{amblard2007,viero2009}; and (iii) a merger model of dark matter halos with a very simple dust evolution model \citep{righi2008}.\\

The angular power spectrum of CIB anisotropies has two contributions, a white-noise component caused by shot noise
and an additional component caused by spatial correlations between the sources of the CIB.
Correlated CIB anisotropies have been measured at 3330~GHz by {\it AKARI\/} \citep{matsuura2010}, 3000~GHz by IRAS/IRIS \citep{penin2011a}, 1875~GHz by {\it Spitzer\/} \citep{lagache2007, grossan2007}, 1200, 857, 600~GHz by BLAST and SPIRE \citep{viero2009, amblard2011}, and 220~GHz by SPT \citep{hall2010} and ACT \citep{dunkley2010}.
Depending on the frequency, the angular resolution and size of the survey these measurements can probe two different clustering regimes.  On small angular scales ($\ell \ge 2000$), they measure the clustering within a single dark matter halo and accordingly the physics governing how dusty, star-forming galaxies form within a halo.
On large angular scales, CIB anisotropies measure clustering between galaxies in different dark matter halos.
These measurements primarily constrain the large-scale, linear bias, $b$, of dusty galaxies, which is usually assumed to be
scale-independent over the relevant range.
Given their limited dynamic range in scale, current measurements are equally consistent with an HOD model, a power-law correlation function or a scale-independent, linear bias.  All models return a value for the large-scale bias that is 2--4 times higher than that measured for local, dusty, star-forming galaxies (where $b\simeq 1$).

Owing to its frequency coverage from 100 to $857\,$GHz, the HFI instrument on-board \Planck\ is ideally suited to probe the dark matter -- star-formation connection.  \Planck\ \citep{tauber2010a, planck2011-1.1} is the third-generation space mission to measure the anisotropy of the cosmic microwave background (CMB).  It observes the sky in nine frequency bands covering 30--857\,GHz with high sensitivity and angular resolution from 31\arcm\ to 5\arcm.  The Low Frequency Instrument (LFI; \citealt{Mandolesi2010, Bersanelli2010, planck2011-1.4}) covers the 30, 44, and 70\,GHz bands with amplifiers cooled to 20\,\hbox{K}.  The High Frequency Instrument (HFI; \citealt{Lamarre2010, planck2011-1.5}) covers the 100, 143, 217, 353, 545, and 857\,GHz bands with bolometers cooled to 0.1\,\hbox{K}.  Polarization is measured in all but the highest two bands \citep{Leahy2010, Rosset2010}.  A combination of radiative cooling and three mechanical coolers produces the temperatures needed for the detectors and optics \citep{planck2011-1.3}.  Two data processing centres (DPCs) check and calibrate the data and make maps of the sky \citep{planck2011-1.5, planck2011-1.6}.  \Planck's sensitivity, angular resolution, and frequency coverage make it a powerful instrument for Galactic and extragalactic astrophysics as well as cosmology.  Early results are given in Planck Collaboration (2011a--u).

\begin{table*}
\begin{center}
\begin{tabular}{c|c|c|c|c|c} 
Field & Galactic Longitude & Galactic Latitude & Size & Mean $N$(\hi)  & $\sigma$ $N$(\hi)\\ 
& degrees & degrees & arcmin$\times$arcmin & 10$^{20}$ cm$^{-2}$ & 10$^{20}$ cm$^{-2}$ \\ \hline
N1 & 85.33     &  44.28 & 308$\times$308 &  1.2 & 0.3\\ 
AG & 164.84 &  65.50 & 308$\times$308  & 1.8 & 0.6 \\
SP & 132.37    &  47.50 & 308$\times$308  & 1.2 & 0.3 \\
LH2 & 152.38 & 53.30 & 241.5$\times$241.5  & 0.7 & 0.2\\
Bootes 1 & 61.29  & 72.32 & 283.5$\times$283.5 & 1.2 & 0.2 \\
Bootes 2 & 58.02 & 68.42 & 283.5$\times$283.5  & 1.1 & 0.2 \\ 
\end{tabular}\\
\caption{CIB field description: centre (Galactic coordinates), size, mean and dispersion of \hi\ column density. \label{tab:Fields}}
\end{center}
\end{table*}

\begin{figure*}
\begin{center}
\includegraphics[width=8cm]{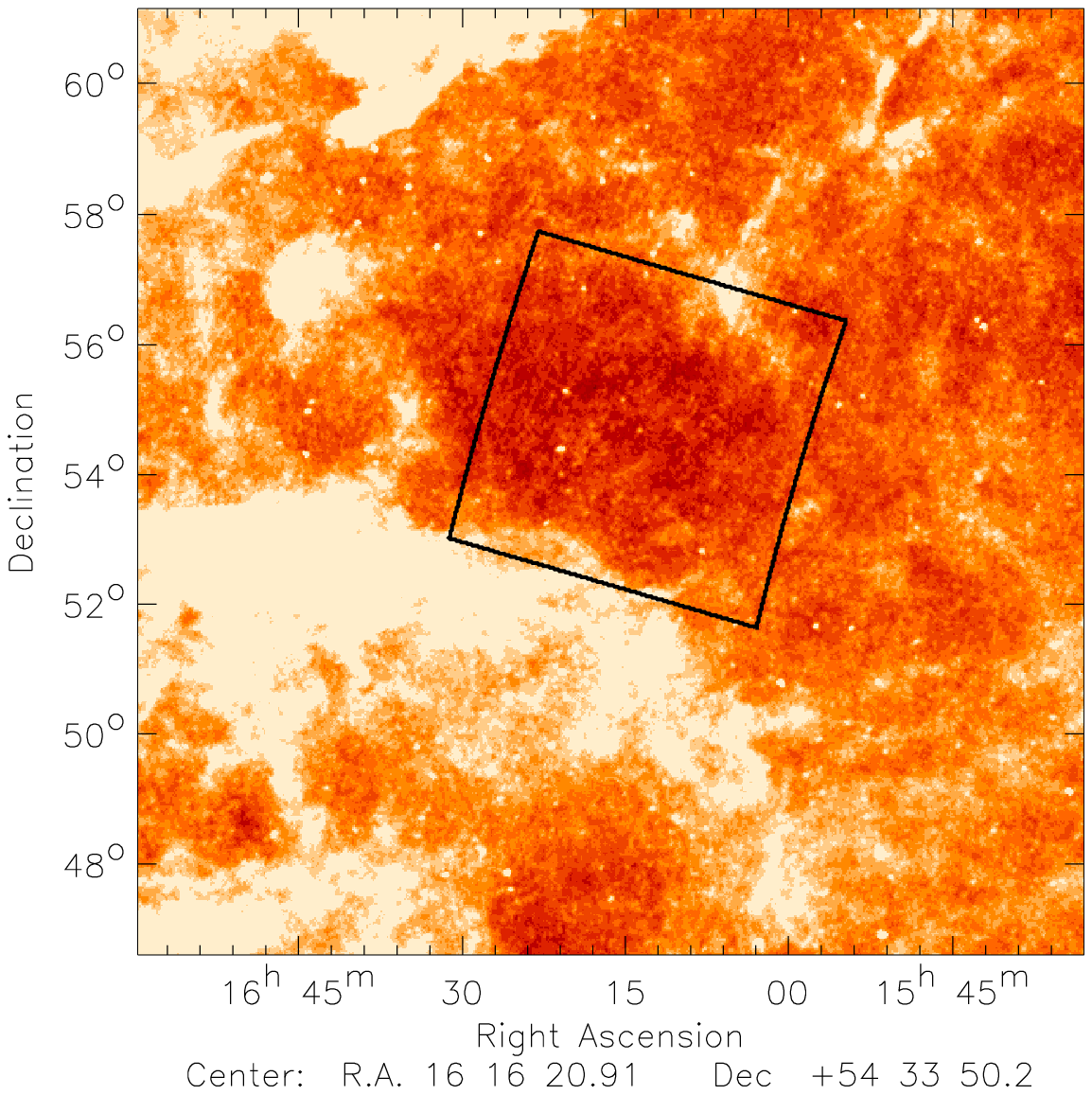}
\includegraphics[width=8cm]{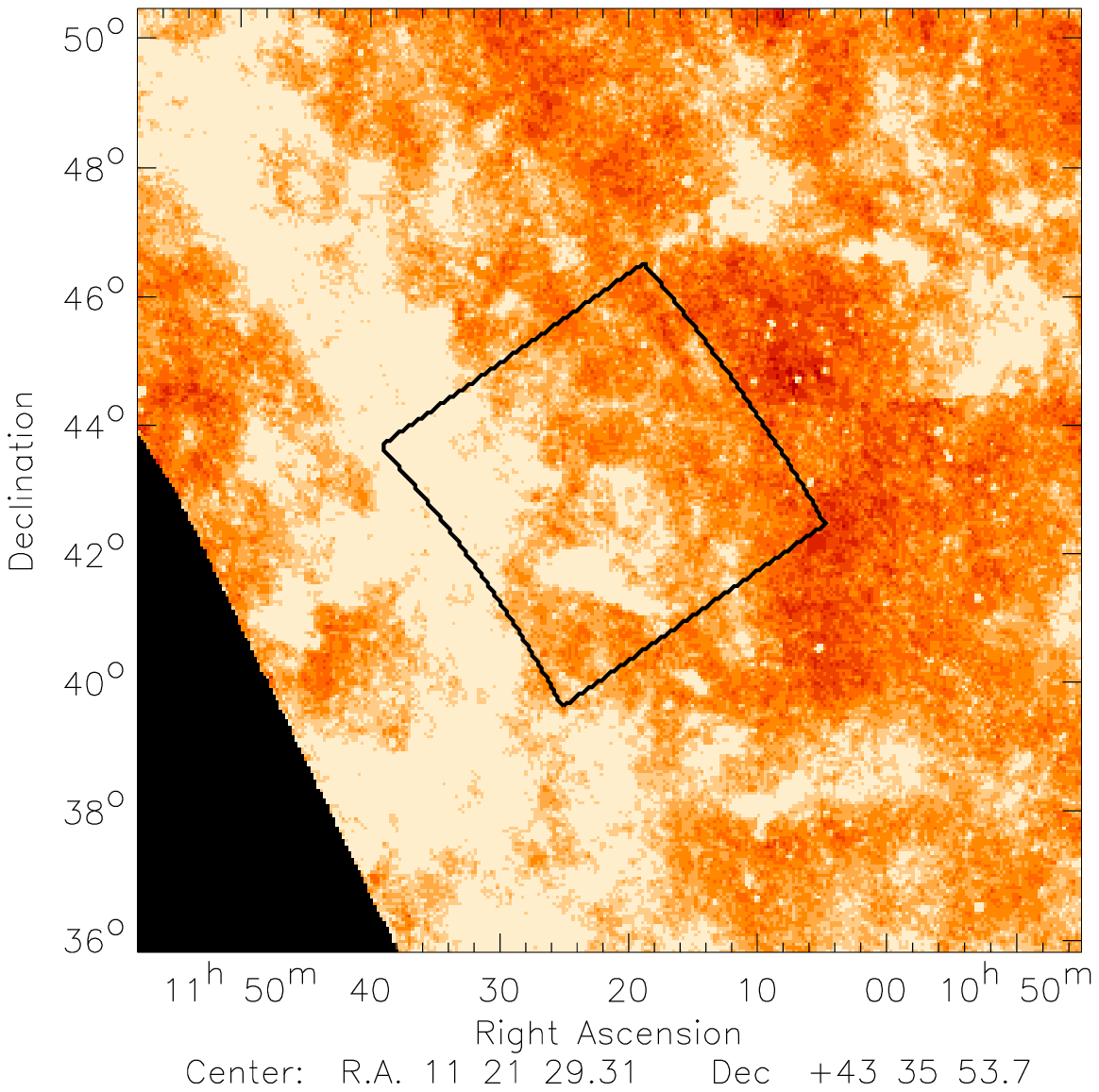}
\includegraphics[width=8cm]{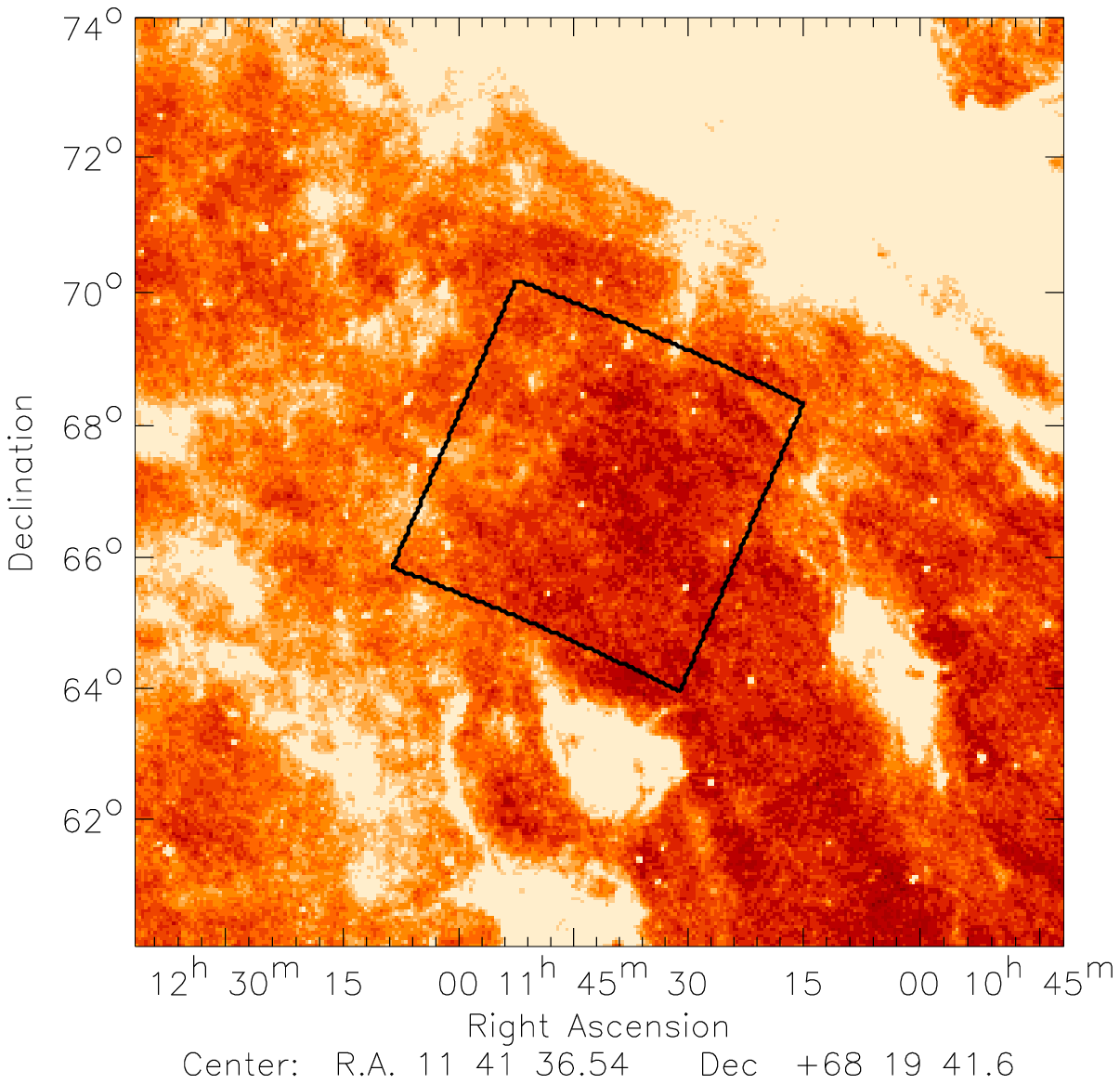}
\includegraphics[width=8cm]{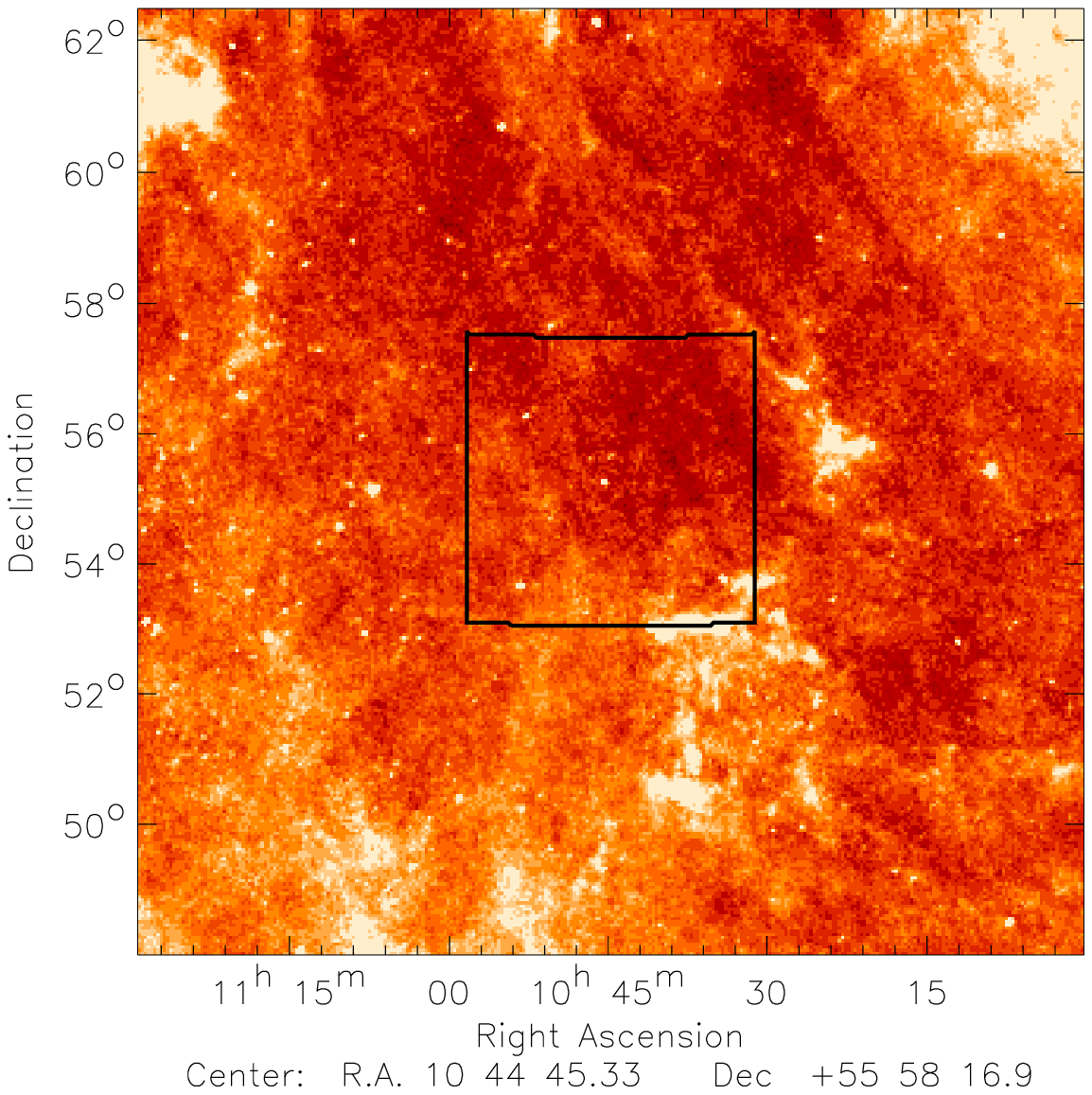}
\includegraphics[width=9cm]{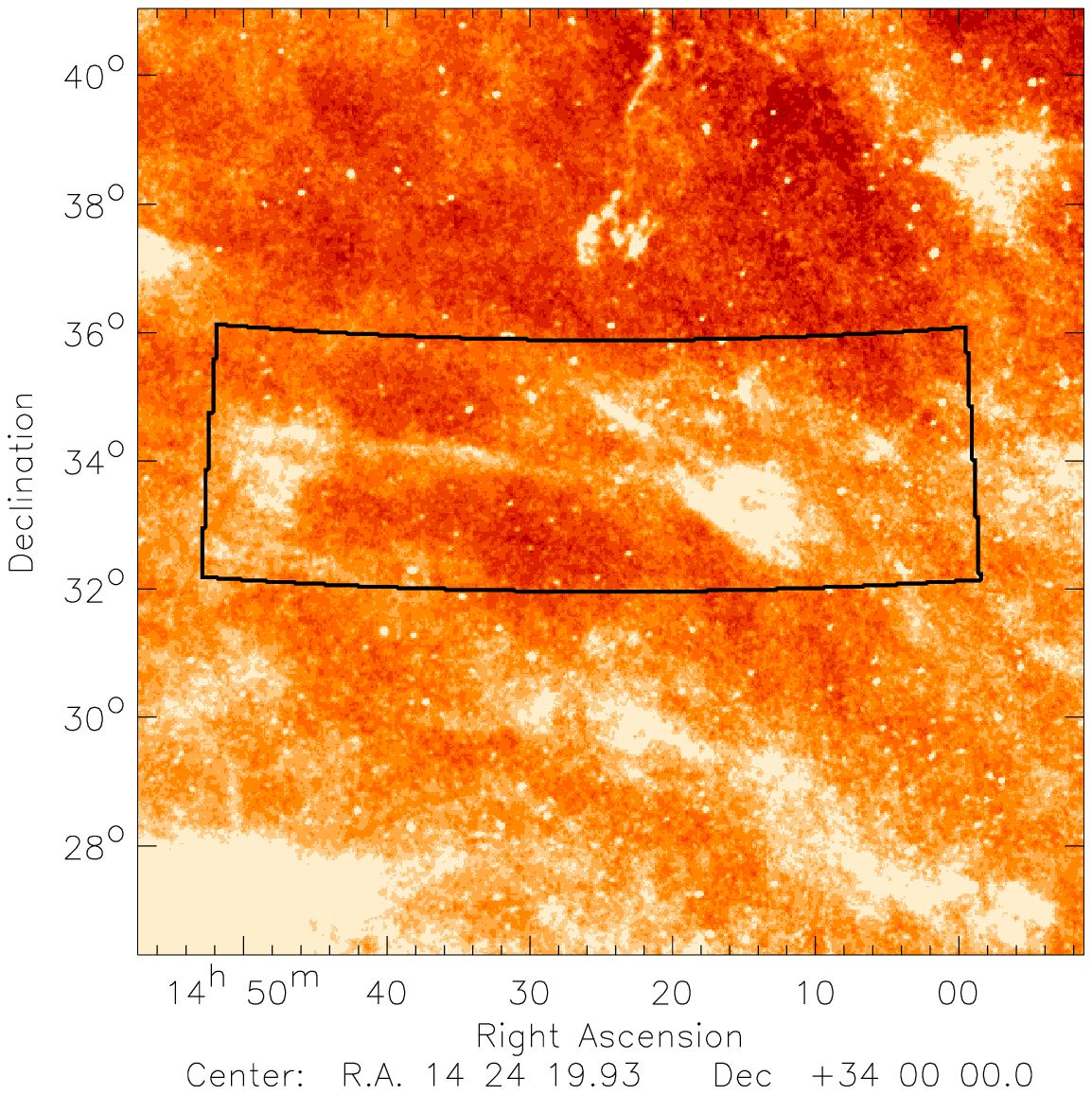}
\end{center}
\caption{\label{fig:overlay_IRIS} From {\it left} to {\it right} and {\it top} to {\it bottom}: {\tt N1, AG, SP, LH2} and {\tt bootes} fields overlaid on IRIS $100\, \mu$m map
\citep{miville2005}. Fields {\tt Bootes 1} and {\tt 2} are both included in the large rectangle. All IRIS images have the same dynamic range, with a linear colour scale ranging from dark red to white from 0 to 2 MJy\,sr$^{-1}$.} 
\end{figure*}

The primary objective of this paper is to measure with \Planck\ HFI the CIB anisotropies caused by the clustering of star-forming galaxies. To achieve this, we analyse small regions of sky with a total area of about 140\,${\rm deg^2}$, where we are able to cleanly separate the foreground (Galactic cirrus) and background (CMB) components from the signal.
Unlike previous CIB anisotropy studies \citep[but see][]{penin2011a}, we do not remove the cirrus by fitting a power-law power spectra at large scales, but use an independent, external tracer of diffuse dust emission (the \hi\ gas).
We accurately measure the instrumental contributions (noise, beam) to the power spectra of CIB anisotropies and use a dedicated optimal method to measure power spectra \citep{poker}.
All these steps allow us to recover for the first time the power spectra of CIB anisotropies from $200\le \ell \le2000$ at four frequencies simultaneously: 217, 353, 545 and 857\,GHz. 

The paper is organized as follows: in Sect. \ref{sect_2} we present the data we are using, the field selection and the removal of foreground and background components (CMB, Galactic cirrus, bright point sources) from the CIB.
In Sect.~\ref{compo_pk} we discuss the different contributions to the power spectra of the residual maps.
Section~\ref{ang_pow} describes how we estimated the power spectrum, its bias, and errors.
Our main results are presented in Sect.~\ref{sect_5}. This section also describes our modelling and discusses the clustering of high-redshift, dusty galaxies. We conclude in Sect.~\ref{sect_cl}. In the appendices we show two flow charts summarizing the data processing and cleaning, and the power spectra measurements (Appendix \ref{A1}), and we give some details about the dusty star-forming galaxy evolution model (Appendix \ref{A2}) and the halo model (Appendix \ref{A3}) we are using.
Throughout the paper we use the WMAP7 cosmological parameters for standard $\Lambda$CDM cosmology \citep{larson2010}.

\section{Selected fields and data cleaning \label{sect_2}}

\subsection{{\it Planck} data}\label{data}

We used \Planck\ channel maps of the HFI at 5 frequencies: 143, 217, 353, 545 and 857\,GHz. Their characteristics and how they were created is 
described in detail in the companion paper on HFI early processing \citep[hereafter HEP;][]{planck2011-1.5}.
In summary, the channel maps correspond to temperature observations for the two first sky surveys by \Planck.
The data are organized as time-ordered information, hereafter TOI.
The attitude of the satellite as a function of time is provided by two star trackers on the spacecraft.
The pointing for each bolometer is computed by combining the attitude with the location of the bolometer
in the focal plane, as determined by planet observations (see below).
Time-ordered informations of raw bolometer data are first processed to produce cleaned timelines and to set flags to mark data we do not
currently fit.  This TOI processing includes (1) signal demodulation and filtering, (2) deglitching, which flags the strong part of any glitch and subtracts the tails, (3)  conversion from instrumental units (volts) to physical units (watts of absorbed power, after a correction for the weak non-linearity of the response), (4) decorrelation of thermal stage fluctuations, (5) removal of the systematic effects induced by 4\,K cooler mechanical vibrations, and (6) deconvolution of the  bolometer time constant.
Focal plane reconstruction and beam shape estimation are made using observations of Mars.
The simplest description of the  beams, an elliptical Gaussian, leads to full-width at half-maximum (FWHM) values, $\theta_{S}$, given in Table~\getsymbol{tabHEP} of the HEP ({\it i.e.\/},  $7.08\arcm, 4.71\arcm, 4.50\arcm, 4.72\arcm$
and $4.42\arcm$ from 143 to 857\,GHz, with an uncertainty between 0.12\arcm and 0.28\arcm).
From the cleaned TOI and the pointing, channel maps were computed using all the bolometers at a given frequency. The path from TOI to maps in the HFI DPC is schematically divided into three steps: ring-making, ring offset estimation, and map-making.  The first step combines the data within a stable pointing period, during which the same circle on the sky is scanned repeatedly to create rings with higher signal-to-noise ratio,  taking full advantage of the redundancy of observations provided by the \Planck\ scanning strategy.  The low-frequency component of the noise is accounted for in a second step by using a destriping technique that models this component as an offset of the ring values. Finally, cleaned maps are produced by coadding the offset-corrected rings.  The maps are produced in Galactic coordinates, using the \healpix\ pixelisation scheme (see http://healpix.jpl.nasa.gov and \citet{gorski2005}).
Photometric calibration is performed either at ring level (using the CMB dipole) for the lower frequency channels or at the map level (using FIRAS data) for the  higher frequency channels (545 and 857\,GHz).
The absolute gain calibration of the HFI \Planck\ maps is known to better than 2\% for the lower frequencies (143, 217 and 353\,GHz) and 7\% for the higher frequencies (545 and 857\,GHz),  as summarised in the HEP
Table~\getsymbol{tabHEP}.
Inter-calibration accuracy between channels is better than absolute calibration.\\

We made use of the so-called DX4 HFI data release, a dataset from which the CMB has not been removed.  We used the 217, 353, 545, and 857\,GHz channels for the CIB analysis and the 143\,GHz channel for CMB removal.
Maps are given either in MJy\,sr${}^{-1}$ (with the photometric convention $\nu I_{\nu}$=constant\footnote{The convention $\nu I_{\nu}$=constant means that the MJy\,sr${}^{-1}$ are given for a source with a spectral energy distribution $I_{\nu}\propto \nu^{-1}$.  For a source with a different spectral energy distribution a colour correction has to be applied (see \citealt{planck2011-1.5}).}) or $\mu$K$_{\rm CMB}$, the conversion between the two was exactly computed using the bandpass filters \citep[see][]{planck2011-1.5}.

\begin{figure}
\begin{center}
\includegraphics[width=7.3cm, draft=false]{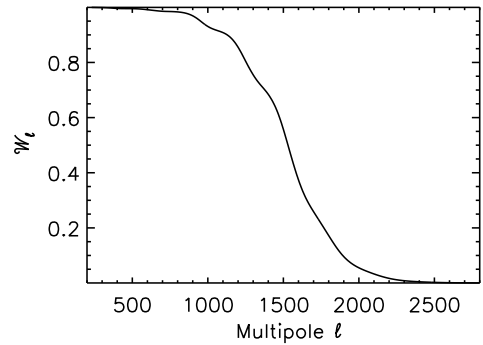}
\caption{\label{fig:wienerfilter} Wiener filter applied to the 143\,GHz map for CMB subtraction. The filter essentially cuts out high multipoles where the CMB-to-noise ratio of the 143\,GHz map is low. Whereas this filter has to be known for estimating and subtracting the contribution of the residual CMB and 143\,GHz instrument noise to the power spectrum of CMB-cleaned channels, the exact value of the filter is not really critical.}
\end{center}
\end{figure}

\subsection{Extragalactic fields with high angular resolution \hi\ data}

Although \Planck\ is an all-sky survey, we restricted our first CIB anisotropy measurements to a few fields at high Galactic latitude to minimize the Galactic dust contamination. The choice of the fields was driven by the availability of \hi\ data at an angular resolution close to HFI.

The 21-cm \hi\ spectra used here were obtained with the 100-m Green Bank Telescope (GBT) over the period 2005 to 2010.  Details of this high-latitude survey are presented by Martin et al. (in prep). The total area mapped is about 825\,${\rm deg^2}$.

The spectra were taken with on-the-fly mapping.  The primary beam of the GBT at 21\,cm has a FWHM of 9.1\arcmin,
and the integration time (4\,s) and telescope scan rate were chosen to sample every 3.5\arcmin, more finely than the Nyquist interval, 3.86\arcmin.  The beam is only slightly broadened to 9.4\arcmin\ in the in-scan direction.  Scans were made moving the telescope in one direction (galactic longitude or right ascension), with steps of 3.5\arcmin\ in the orthogonal coordinate direction before the subsequent reverse scan.

Data were recorded with the GBT spectrometer by in-band frequency switching, yielding spectra with a velocity coverage $-450 \leq V_{\rm LSR} \leq +355\,$\kms\ at a resolution of 0.80\,\kms.  Spectra were calibrated, corrected for stray radiation, and placed on a brightness temperature ($T_b$) scale as described in \citet{blagrave2010},
Boothroyd et al. (in prep), and Martin et al. (in prep).
A third-order polynomial was fitted to the emission-free regions of the spectra to remove any residual instrumental baseline.  The spectra were gridded on the natural GLS (SFL) projection to produce a data cube.  Some regions were mapped two or three times. With the broad spectral coverage, all \hi\ components from local gas to high-velocity clouds are accessible.

We selected from this GBT cirrus survey the six faintest fields in terms of \hi\ column densities. Their main characteristics are given in Table \ref{tab:Fields} and the IRAS $100\,\mu$m maps are shown in Fig. \ref{fig:overlay_IRIS}.
They all have very low dust contamination and consequently \hi\ column densities, including the faintest all-sky sight line (referenced as LH2 in the Table).
The field areas are between 16 to 25\,${\rm deg^2}$ for a total coverage of about 140\,${\rm deg^2}$. Going to higher average \hi\ column densities (N(\hi)$>2.5\times10^{20}$ cm$^{-2}$) is not recommended because dust emission associated with molecular gas starts to contaminate the signal
\citep[see Fig. 9 of][]{planck2011-7.12} and \hi\ is no longer a good tracer of dust emission.

The \healpix\  HFI maps were reprojected onto the small \hi\ maps by binning the original \healpix\  data (sampled with \healpix\  Nside of 2048, corresponding to a pixel size of 1.72\arcm) into \hi\ map pixels (pixel size 3.5\arcm\ for all fields).  An average of slightly more than four \healpix\  pixels were averaged for each small map pixel.

\subsection{Removing the bright sources from HFI maps \label{se:source_remove}}

We removed from the maps all sources listed in the \Planck\ Early Release Compact Source Catalog (ERCSC) \citep{planck2011-1.10}. This represents only a few sources per field (if any), but the bright source removal is important for both power spectrum analysis and CMB map construction. It is also important to know the flux limit to compute the radio and dusty galaxy shot-noise contribution to the power spectra. Since our fields have roughly the same (very low) dust contamination, source detection is not limited by cirrus. Indeed, the flux cut is set by extragalactic source confusion at high frequencies and CMB contamination at low frequencies. The same flux cut can therefore be applied to all our fields. We took the minimum ERCSC flux densities in our fields as the flux cuts. They are given in Table \ref{tab:fluc_cut}.

In practice, point source removal is performed in the original HFI \healpix\ data prior to reprojection.
For each source, a disc of size equal to the FWHM of the beam centred on the source position is blanked. Holes caused by missing data are then filled by a gap-filling process, which interpolates/extrapolates into the hole the values of neighbouring pixels. 

\begin{figure}
\begin{center}
\includegraphics[width=6.5cm, draft=false]{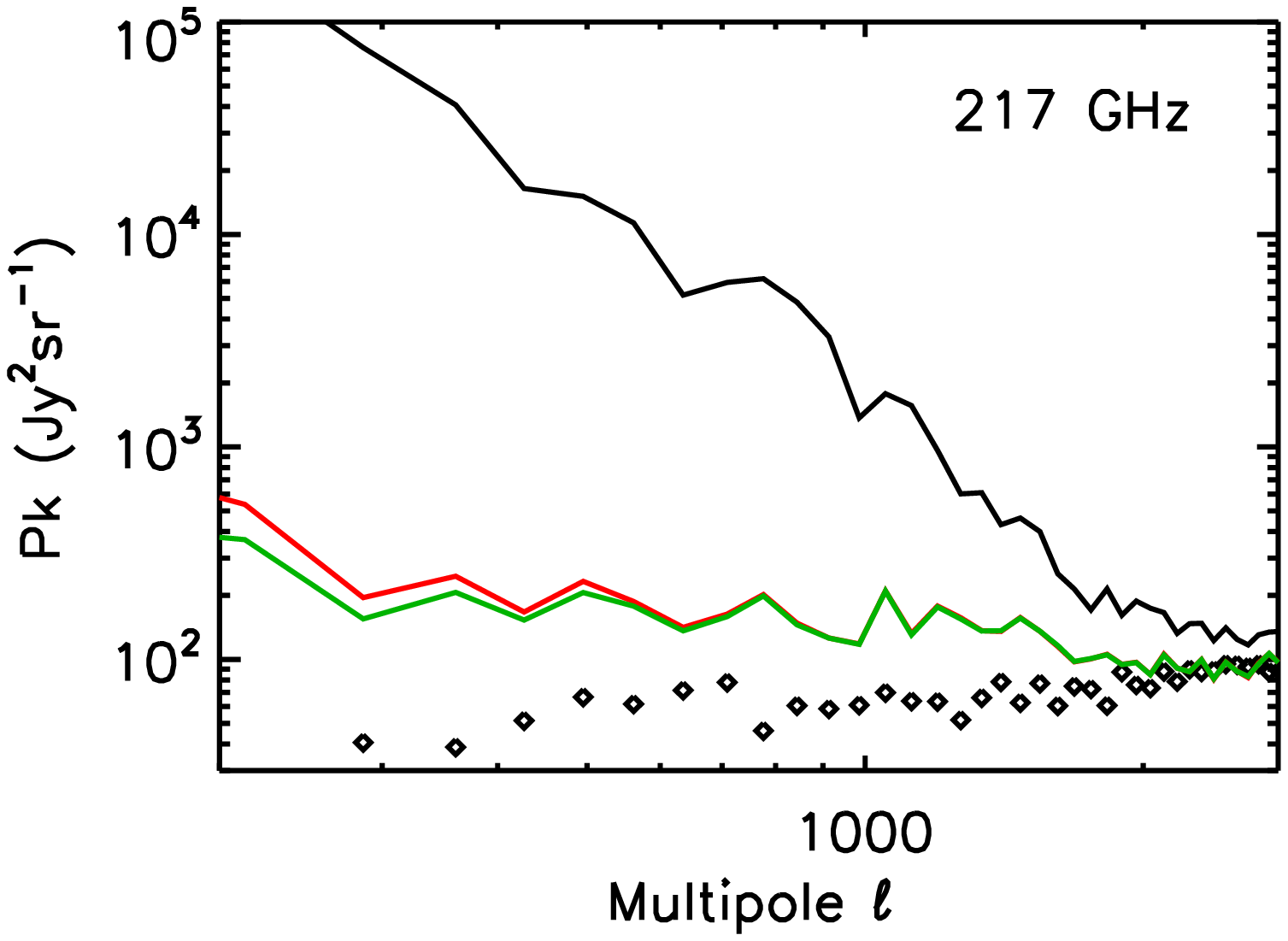}
\includegraphics[width=6.5cm, draft=false]{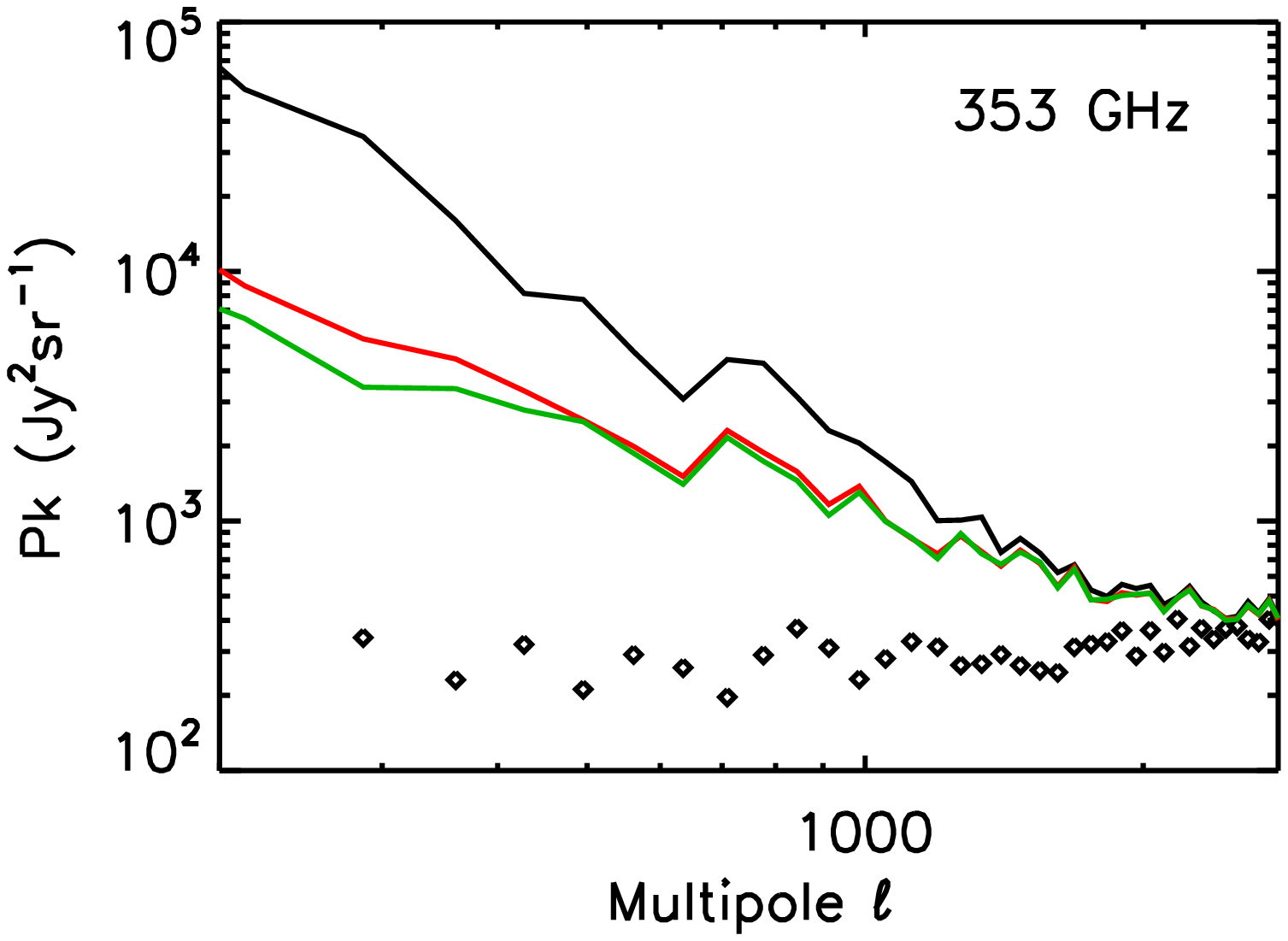}
\includegraphics[width=6.5cm, draft=false]{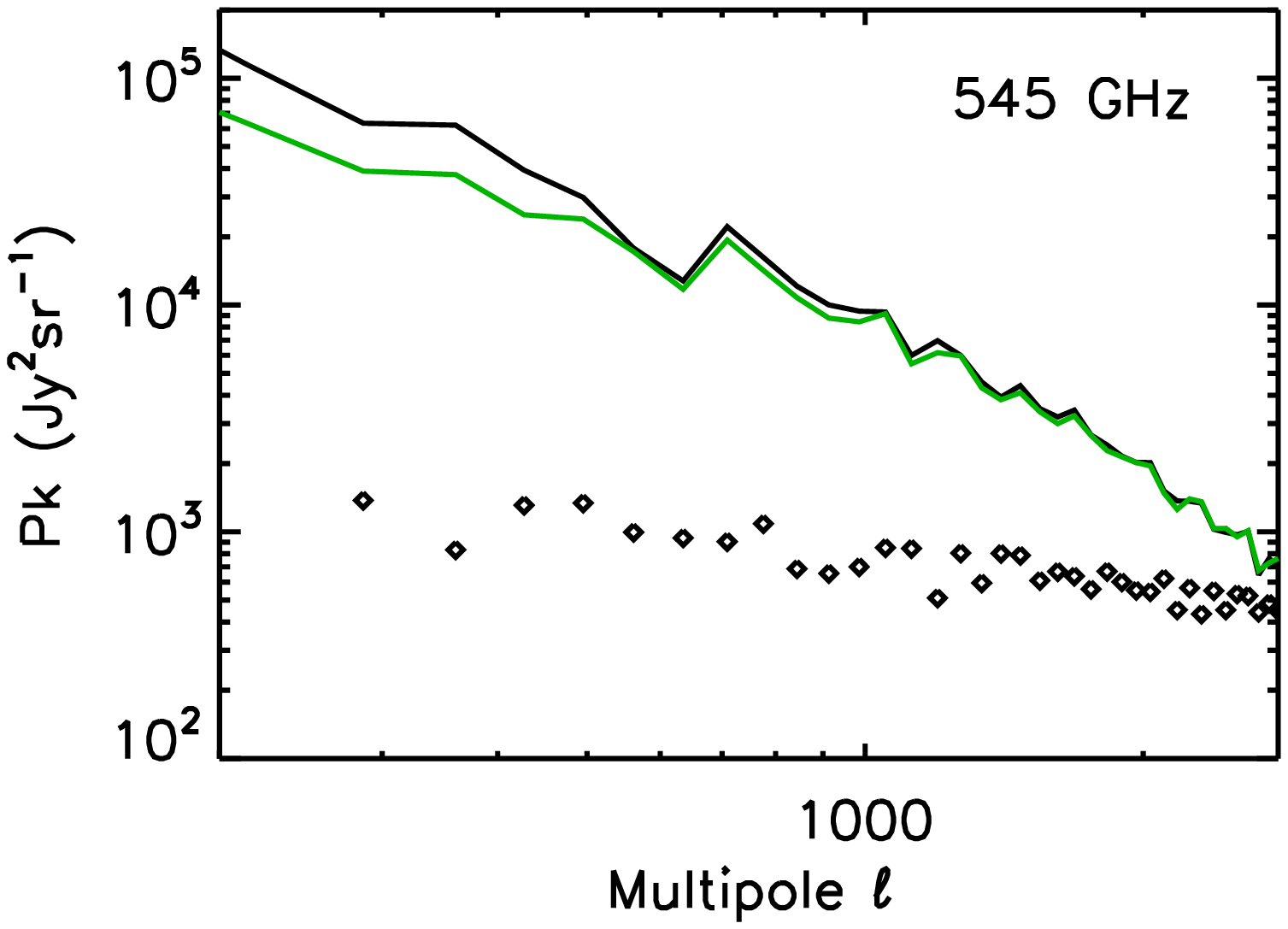}
\includegraphics[width=6.5cm, draft=false]{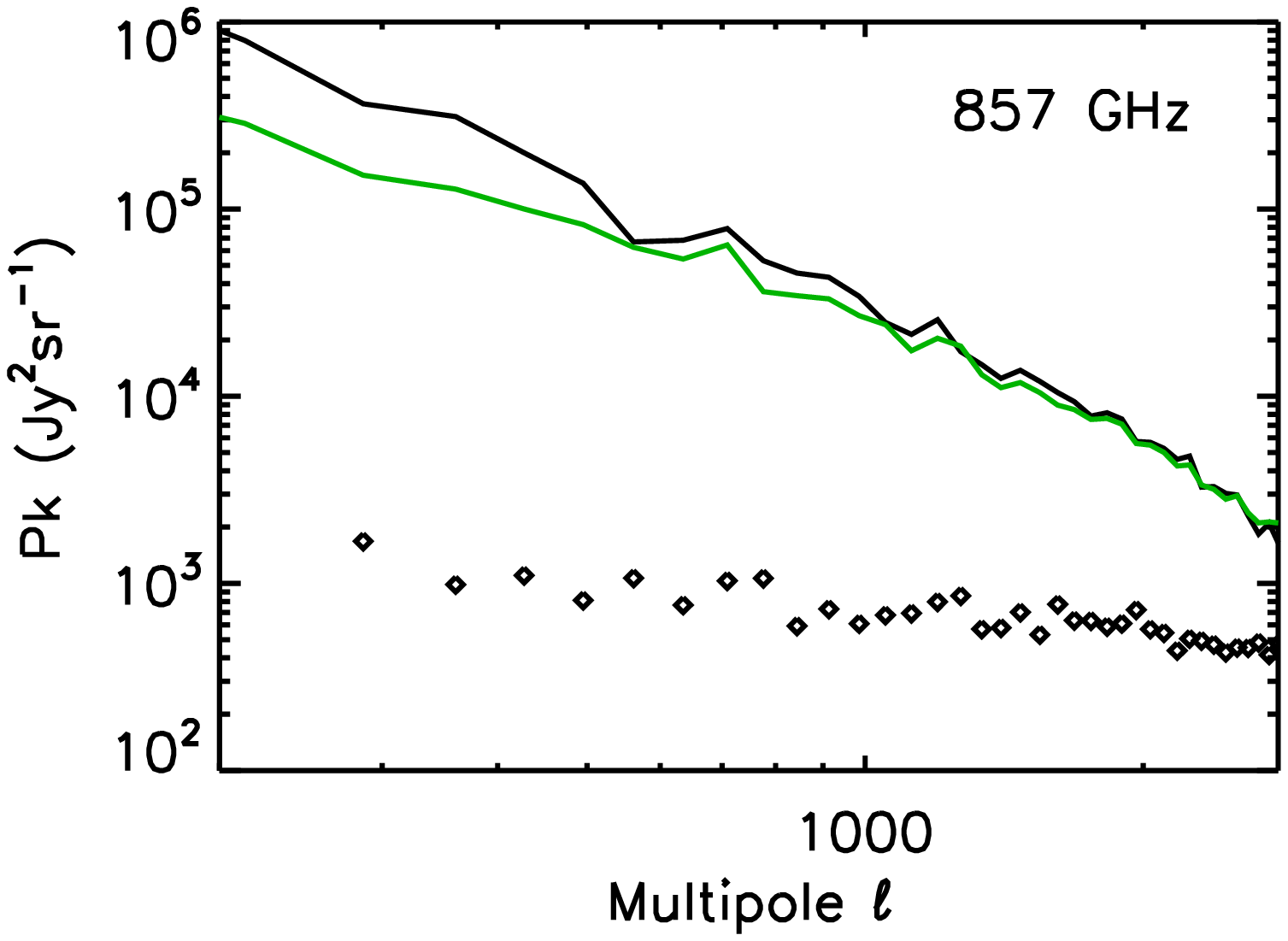}
\end{center}
\caption{\label{Fig_Pktot_pknoise} Power spectra of the different components for field {\tt SP} (the figure is
similar for the other fields). Power spectra of the 217, 353, 545, and 857\,GHz \Planck\ maps (continuous black line) are compared to the noise power spectra (diamonds), to the CMB-cleaned power spectra (red), and to the CMB- and interstellar dust-cleaned power spectra (green). In this plot  signal power spectra have not been corrected for the beam window function. Noise power spectra are computed using  half-pointing period maps, as explained in Sect. \ref{se:inst_noise}.}
\end{figure}

\begin{figure*}
\includegraphics[width=4.5cm, draft=false]{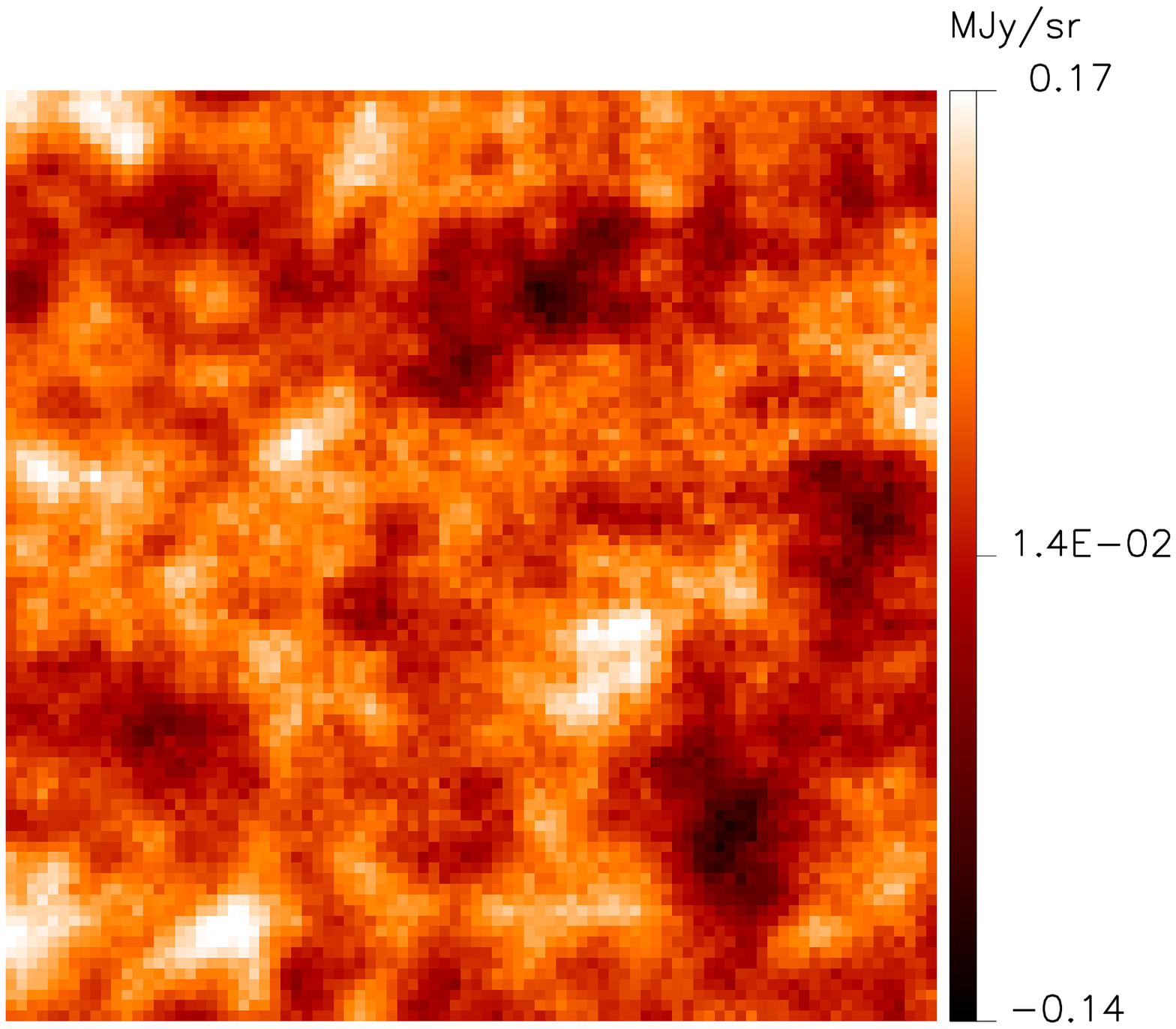}
\includegraphics[width=4.5cm, draft=false]{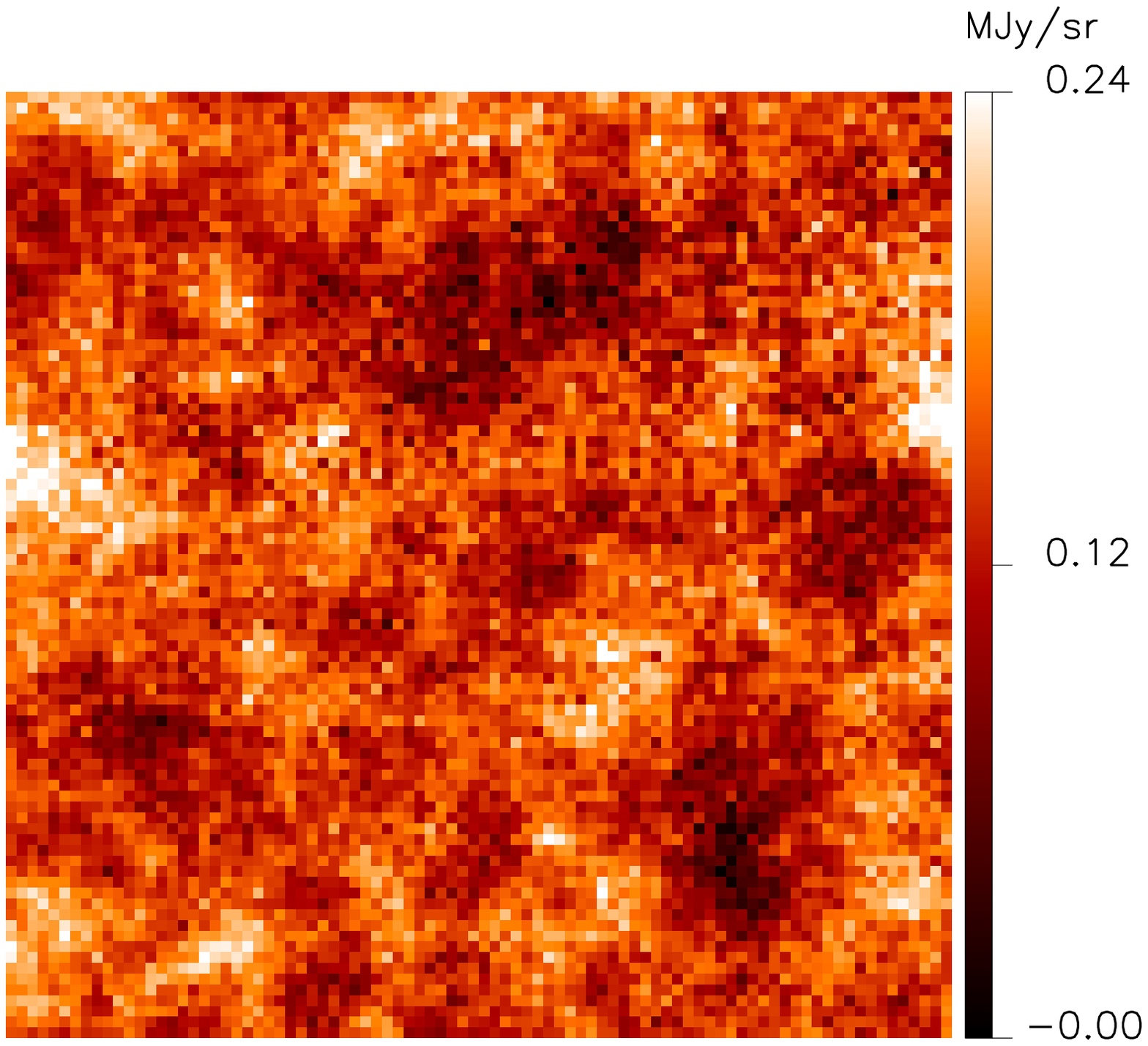}
\includegraphics[width=4.5cm, draft=false]{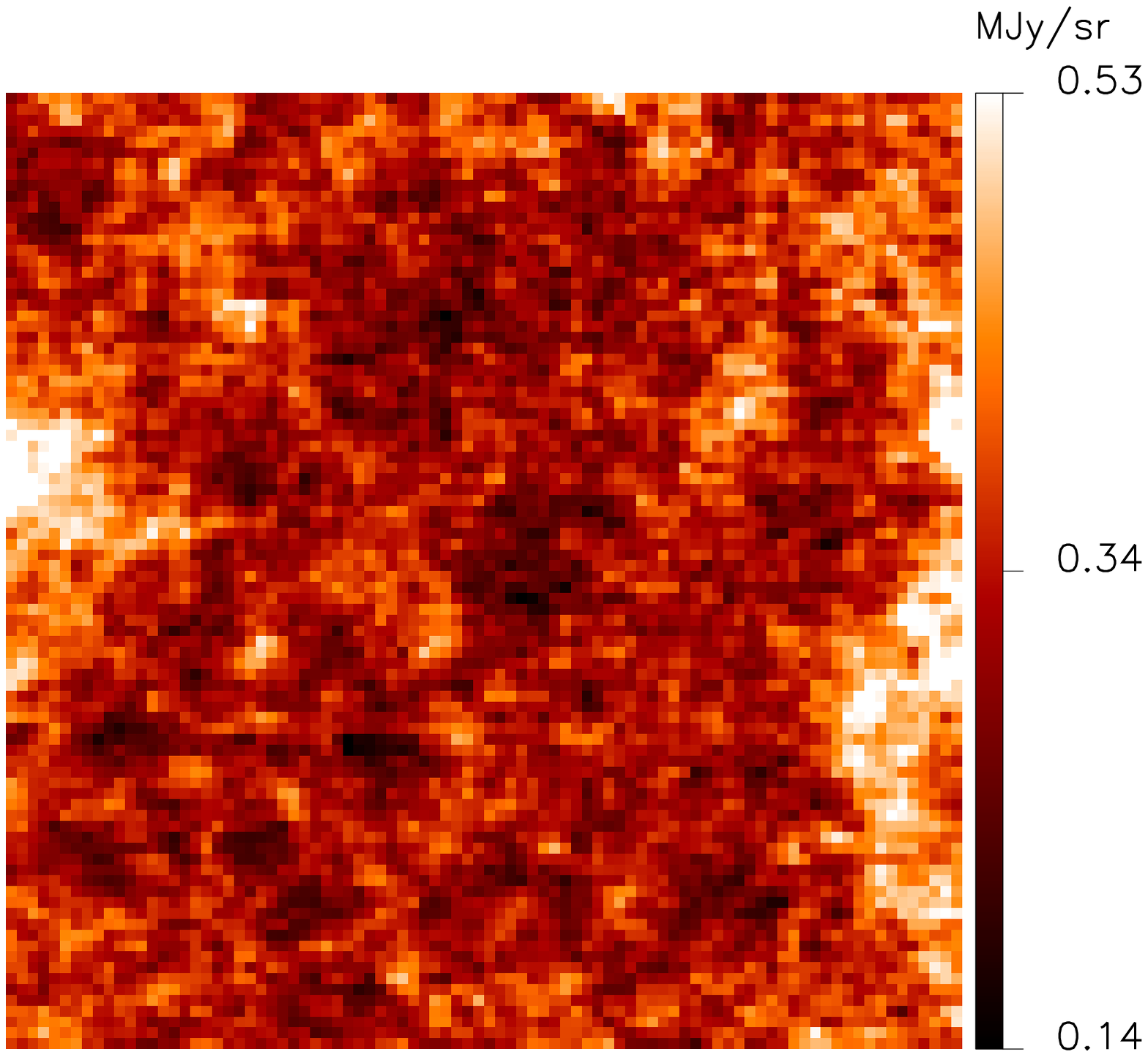}
\includegraphics[width=4.5cm, draft=false]{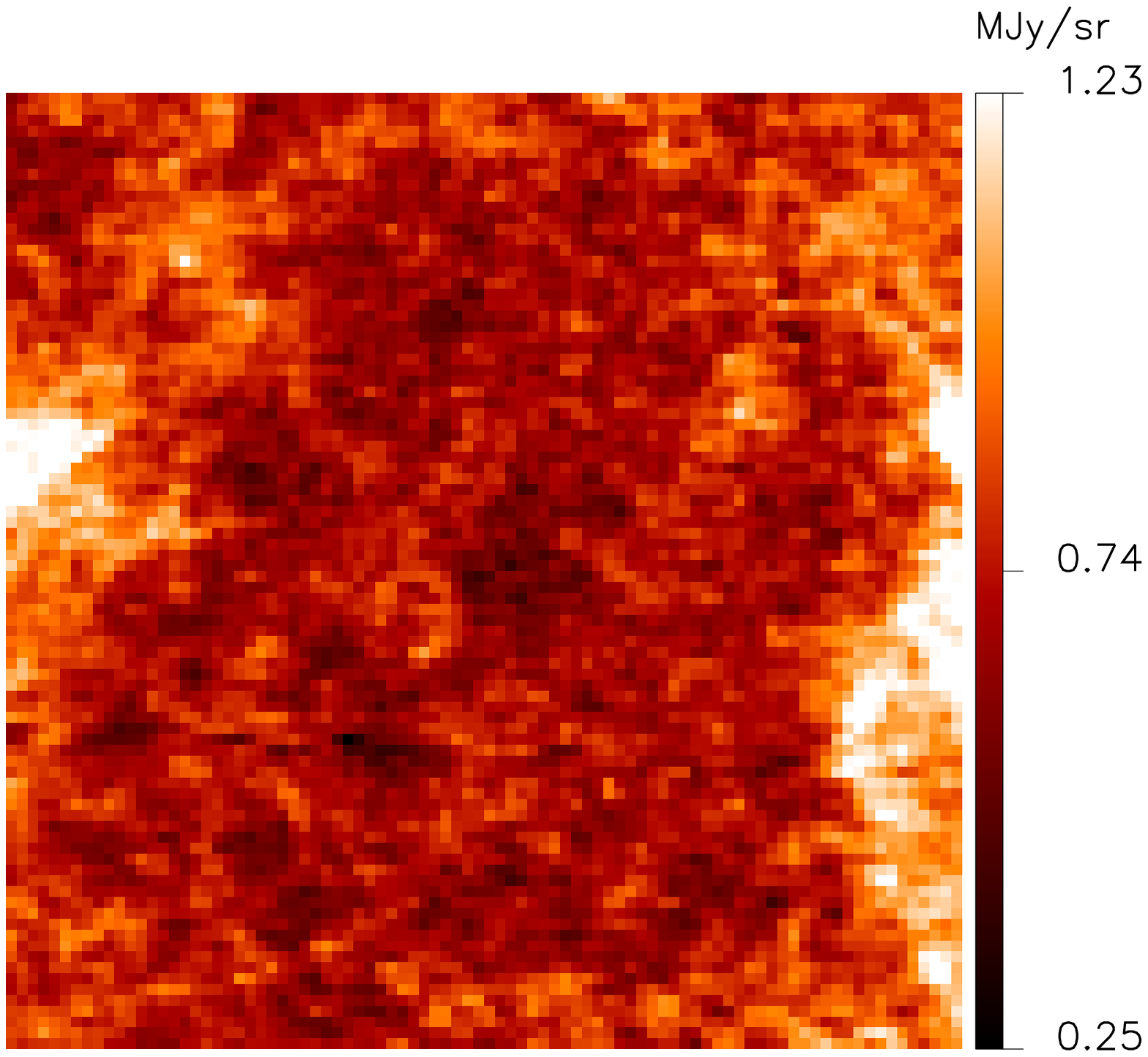}\\
\includegraphics[width=4.5cm, draft=false]{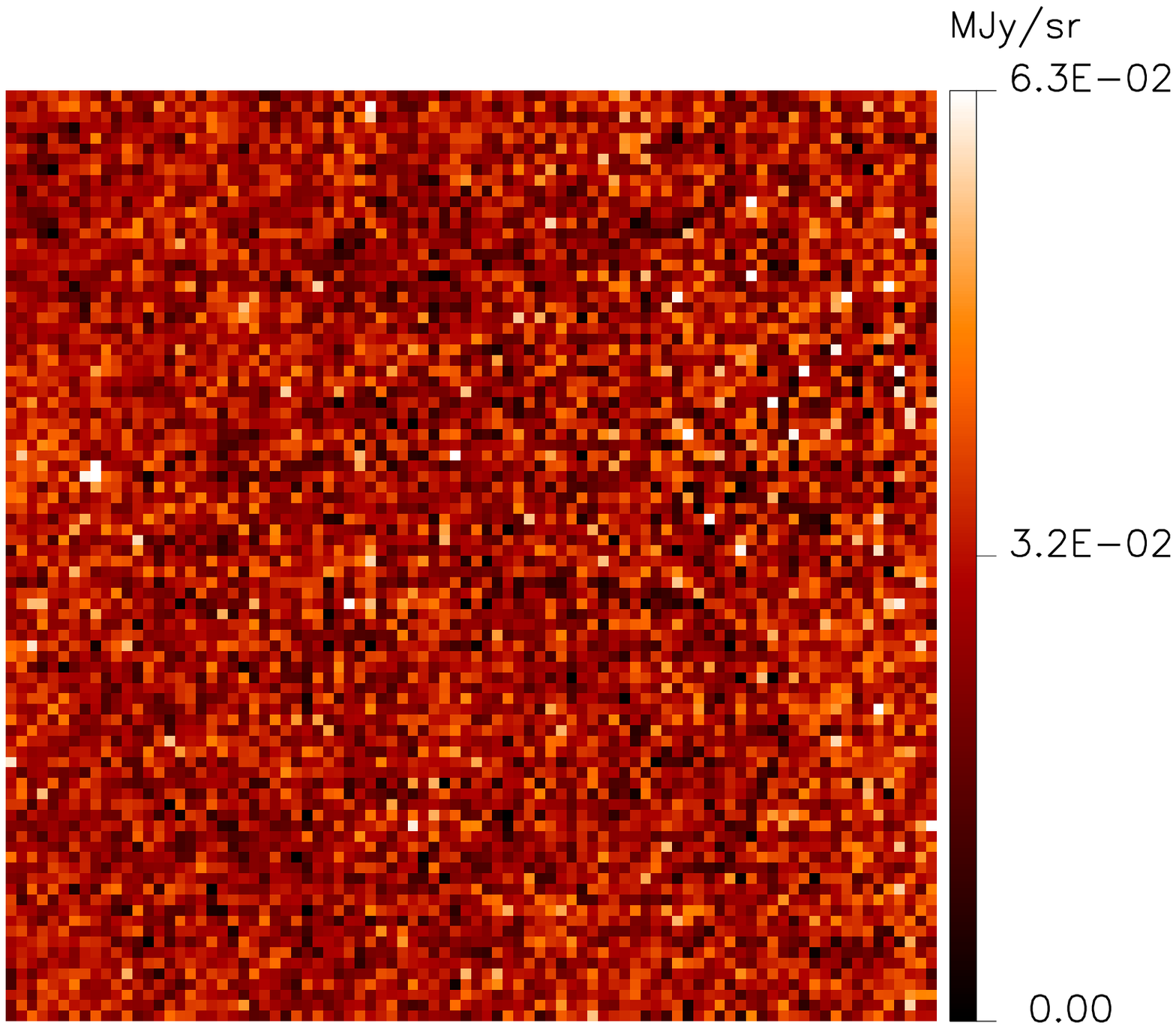}
\includegraphics[width=4.5cm, draft=false]{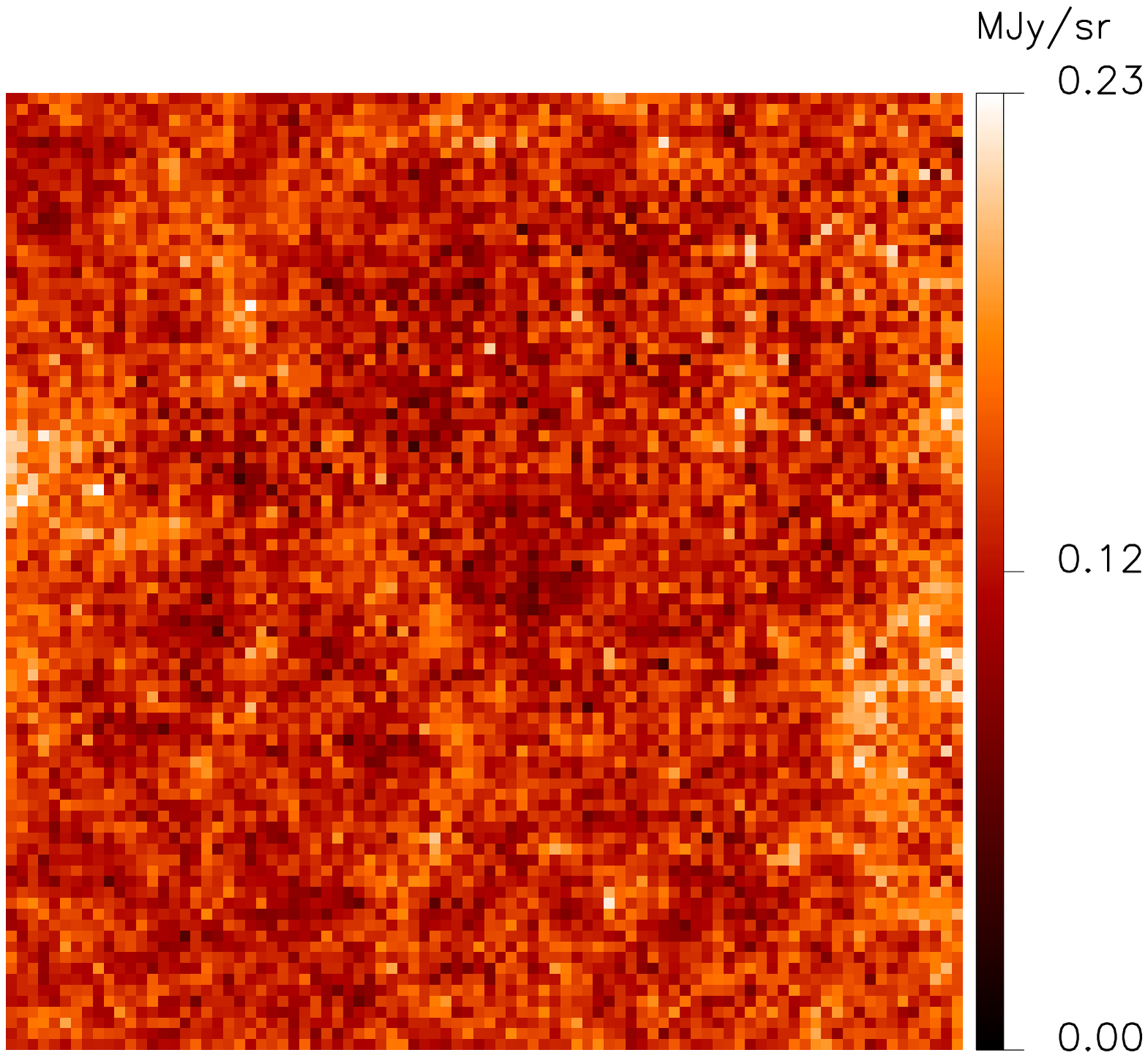}
\includegraphics[width=4.5cm, draft=false]{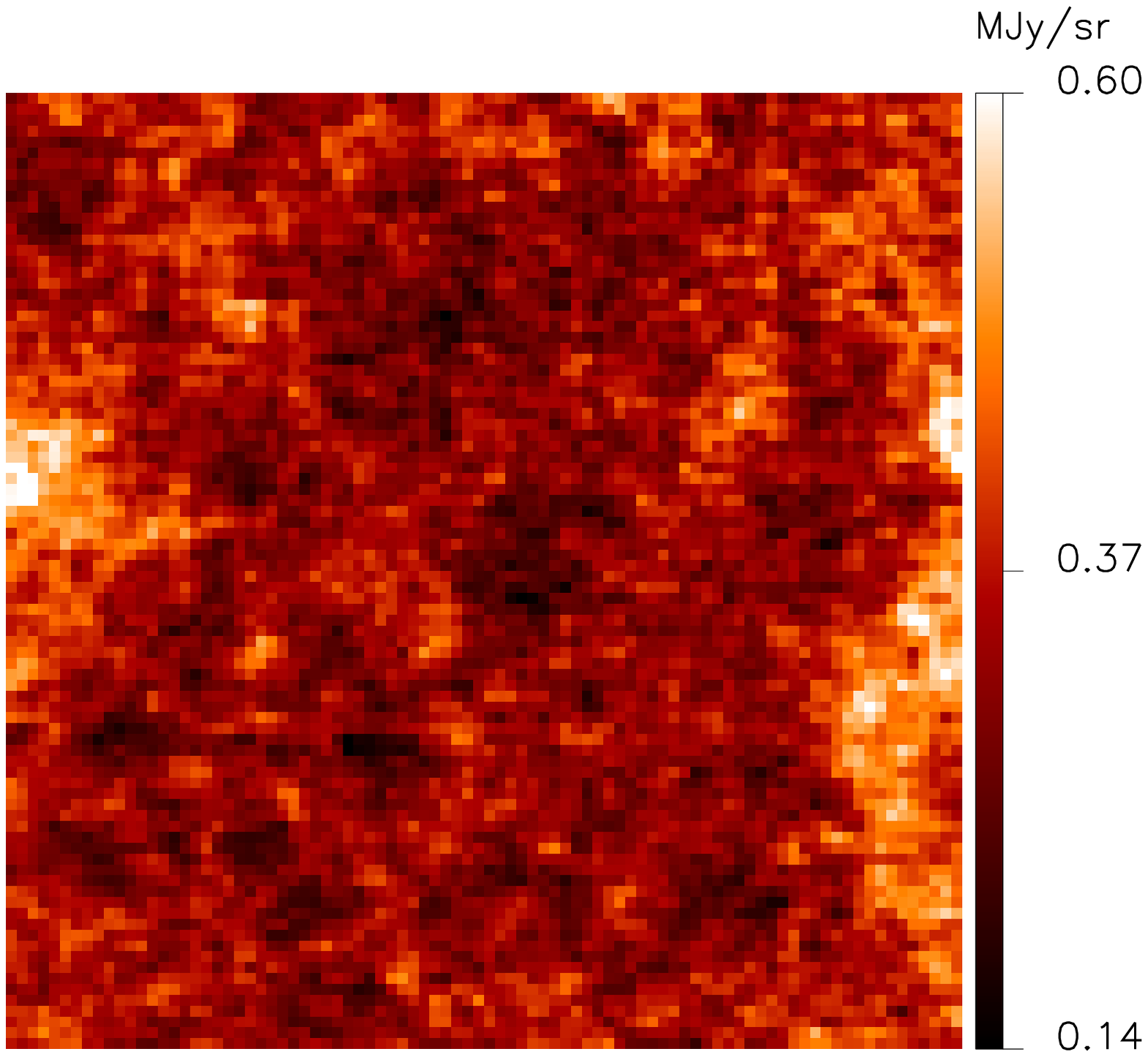}
\includegraphics[width=4.5cm, draft=false]{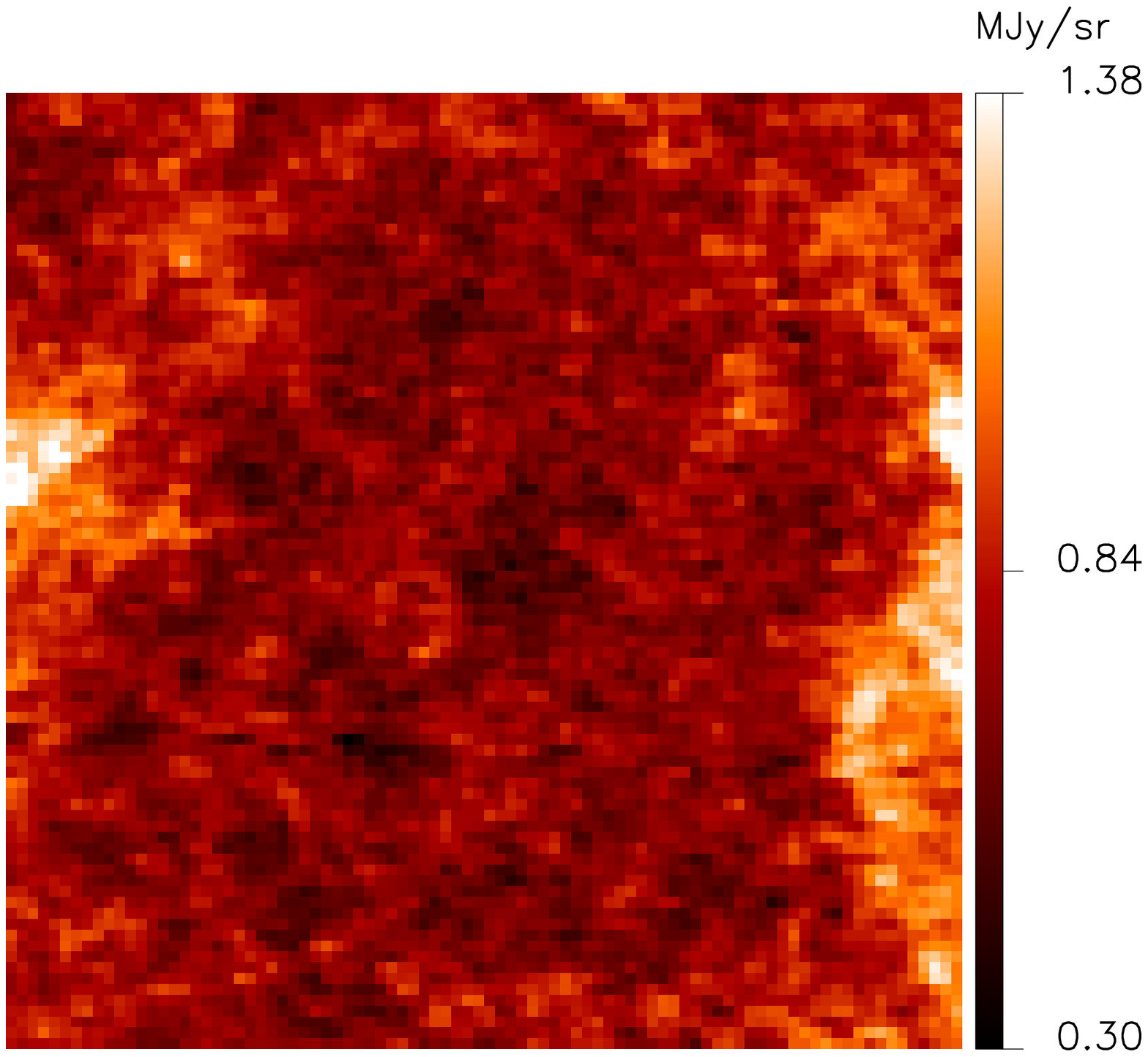}\\
\includegraphics[width=4.5cm, draft=false]{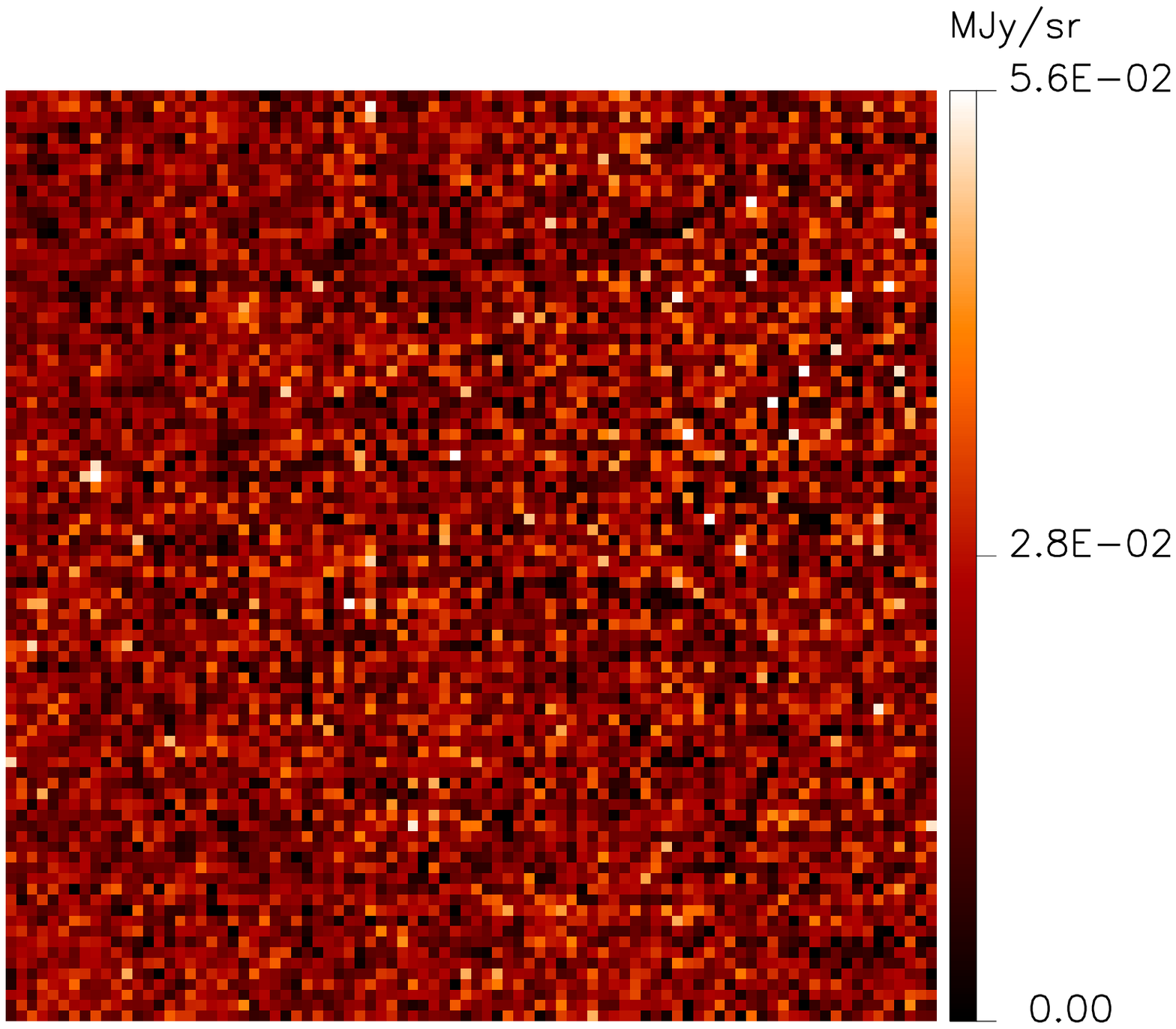}
\includegraphics[width=4.5cm, draft=false]{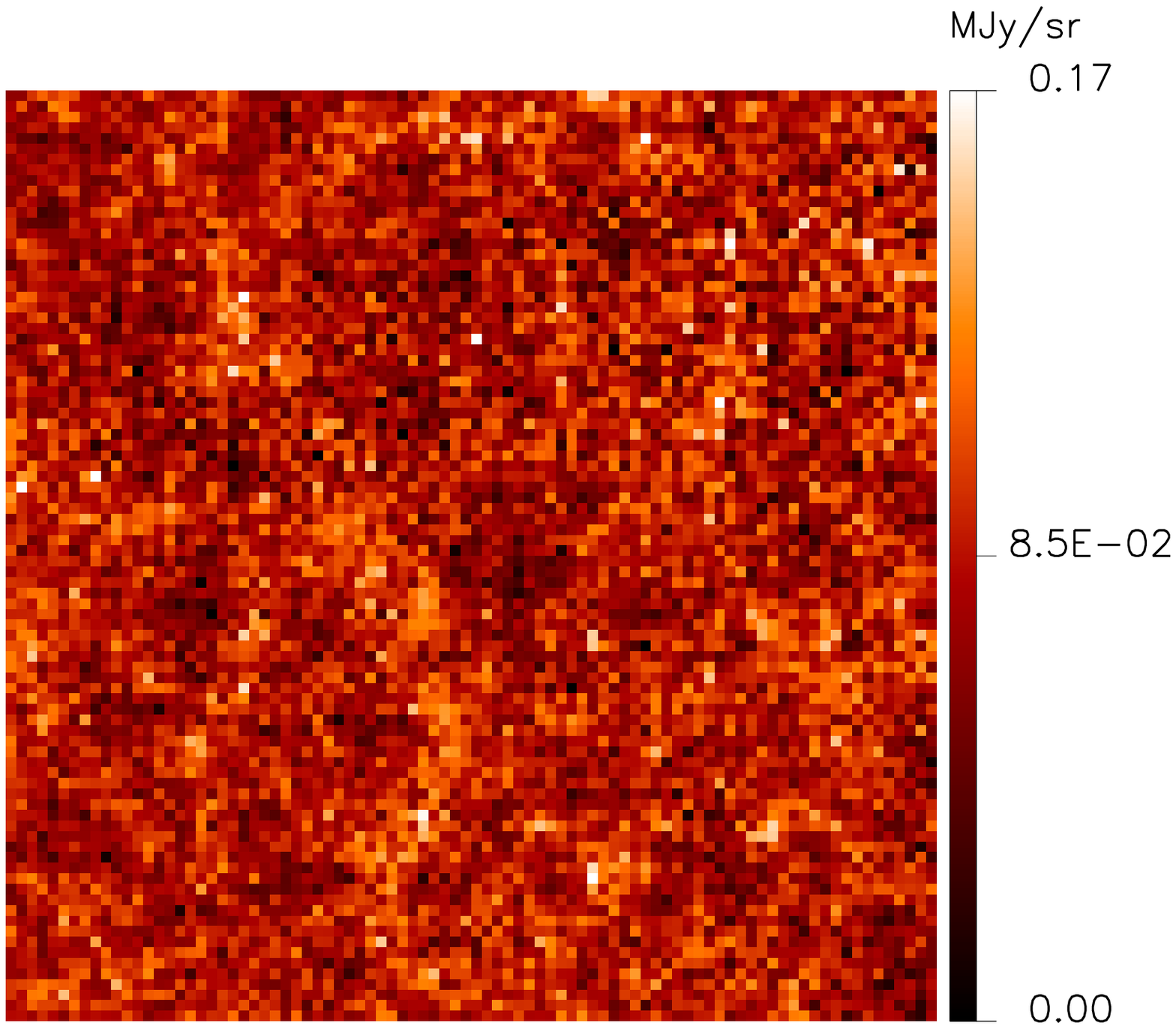}
\includegraphics[width=4.5cm, draft=false]{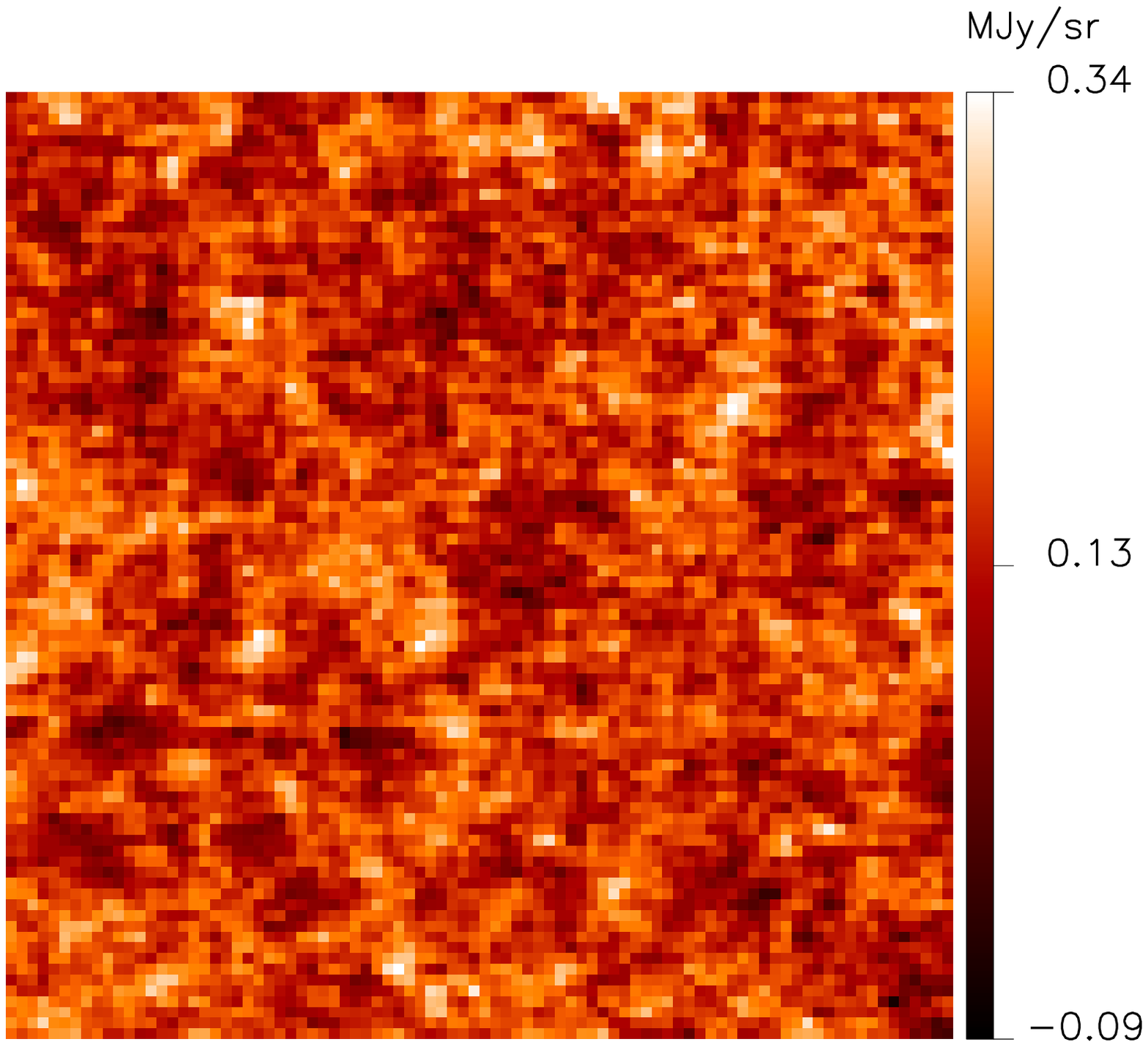}
\includegraphics[width=4.5cm, draft=false]{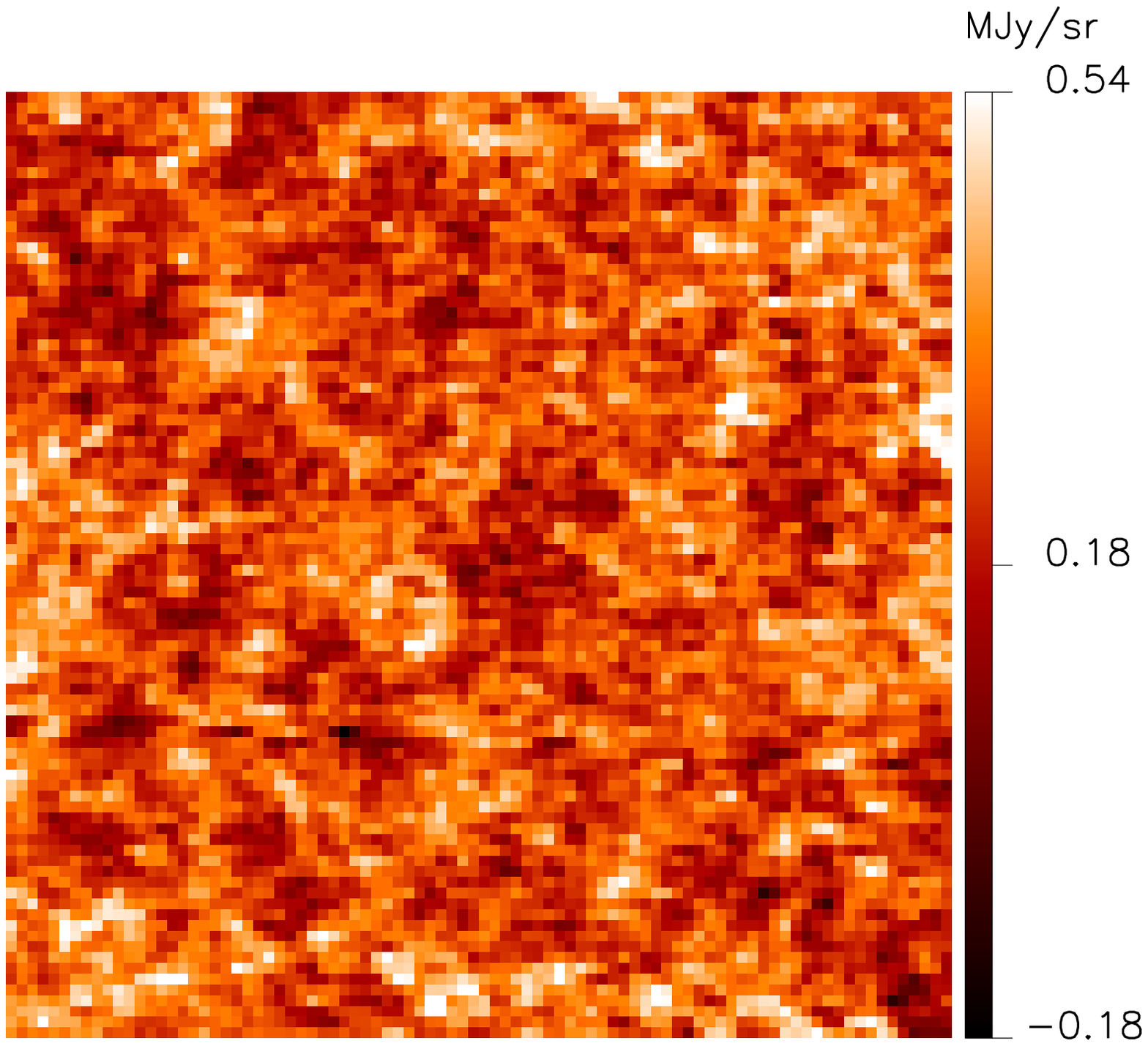}\\
\includegraphics[width=4.5cm, draft=false]{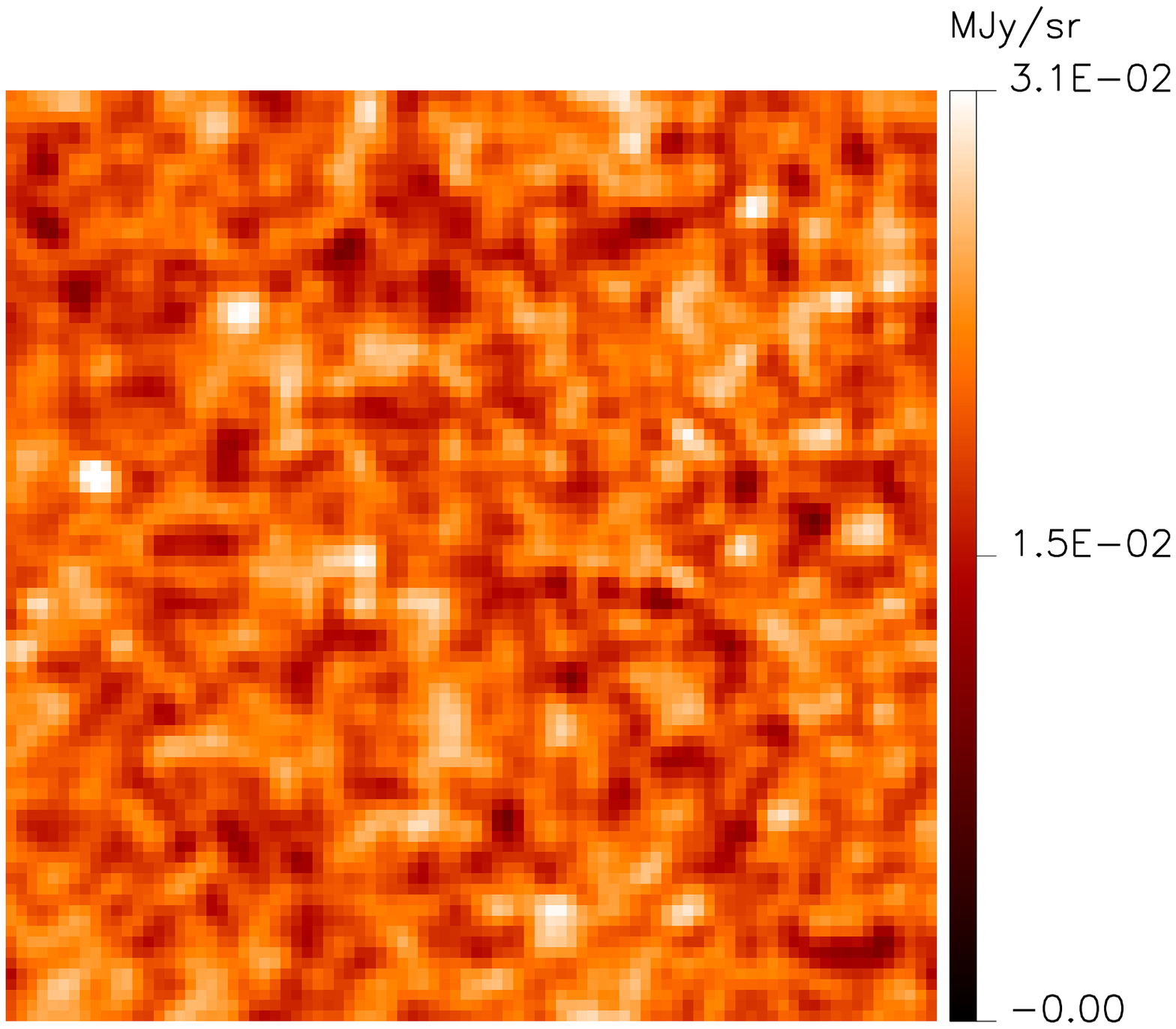}
\includegraphics[width=4.5cm, draft=false]{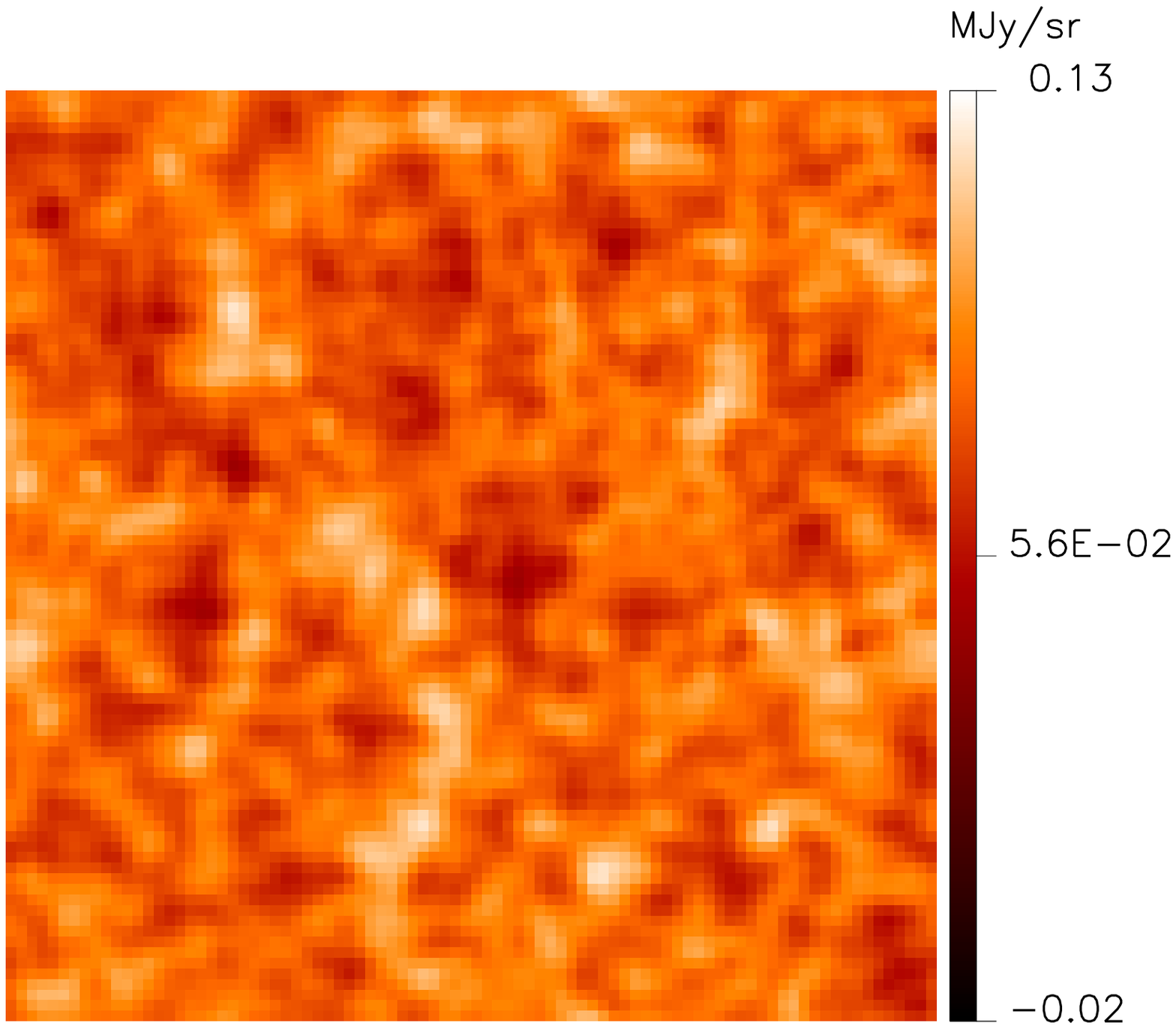}
\includegraphics[width=4.5cm, draft=false]{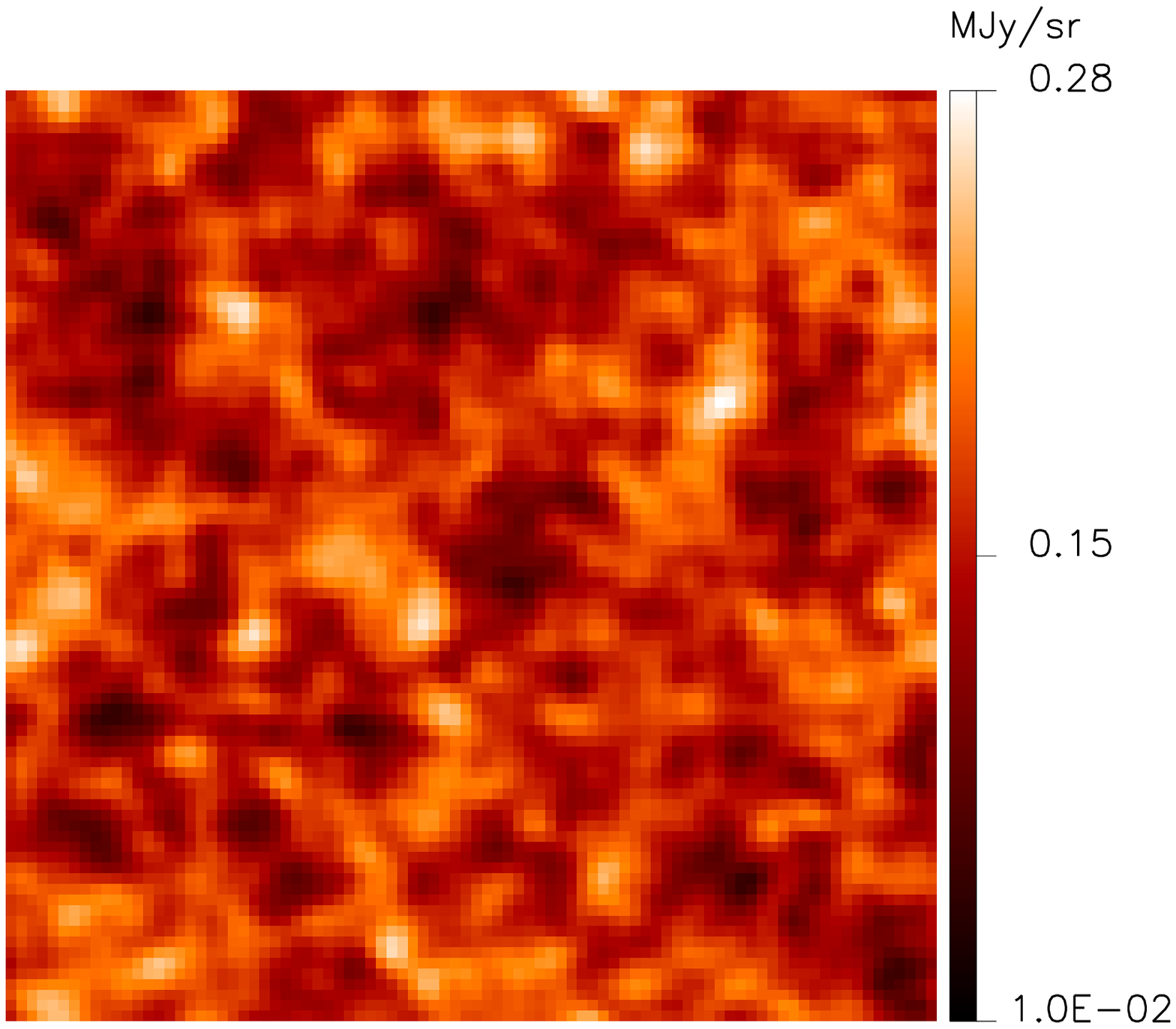}
\includegraphics[width=4.5cm, draft=false]{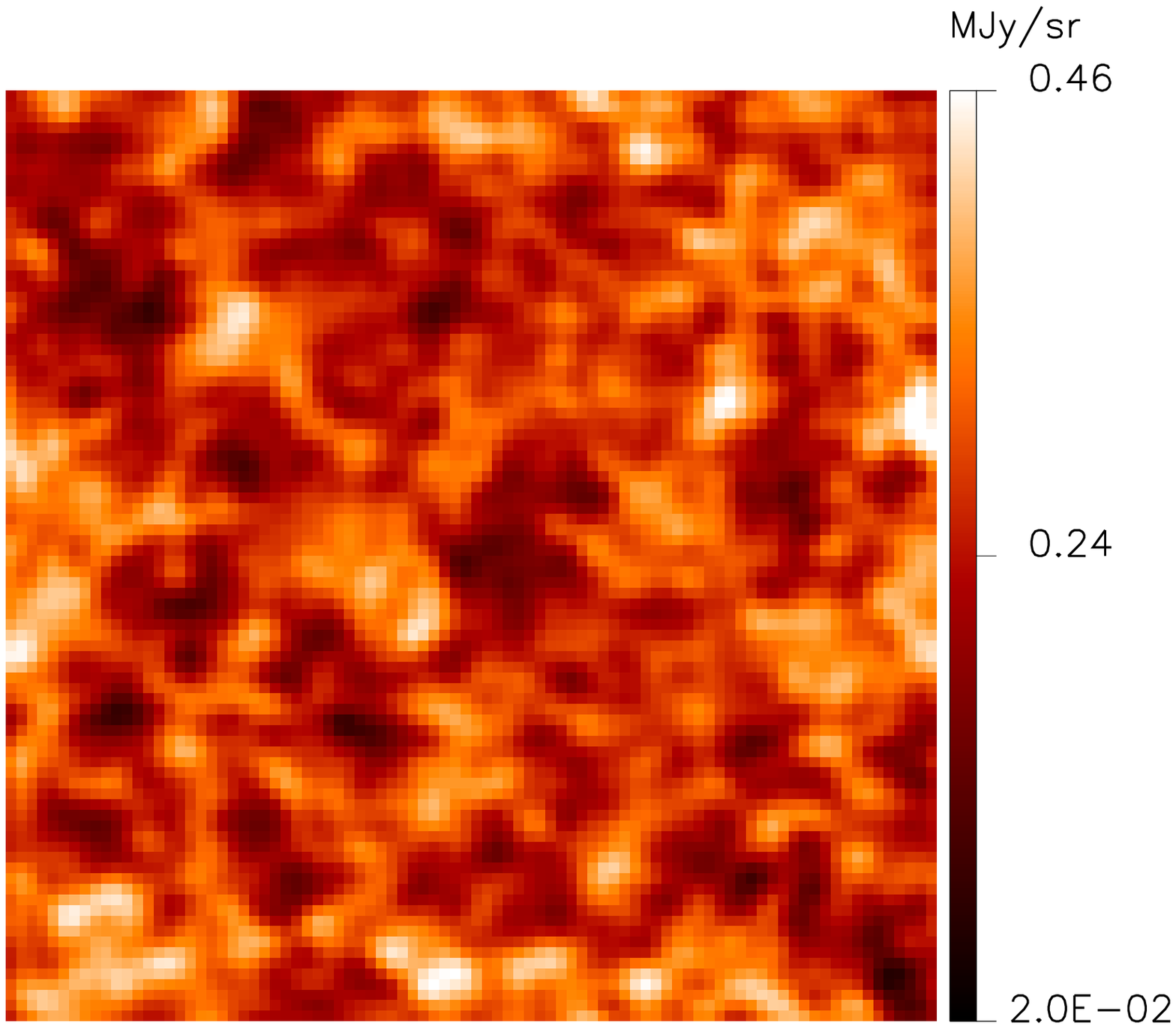}
\caption{\label{Fig_CIB}  Maps of the 26 ${\rm deg^2}$ of the {\tt N1} field, from {\it left} to {\it right}: 217, 353, 545 and 857\,GHz. From {\it top} to {\it bottom}: raw HFI maps; CMB- and ERCSC source-cleaned maps; residual maps (CMB-, sources-, and cirrus-cleaned); residual maps smoothed at 10\arcm\ to highlight the CIB anisotropies. The joint structures clearly visible (bottom row) correspond to the anisotropies of CIB. 
Residual point sources are also visible. They have fluxes lower than the fluxes of the ERCSC removed sources. They have no impact on our analysis.}
\end{figure*}

\subsection{Removing the CMB contamination from HFI maps \label{se:cmb_remove}}

Cosmic Microwave Background anisotropies contribute significantly to the total HFI map variance in all channels at frequencies up to and including 353\,GHz. The detection and characterisation of CIB anisotropies at these frequencies requires the separation of the contribution from the CMB.

The present work focuses on very clean regions of the sky, for which Galactic foregrounds are very faint and are monitored using ancillary \hi\  observations.
To remove the CMB in the fields retained for our analysis, we used a simple subtraction technique.  While this simple
method could be improved in future, it enables us to reliably evaluate CMB residuals, noise contamination, and to propagate errors due to imperfect instrumental knowledge. It also guarantees that high-frequency CIB anisotropy signals will not leak into lower frequency, CMB-free maps.\\

We removed the CMB contamination in the 217 and 353\,GHz channels by subtracting a CMB template obtained from the lower frequency data.  We modelled the data for each frequency $\nu$ as
\begin{equation}
x_{\ell m}(\nu) = b_{\ell}(\nu) \left [ a^{\rm CMB}_{\ell m} + a^{\rm CIB}_{\ell m}(\nu) \right ] + n_{\ell m}(\nu),
\end{equation}
where $x_{\ell m}(\nu)$ represents the channel data at frequency $\nu$, $a^{\rm CMB}_{\ell m}$ the CMB map,
$a^{\rm CIB}_{\ell m}(\nu)$ the CIB map and $n_{\ell m}(\nu)$ is a noise term comprising (if needed) any other
astrophysical contaminant.
The effect of the beam was accounted for with a (channel dependent) multiplicative factor, $b_{\ell}(\nu)$ (see Sect. \ref{sec:beam}).
For the purpose of CMB removal,  $b_{\ell}(\nu)$ was obtained from the Gaussian best-fit to the effective HFI beam of the channel maps.

At 100 and 143\,GHz, we assumed that in the fields of interest only CMB and noise is present.
Cosmic infrared background anisotropies are very small, and in the selected fields the contamination by other sources (e.g., cirrus) is negligible (see Sect. \ref{sect_conta}).
In principle, both channels can be used to make a template of CMB emission.
However, the 100\,GHz channel is significantly less sensitive than 143\,GHz and has an angular resolution two times worse than the 217 and 353\,GHz channels.
Therefore, we only used the 143\,GHz channel as a CMB template. We corrected the 217\,GHz maps for CMB contamination as follows

\begin{align}
y_{\ell m}(\nu_{217})	& =  x_{\ell m}(\nu_{217})	 - \frac{b_{\ell}(\nu_{217})	 }{b_{\ell}(\nu_{143}) }w_\ell x_{\ell m}(\nu_{143}) \nonumber \\ 
& = b_{\ell}(\nu_{217}) \left [ a^{\rm CIB}_{\ell m}(\nu_{217}) + (1-w_\ell) a^{\rm CMB}_{\ell m} \right ]  \nonumber \\
&  + n_{\ell m}(\nu_{217}) - \frac{b_{\ell}(\nu_{217})	 }{b_{\ell}(\nu_{143}) }w_\ell n_{\ell m}(\nu_{143}),
\label{eq:cmb-subtraction}
\end{align}

where $w_\ell$ is a Wiener filter, designed to minimize the total contamination of the new map, $y(\nu_{217}) $, by CMB and instrument noise. The 353\,GHz map is corrected from CMB contamination in a similar way.
Note that this cleaning was performed on a large region comprising all small fields used in the present analysis. The Wiener filter is obtained as
\begin{equation}
w_\ell = \frac{b_{\ell}(\nu_{143}) C_\ell^{\rm CMB} }{Y_\ell(\nu_{143})},
\end{equation}
where $C_\ell^{\rm CMB}$ is the current best-fit CMB model spectrum, and $Y_\ell(\nu_{143})$ is the power spectrum of the 143\,GHz map.
The Wiener filter $w_\ell$ is close to 1 at low $\ell$, and close to 0 at large $\ell$ (see Fig. \ref{fig:wienerfilter}).
Note that errors on the beam estimate, on the assumed CMB power spectrum, or on the estimation of the 143\,GHz power spectrum would result in sub-optimal filtering rather than in biases.
We checked that the CMB remaining in the CMB-cleaned maps does not change significantly with different assumptions leading to different $w_\ell$.

Errors in photometric calibration between channels are a problem.
Although these errors are estimated to be small (2\% at 143, 217, and 353\,GHz),
they may result in residual CMB at low $\ell$.
They are accounted for in the processing, as detailed in Sect. \ref{sec:stat}.

Fig.~\ref{Fig_Pktot_pknoise} shows the HFI  power spectra of the raw and CMB-cleaned maps for one of the six fields. The CMB correction is very large at 217\,GHz: the residual is a factor $\sim$100 below the raw power spectrum at $\ell \simeq 430$ (it is a factor $\sim$2 below at 353\,GHz). Note that whereas this illustrates the effectiveness of CMB removal, it is also a source of worry about the impact of relative calibration errors for the 217\,GHz channel. However, the power spectrum after CMB cleaning is $\sim$1\% of the original map power spectrum only for $\ell \leq 600$. Cosmic microwave background-cleaned maps are shown on Fig \ref{Fig_CIB}. We see that the CMB has been efficiently removed.\\

Finally we remark that an alternative method of removing CMB contamination, based on an internal
linear combination of frequency maps and a needlet analysis \citep{delabrouille2009}, was extensively
studied and used in some of the \Planck\ early papers, but it was not well suited to our purposes.  
The method tended to perform well over large patches of sky but left visible, large-scale residuals in the sky patches of interest, and had leakage between the faint CIB and the CMB when other components (noise and Galactic cirrus) are present.

\subsection{Removing the cirrus contamination from HFI maps \label{se:cirrus_remove}}

From 100~$\mu$m to 1\,mm, at high Galactic latitude and outside molecular clouds a tight correlation is observed between far-infrared emission from dust and the 21-cm emission from gas\footnote{The Pearson correlation coefficient is $>0.9$ \citep{lagache2000}.}  \citep[e.g.][]{boulanger1996, lagache1998}. \hi\ can thus be used as a tracer of cirrus emission in our fields, and indeed it is the best tracer of diffuse interstellar dust emission. 

\begin{figure*}
\includegraphics[width=4.5cm, draft=false]{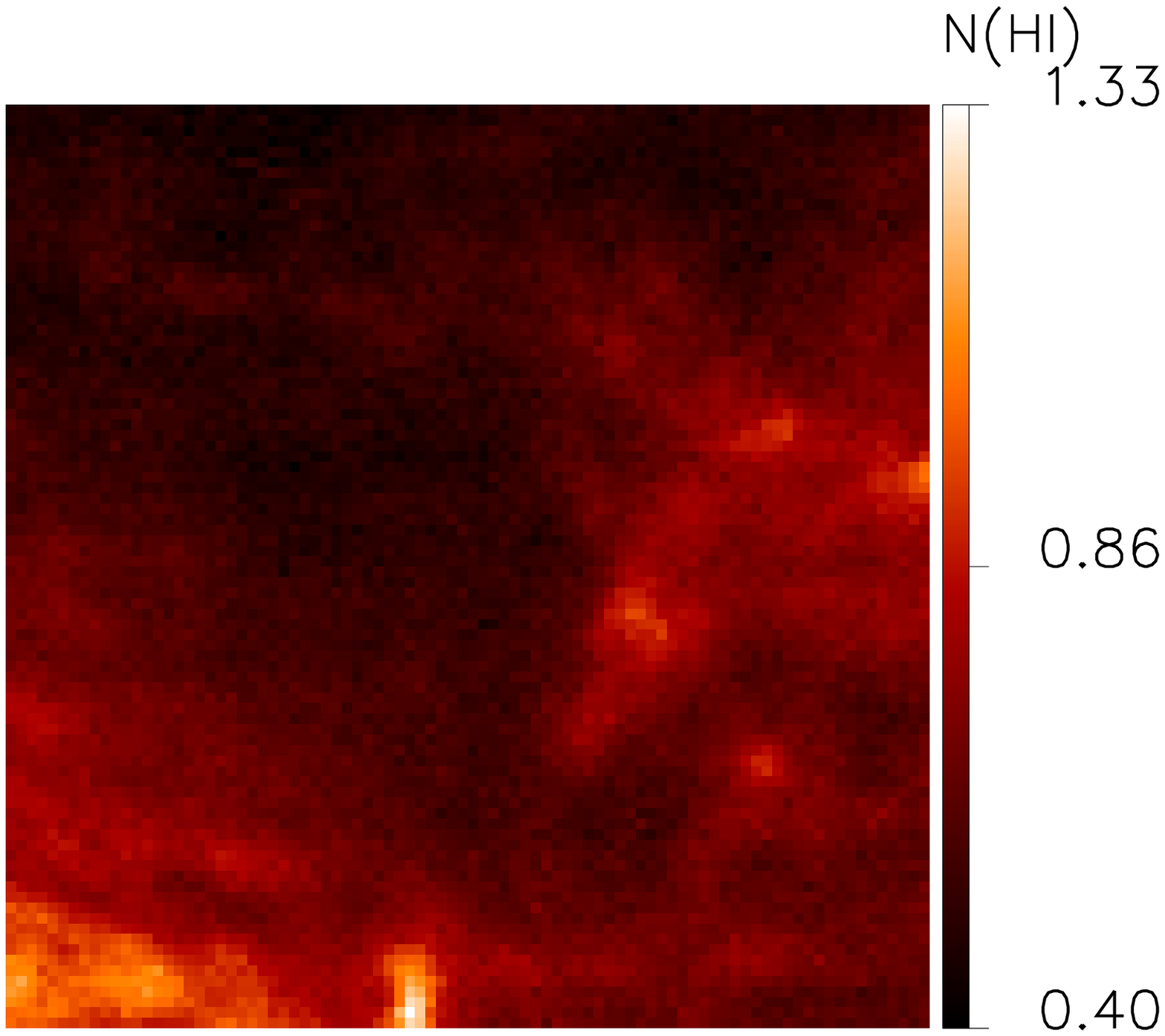}
\includegraphics[width=4.5cm, draft=false]{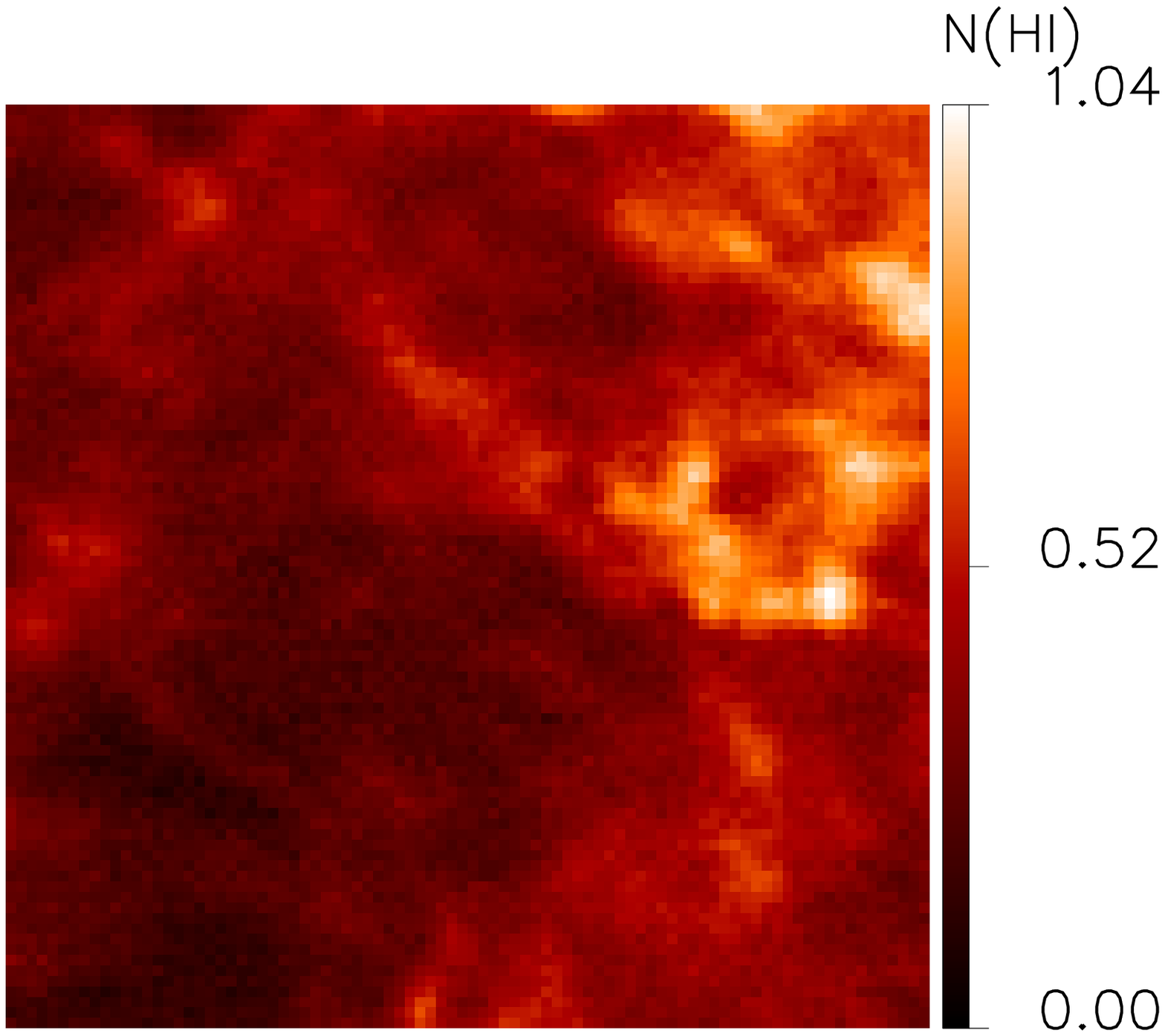}
\includegraphics[width=4.5cm, draft=false]{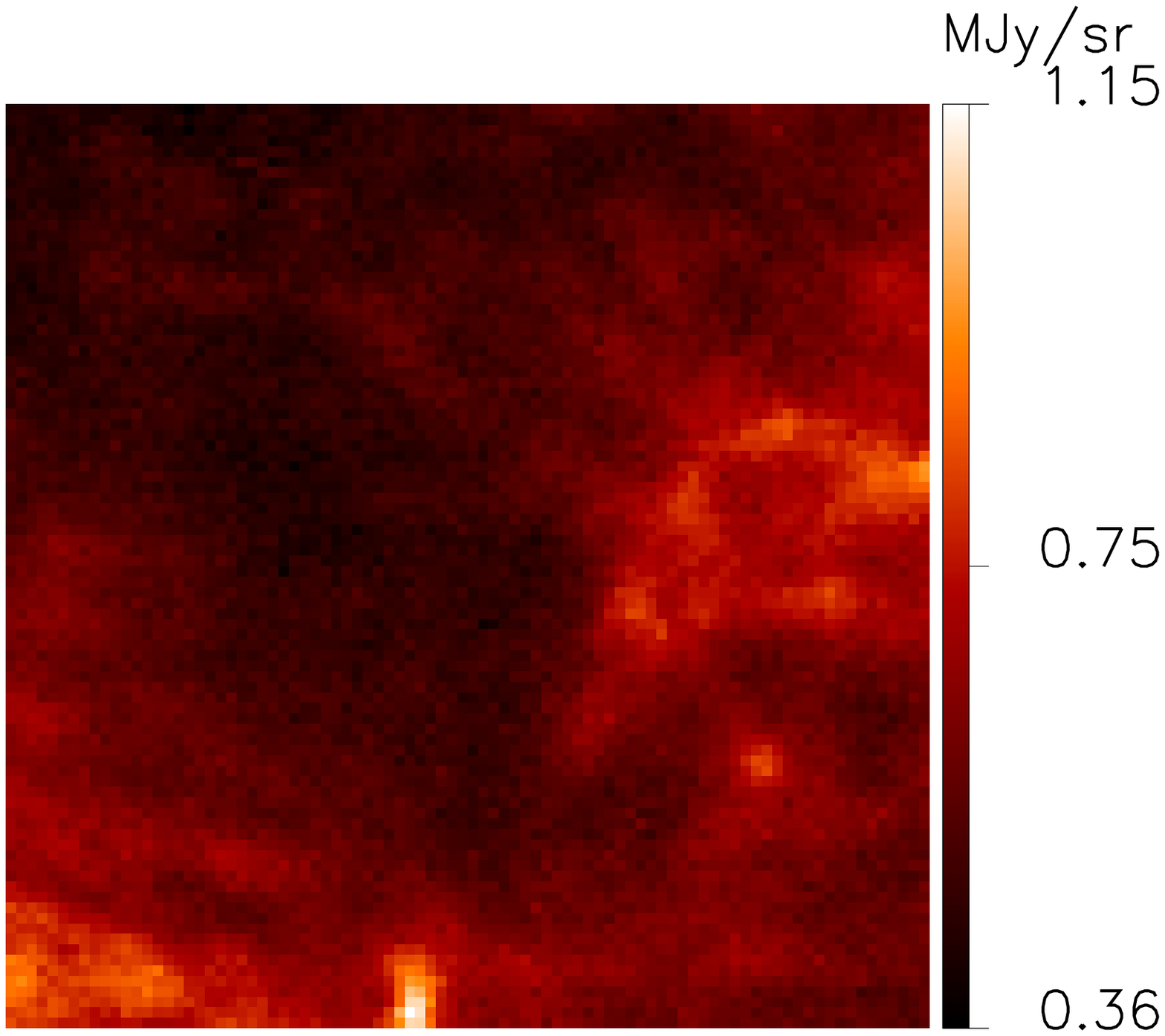}
\includegraphics[width=4.5cm, draft=false]{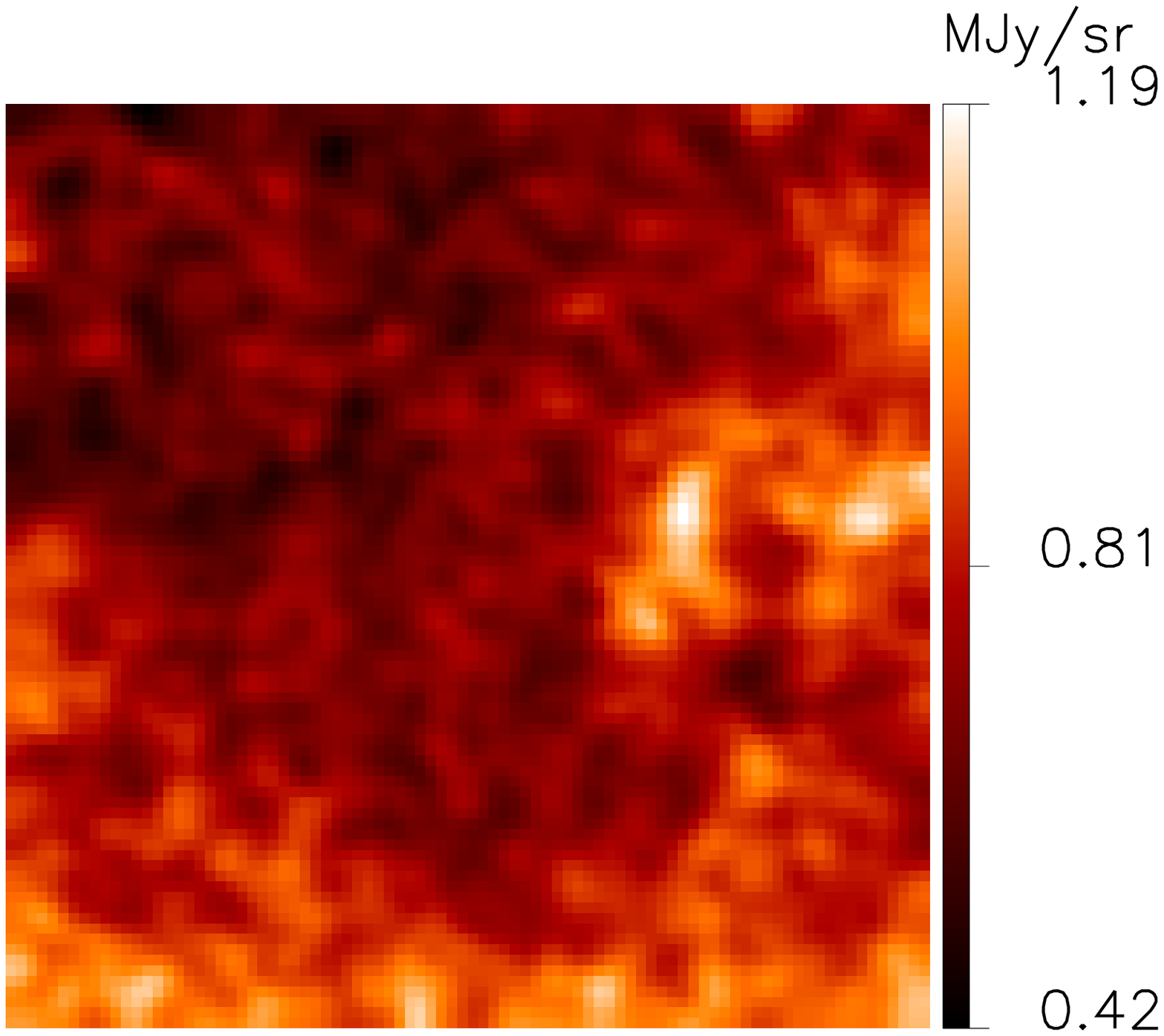}\\
\includegraphics[width=4.5cm, draft=false]{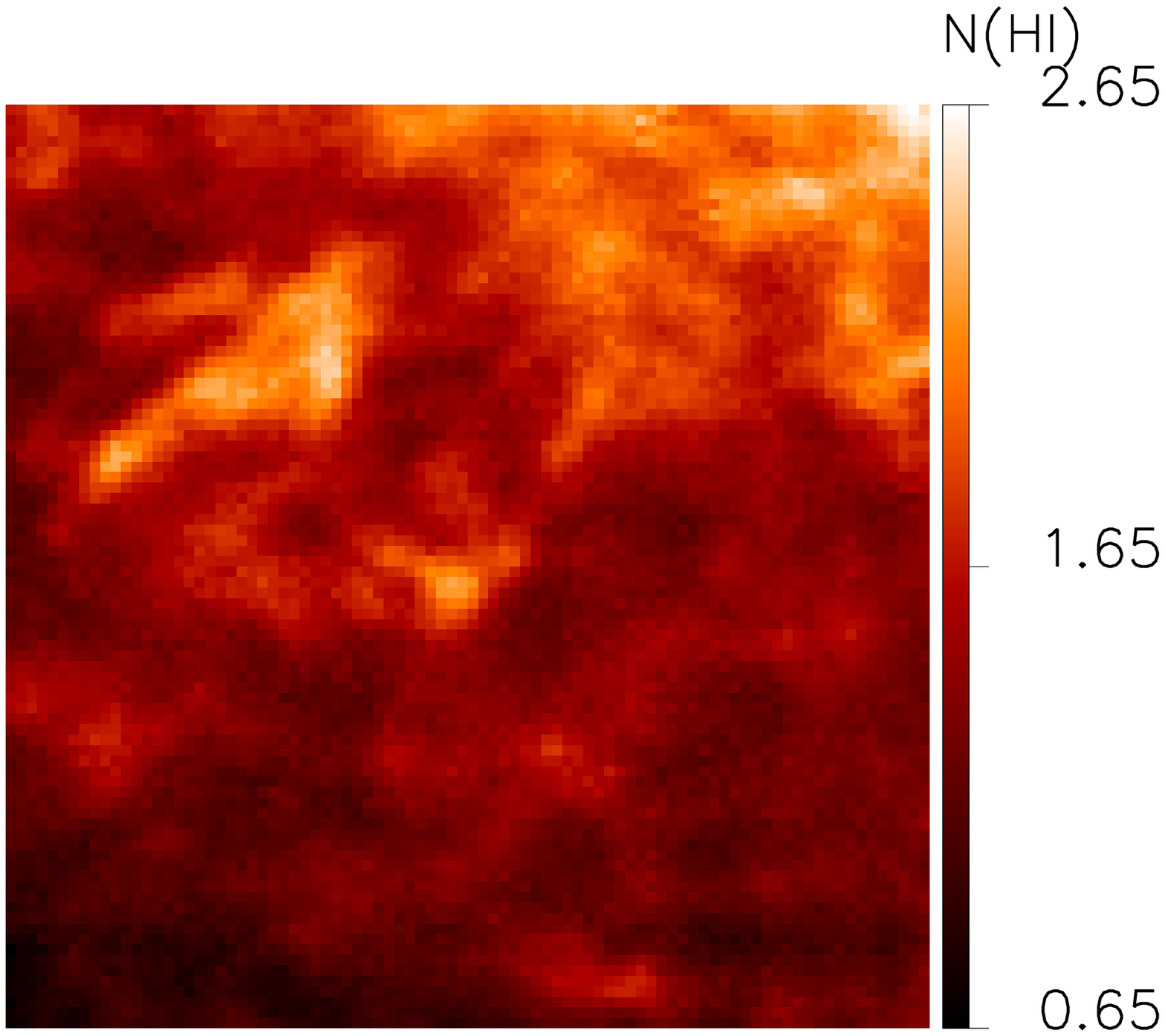}
\includegraphics[width=4.5cm, draft=false]{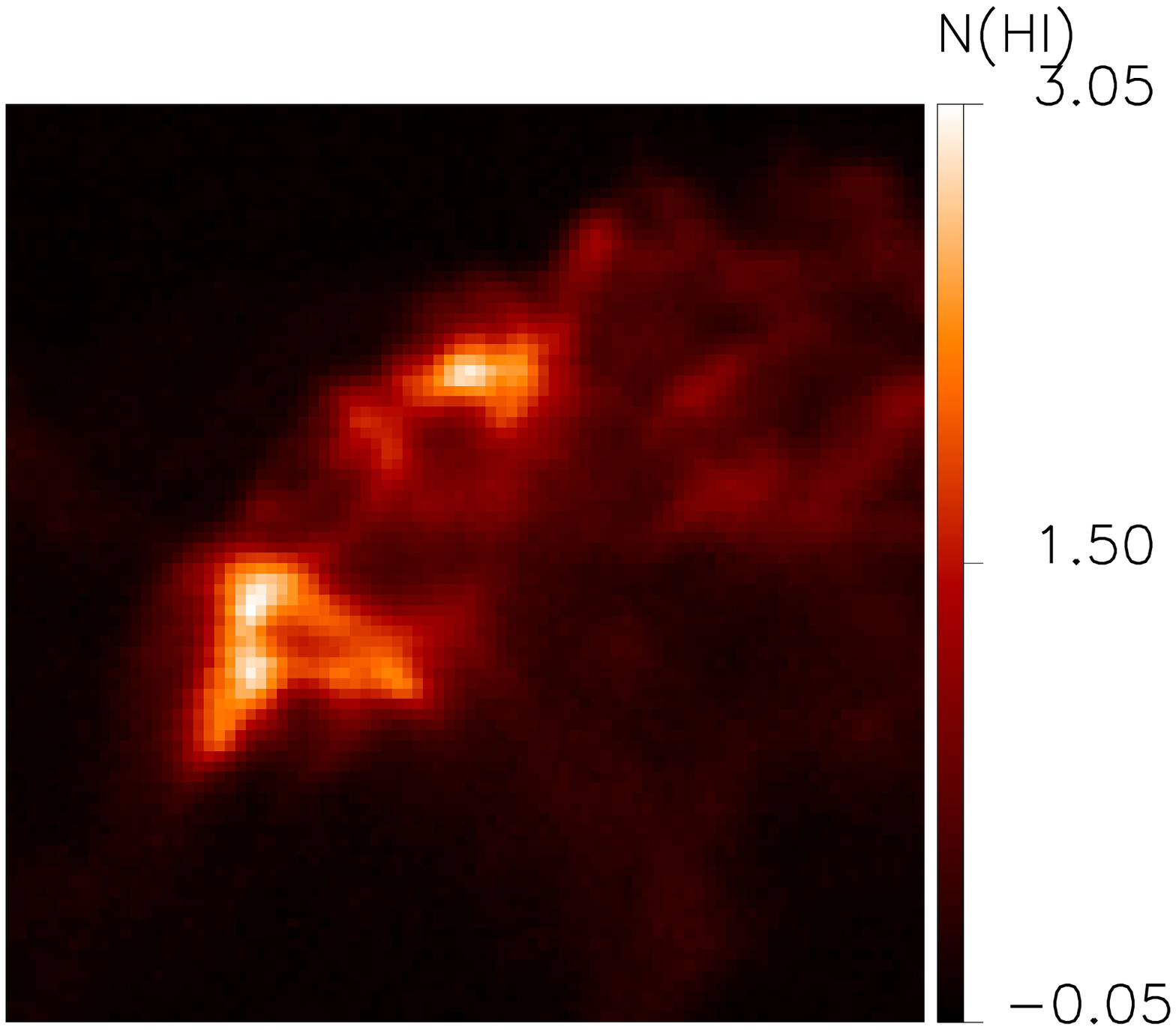}
\includegraphics[width=4.5cm, draft=false]{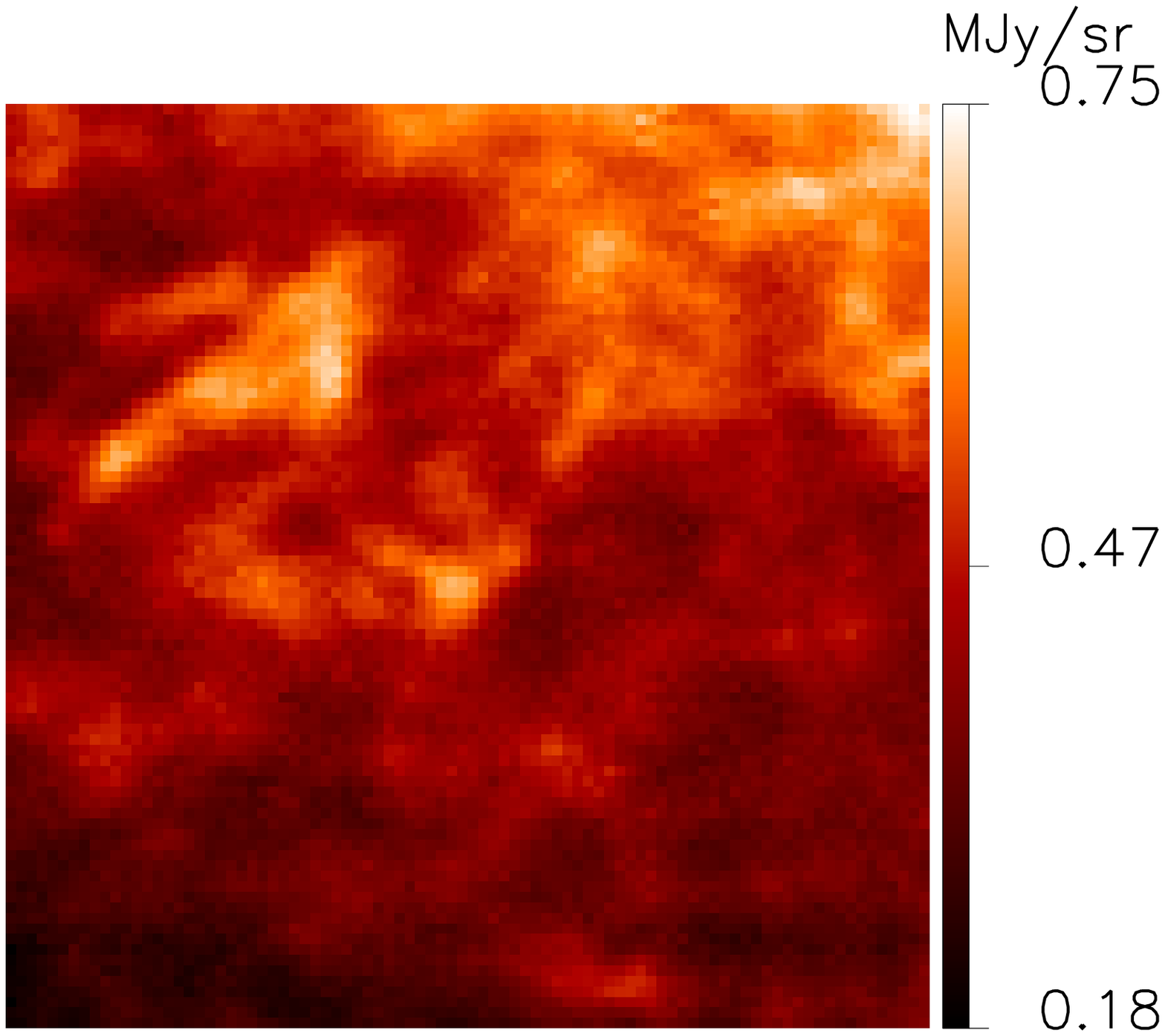}
\includegraphics[width=4.5cm, draft=false]{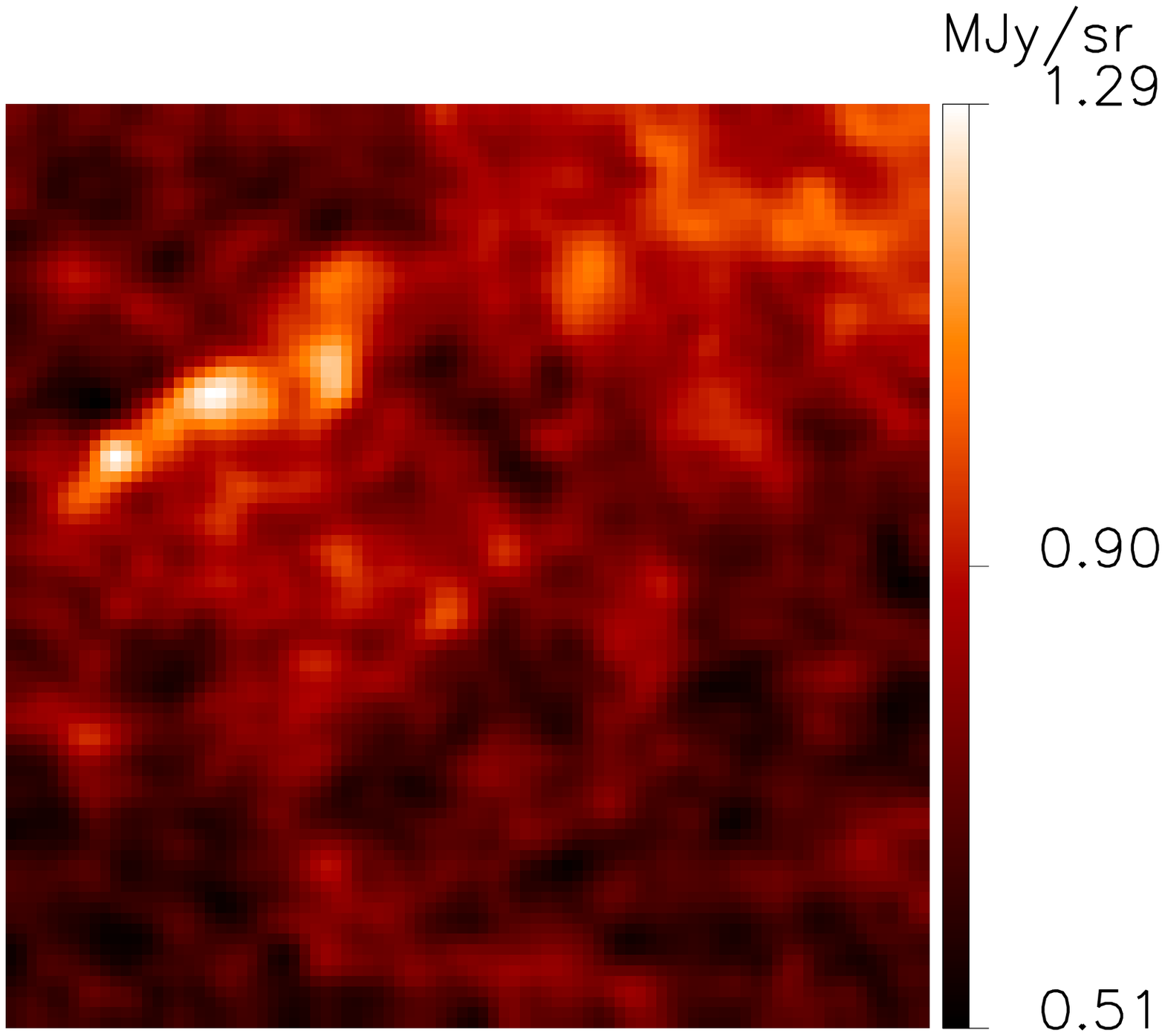}
\caption{\label{Fig_HI} \hi\ and dust maps for two fields: {\tt SP} {\it (top) } and {\tt AG} {\it (bottom).} The first two maps on the left show the \hi\ components (Local and IVC for {\tt SP}, IVC and HVC for {\tt AG}), the third maps show the 857\,GHz emission associated with \hi\ ($\sum_i \alpha^{i}_\nu N_{\rm HI}^{i}$) and the maps on the right side show the HFI 857\,GHz maps.
Those HFI maps have been convolved by the GBT beam to allow a better comparison by eye. \hi\ maps are given in
units of $10^{20}$ atoms cm$^{-2}$.  Note the correlation of the dust emission with the different \hi\ velocity components and its variation from field to field.}
\end{figure*}

\paragraph*{\hi\ components --} 
The \hi\ data in each field show different velocity components: a local component, typical of high-latitude \hi\ emission, intermediate-velocity clouds (IVCs) and high-velocity clouds (HVCs). These clouds are typically defined as concentrations of neutral hydrogen at velocities inconsistent with a simple model of differential Galactic rotation. The distinction between IVCs and HVCs is loosely based on the observed radial velocities of the clouds; IVCs have radial velocities with respect to the local standard of rest (LSR) of $30 \le |V_{\rm LSR}| \le 90\,$km s$^{-1}$, while HVCs have velocities $|V_{\rm LSR}|>90\,$km s$^{-1}$. High-velocity clouds might be infalling clouds fueling the Galaxy with low-metallicity gas, whereas IVCs might have a Galactic origin \citep[e.g.][]{richter2001}. For each field, we constructed integrated \hi\ emission maps of the three velocity components. The selection of the velocity range for each component was based on
inspection of the median 21-cm spectrum and of the rms 21-cm spectrum ({\it i.e.\/}, the standard deviation of each channel map). It is fully described in the \citet{planck2011-7.12}. The \hi\ maps were then converted to \hi\ column density using the optically thin approximation:
\begin{equation}
  N({\rm HI})(x, y)=1.823 \times10^{18}\ \sum_{\rm{v}} T_{\rm b}(x, y, \rm{v})  \delta \rm{v} ,
\end{equation}
where $T_{\rm b}$ is the 21-cm brightness temperature and $\rm{v}$ the velocity.
Corrections have been applied for opacity \citep[see][]{planck2011-7.12}, they are lower than 5\% for our CIB fields.
As illustrated in Fig.~\ref{Fig_HI}, the different fields have clearly distinct \hi\ contributions, with, e.g., no local component in the direction of the {\tt AG\/} field.

\begin{table}
\begin{center}
\begin{tabular}{c|c|c|c|c} 
{\tt N1} & 217\,GHz & 353\,GHz & 545\,GHz & 857\,GHz \\ \hline
217  GHz & 1     &  0.56 & 0.53 &  0.49 \\ 
353  GHz &   & 1 & 0.84 & 0.77 \\
545  GHz &  &  & 1 & 0.91 \\ \hline
{\tt Bootes 1} & 217 GHz & 353  GHz & 545  GHz & 857  GHz \\ \hline
217  GHz & 1     &  0.44 & 0.39 &  0.39 \\ 
353  GHz &   & 1 & 0.75 & 0.74 \\
545  GHz &  &  & 1 & 0.89 \\ 
\end{tabular}\\
\caption{\label{tab:coeff_corr} Pearson correlation coefficient between CIB anisotropy maps (values are given for the {\tt N1} and {\tt Bootes 1} fields to illustrate the range of coefficients).  The high-frequency maps are highly correlated. A decorrelation is seen when going to lower frequencies. We interpret this decorrelation as reflecting the redshift distribution of CIB anisotropies (see text, \cite{fernandez-conde2008} and \cite{penin2011b}).}
\end{center}
\end{table}

\paragraph*{\hi-dust correlation --}  To remove the cirrus contamination from the HFI maps, we need to determine the far-IR to mm emission of the different \hi\ components identified with the 21\,cm observations.
We assumed that HFI maps, $I_\nu(x,y)$, at frequency $\nu$ can be represented by the following model
\begin{equation}
\label{eq_regress}
  I_\nu(x,y) = \sum_i \alpha^{i}_\nu N_{\rm HI}^{i}(x,y) + C_\nu(x,y),
\end{equation}
where $N_{\rm HI}^i(x,y)$ is the column density of the $i$th \hi\ component, $\alpha^i_\nu$ is the far-IR to mm -- \hi\ correlation coefficient of component $i$ at frequency $\nu$ and $C_\nu(x,y)$ is a residual. 
The correlation coefficients $\alpha_\nu^i$ (often called emissivities) were estimated using a $\chi^2$ minimization
given the \hi\ and HFI data and the model (Eq.~\ref{eq_regress}). Although the \hi\ column densities of the different velocity components are quite similar (see Fig. \ref{Fig_HI}), the emissivities may vary by factors of more than 10 between the local/IVC and HVC \citep[see][]{planck2011-7.12}, so it is important to consider them separately. 
The emissivities can be used to characterise the opacity and temperature of the dust emission in the different components. This is beyond the scope of this paper, but it is extensively discussed in the \citet{planck2011-7.12}.

\paragraph*{Cirrus contamination removal --} We removed from the HFI maps the \hi\ velocity maps multiplied by the correlation coefficients. Maps are shown in the last two columns of Fig.~\ref{Fig_HI} for two fields. The removal was made at the HFI angular resolution, even though the \hi\ map is of lower resolution.
This is not a problem because cirrus, with a $k^{-3}$ power-law power spectrum \citep{miville2007}, has negligible power between the GBT and HFI angular resolutions, in comparison to the power in the CIB. 

\paragraph*{Residual maps and power spectra --} 
Fig.~\ref{Fig_Pktot_pknoise} shows the HFI  power spectra before and after the dust removal in the {\tt SP} field. Cirrus removal has more impact for the two high-frequency channels. At 217\,GHz, the correction is very small (13\% at $\ell$=500). 
This method of using \hi\ data to remove the cirrus contamination from power spectra has also been successfully applied by \citet{penin2011a} at higher frequencies than ours, where the cirrus contamination is higher.
The authors have been able to isolate precisely the CIB anisotropies power spectra at 1875 and 3000\,GHz with {\it Spitzer\/} and IRAS/IRIS, in the {\tt N1\/} field.

\begin{figure}
\includegraphics[width=\linewidth, draft=false]{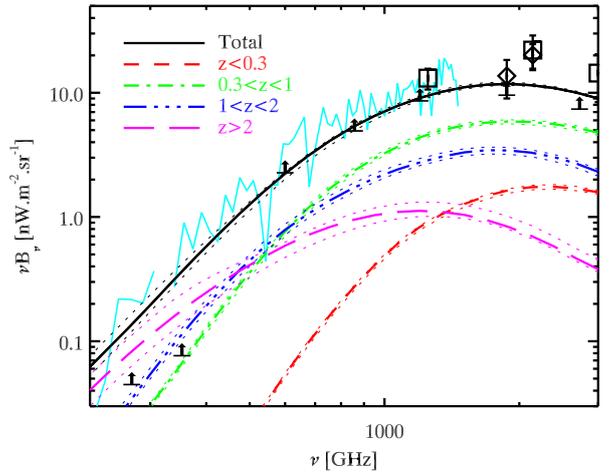}
\caption{\label{cib_z_distrib} Contribution to the CIB per redshift slice, extracted from \citet{bethermin2010}.
The black solid line is the CIB spectrum predicted by the model. The contribution to the CIB from $0<z<0.3$, $0.3<z<1$, $1<z<2$ and $z>2$ galaxies is given by the red short-dashed, green dot-dashed, blue three dot-dashed and purple long-dashed lines, respectively.  Lower limits coming from
the stacking analysis at 100~$\mu$m, 160~$\mu$m \citep{berta2010}, 250~$\mu$m, 350~$\mu$m, 500~$\mu$m \citep{marsden2009}, 850~$\mu$m \citep{greve2009} and 1.1~mm \citep{scott2010} are shown as black arrows.
The black diamonds give the \citet{matsuura2010} absolute measurements with {\it AKARI\/}.
The black square the \citet{lagache2000} absolute measurements with DIRBE/WHAM and the cyan line the \citet{lagache2000} FIRAS measurement.}
\end{figure}

The residual maps at the HFI angular resolution are shown in Fig.~\ref{Fig_CIB} for the {\tt N1\/} field. We clearly see that the cirrus has been efficiently removed. The bottom row shows the residual maps, smoothed at 10\arcm.
Common structures, corresponding to the CIB anisotropies, are clearly visible at the four frequencies.
Table \ref{tab:coeff_corr} gives the Pearson correlation coefficients between the CIB anisotropy maps.
They are about 0.9 between the 545 and 857\,GHz maps and 0.5 between the 217 and 857\,GHz CIB maps.
The decrease when the frequency difference between the maps is larger is expected because the contribution of
high-redshift galaxies to the CIB (and its anisotropies) increases with wavelength.
This is illustrated in Fig.~\ref{cib_z_distrib}, extracted from \citet{bethermin2010}, where we show the redshift distribution of the CIB.
The redshift distribution of correlated CIB anisotropies is discussed in
\citet{fernandez-conde2008, fernandez-conde2010} and \citet{penin2011b}.

\begin{center}
\begin{figure}
\includegraphics[width=8.cm, draft=false]{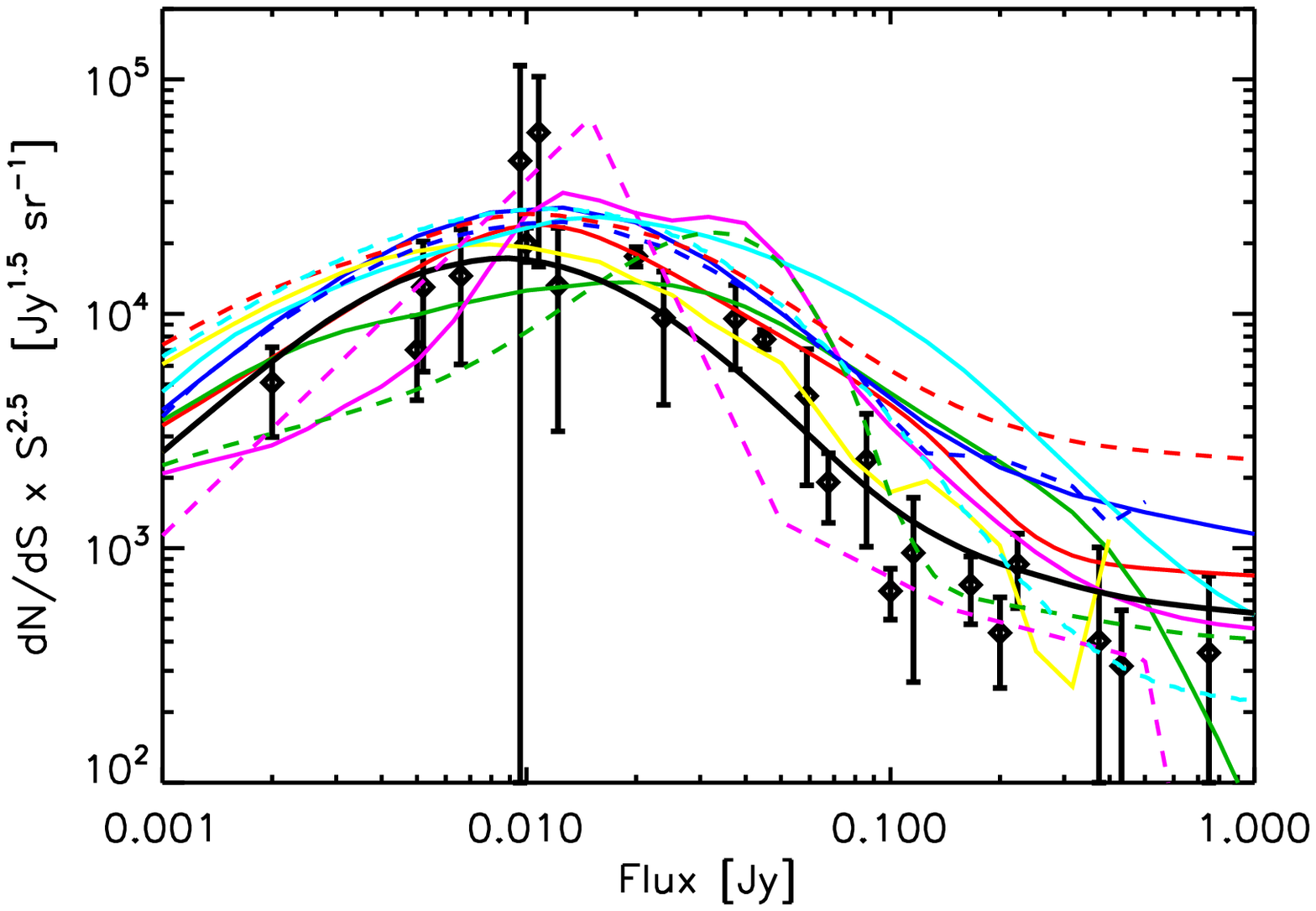}
\includegraphics[width=8.cm, draft=false]{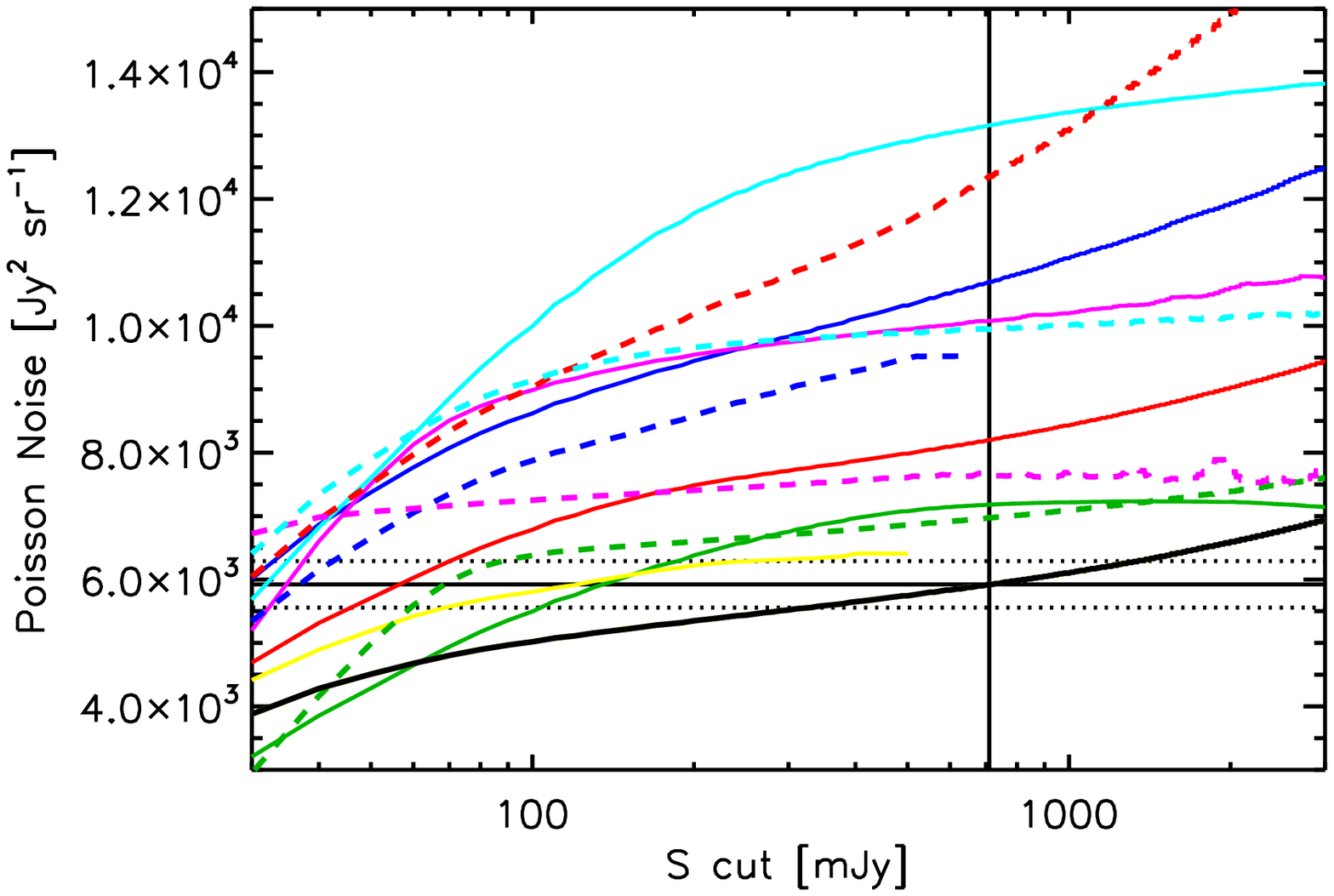}
\caption{\label{fig_poisson_350} A number of recent models of dusty-galaxy evolution and their associated shot noise for different flux cuts at 857\,GHz. {\it Top:} Comparison of the models with the {\it Herschel\/} and BLAST differential numbers counts. Models are from \cite{lagache2004, negrello2007, leborgne2009,patanchon2009, pearson2009, valiante2009, bethermin2010,franceschini2010, lacey2010, marsden2011, rowan2009, wilman2010}. Data points are from \cite{oliver2010, bethermin2010b, Glenn2010}. {\it Bottom:} Shot-noise level as a function of the flux cut for the same models (same colour and line coding between the two figures). The vertical and horizontal continuous dark lines show the \Planck\ flux cut and shot-noise level from Table \ref{tab:fluc_cut}, respectively. The \cite{bethermin2010} model is shown by the continuous dark line. This figure shows that models predicting a very high shot noise (e.g. continuous and dashed light-blue, red-dashed, continuous and dashed dark-blue lines) are incompatible with the measured number counts.}
\end{figure}
\end{center}

\section{Astrophysical and instrumental components of residual HFI maps power spectra \label{compo_pk}}

Once the CMB and cirrus have been removed, there are three main astrophysical contributors to the power spectrum at the HFI frequencies: two from dusty star-forming galaxies (with both shot noise, $C^{{\rm d, shot}}_\ell(\nu)$, and clustering, $C_{\ell}^{{\rm d, clust}}(\nu)$, components), and one from radio galaxies (with only a shot-noise component, $C^{{\rm r, shot}}_\ell(\nu)$, the clustering of radio sources being negligible, see \cite{hall2010}).
If the instrument noise and the signal are not correlated, the measured power spectrum $C_\ell(\nu)$ is
\begin{equation}
\label{eq:pow_spec_all_compo}
\begin{array}{lcl}
  C_\ell(\nu) &=& b_{\ell}^2(\nu) \left[ C_{\ell}^{{\rm d, clust}}(\nu) + C^{{\rm d, shot}}_\ell(\nu) + C^{{\rm r, shot}}_\ell(\nu) \right] \\
  && \\
  &+& N_{\ell}(\nu),
\end{array}
\end{equation}
where $b_{\ell}(\nu)$ is the beam window function, and $N_{\ell}(\nu)$ the power spectrum of the instrument noise.
Note that we neglect here the Sunyaev-Zeldovich \citep[SZ;][]{sz80} contribution to the power spectra.
Extrapolation of SPT \citep{lueker2010} and ACT \citep{dunkley2010} constraints show that SZ is negligible compared to CIB anisotropies at $\nu\ge 217\,$GHz.
Our goal is to accurately measure $C_{\ell}^{{\rm d, clust}}(\nu)$, which we present and extensively discuss in
Sect.~\ref{sect_5}.  We begin by discussing all other components of Eq.~\ref{eq:pow_spec_all_compo}
in this section.

\subsection{Shot noise \label{sect_sn}}

The shot noise arises from sampling of a background composed of a finite number of sources.  We assumed the
distribution is Poisson, so that its power spectrum is independent of $\ell$.
If we identify and remove all sources brighter than $S_{\rm cut}$, the shot noise from the remaining sources
fainter than $S_{\rm cut}$ is given by \citep[e.g.,][]{ScoWhi99}
\begin{equation}
C_\ell^{\rm shot}= \int_{0}^{S_{\rm cut}} S^2 \frac{dN}{dS}\ dS,
\label{eq:poisson}
\end{equation}
where $S$ is the source flux and $dN/dS$ the differential number counts.
These counts can be directly measured or derived from evolution models of the relevant population of
galaxies (dusty, star-forming and radio galaxies in our case).

\subsubsection{Star-forming, dusty galaxy shot noise, $C_\ell^{{\rm d, shot}}$ }

We used the recent model of \citet{bethermin2010} to compute the IR galaxy shot-noise power.
This is an updated version of the \citet{lagache2004} model that better reproduces new observational constraints
(e.g., from {\it Herschel}).
This new, empirical model uses the same galaxy spectral energy distribution (SED) templates as \citet{lagache2004},  but a
fully parametric evolution of the luminosity function.
The parameters of the model were determined by fitting the infrared/sub-mm number counts, and some mid-IR luminosity functions, with a Monte-Carlo Markov Chain (MCMC). More details on the model are given in Appendix \ref{A2}.
The derived shot-noise power is given in Table \ref{tab:fluc_cut}, with uncertainties computed from the MCMC.
The quoted numbers include statistical and photometric calibration uncertainties. 
This model has less energy output at high redshift ($z\simeq 2$) and consequently lower shot-noise power at long wavelength
than the \citet{lagache2004} model.
The shot noise levels depend on the flux cut, which itself has an uncertainty linked to the flux uncertainty in the ERCSC.  If we change the flux cut $S_{\rm cut}$ by 30\% in Eq. \ref{eq:poisson} based on the uncertainty in ERCSC fluxes,
the power spectra change by less than 5\% at all frequencies (and less than 1\% at 217\,GHz). 

As we will discuss in Sect.~\ref{sect_5}, the dusty galaxy shot-noise level will be a major factor in the interpretation of CIB anisotropy power spectra.
Because we are obtaining this value from a model, not measuring it directly in this paper (see Sect.~\ref{sect_5}), we briefly discuss here the constraints on the model and the plausible range of values using the 857\,GHz channel
as an example (the same conclusions are reached for the other \Planck\ channels).
Fig. \ref{fig_poisson_350} shows a compilation of models from the literature superimposed on the latest number counts observed by BLAST and {\it Herschel} and the expected shot noise as a function of $S_{\rm cut}$.
First we see, as stated above, that a small variation in $S_{\rm cut}$ leads to only a small variation in shot-noise power.
Second, we see that the highest shot-noise level is around 13{,}500\,Jy${}^2\,$sr${}^{-1}$, a factor $\sim 2.3$ above
our nominal value, but it comes from a model that overestimates the observed number counts by a large factor
(3--4 for $50 \le S\le 300\,$mJy).
Models that agree reasonably well with the number counts have a shot-noise level below $8000\,$Jy$^2\,$sr${}^{-1}$.
The \citet{bethermin2010} model has the lowest shot noise.
However, it is currently the model that best reproduces all available constraints, from the mid-infrared to the millimeter, including the differential contribution of the $S_{24}\ge 80\,\mu$Jy sources to the CIB as a function of redshift, which is a difficult observation to predict.  Eventually, the shot noise derived from this model agrees well also with the {\it Herschel}/SPIRE measurements in the Lockman-SWIRE field, when none of the point sources is removed (Amblard, private communication), as detailed in Sect.~\ref{compar_blast_spire}.

\begin{table*}
\begin{center}
\begin{tabular}{c|c|c|c|c|c} 
Frequency (GHz) & 143 & 217 & 353 & 545 & 857 \\ \hline
Flux cut (mJy) & 245 & 160 & 325 & 540 & 710 \\ \hline  \hline
IR shot noise$^{1}$ & 1.4$\pm$0.3 & 12.2$\pm$2.9 & 138$\pm$22 & 1150$\pm$92 & 5923 $\pm$367\\ 
(Jy${}^2$ sr${}^{-1}$)  & & & & & \\ \hline
Radio shot noise$^{2}$ & 7.1   &     4.0        &    $<$3.4       &    $<$5.7    &    $<$7.4  \\ 
(Jy${}^2$ sr${}^{-1}$)  & & & & & \\ \hline \hline
IR shot noise$^{1}$ & $(1.0\pm 0.2)\times 10^{-5}$ & $(5.3\pm1.2)\times 10^{-5}$ & $(1.7\pm 0.3)\times 10^{-3}$ & $0.34\pm 0.03$ & $1187 \pm 74$ \\ 
($\mu K_{\rm CMB}^2$)  & & & & & \\ \hline
Radio shot noise$^2$ & $5.2\times 10^{-5}$  &  $1.7\times 10^{-5}$  & $<4.1\times 10^{-5}$  &  $<1.7\times 10^{-3}$  &  $< 1.5$  \\ 
($\mu K_{\rm CMB}^2$)  & & & & & \\
\end{tabular}\\ 
\caption{Flux cut from the ERCSC for our six fields, and the shot-noise power for dusty and radio galaxies
appropriate to those cuts (see text).  Values for shot noise$^{1}$ are derived from the dusty galaxy evolution model of \cite{bethermin2010}, while those for shot noise$^2$ are from the radio galaxy evolution model of \cite{de_zotti2005} (see text for more details). \label{tab:fluc_cut}}
\end{center}
\end{table*}
\endPlancktablewide

\subsubsection{Radio galaxy shot noise, $C_\ell^{{\rm r, shot}}$ }
The shot-noise power from radio galaxies is subdominant to that from dusty sources at the frequencies relevant
to CIB anisotropy analysis.
The radio galaxy shot-noise power can be estimated from the model of \citet{de_zotti2005}.
At frequencies $\leq$~100\,GHz, the model agrees with the source counts computed using the extragalactic
radio sources from the ERCSC.
At 143 and 217\,GHz, and for fluxes below 300\,mJy ({\it i.e.\/}, the case listed in Table~\ref{tab:fluc_cut})
the \cite{de_zotti2005} model agrees with the source counts of \citet{vieira2010}.
At higher fluxes the model needs to be scaled to reproduce the number counts obtained using the ERCSC.
The estimated scaling factors are 2.03 and 2.65 at 143 and 217\,GHz, respectively \citep[see][]{planck2011-6.1}.
At even higher frequencies the number counts by the ERCSC are no longer complete.  We therefore use the 217\,GHz scaling factor to set upper limits for the shot noise.  It is negligible compared to $C_\ell^{d,{\rm shot}}$ at these frequencies (see Table~\ref{tab:fluc_cut}).
Changing the flux cut by 30\% affects the shot noise by 30\%, but because the radio contribution is subdominant at the frequencies relevant for CIB anisotropy analysis, it has little impact on our results.

\subsection{The beam window function, $b_{\ell}(\nu)$ \label{sec:beam}}

\begin{figure}
\begin{center}
\includegraphics[width=8cm]{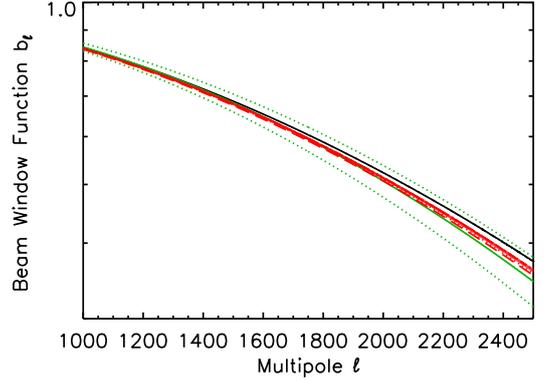}
\end{center}
\caption{\label{compar_beam}  Effective beam window functions ($b_{\ell}$) from FICSBell (black) and FEBeCoP (red) at 545\,GHz (see Sect. \ref{sec:beam} for more details). The six FEBeCoP beam window functions from each field are superimposed (red lines). Also shown for comparison is the Gaussian beam with a FWHM of $4.72^\prime \pm 0.2^\prime$ (green lines), which is the equivalent FWHM of the beam determined on Mars.}
\end{figure}

Because the HFI beams are not azimuthally symmetric, the scanning strategy has to be
taken into account in modelling the effective beam response. 
We used two different methods to compute the effective beam: FEBeCoP and FICSBell. With FEBeCoP, we computed one 
effective beam per field, with FICSBell, one effective beam for the entire sky.

\paragraph*{FICSBell --}  The FICSBell method (Hivon et al, in prep) generalizes the approach of
\citet{hinshaw2007} and \citet{smith2007} to polarization and to include other sources of systematics.
The different steps of the method used for this study can be summarized as follows:
\begin{enumerate}
\item The scanning-related information ({\it i.e.\/}, statistics of the orientation of
each detector within each pixel) is computed first, and only once for a given
observation campaign. The hit moments are only computed up to degree 4, for
reasons described below.
\item The (Mars-based) beam map or beam model of each detector, $d$, is decomposed
into its spherical harmonic coefficients
\begin{equation}
	b^d_{\ell s} = \int d{\bf r}\ B_d({\bf r}) Y_{\ell s}({\bf r}),
\end{equation}
where $B_d(\bf{r})$ is the beam map centred on the North pole, and
$Y_{\ell s}(\bf{r})$ is a spherical harmonic.
Higher $s$ indices describes higher degrees of departure from azimuthal symmetry
and, for HFI beams, the coefficients $b^d_{ls}$ are decreasing functions of $s$ at
most $\ell$ considered. It also appears that for $\ell<3000$, the coefficients with
$|s|>4$ account for $\le 1\%$ of the beam throughput. 
For this reason, only modes with $|s| \le 4$ are considered in the present analysis
(\citet{armitage-caplan2009} reached a similar conclusion in their
analysis of \Planck-LFI beams).
\item  The $b^d_{\ell s}$ coefficients computed above are used to generate $s$-spin
weighted maps for a given CMB sky realisation. 
\item The spin-weighted maps and hit moments of the same order, $s$, are
combined for all detectors involved, to provide an ``observed'' map. 
\item The power spectrum of this map can then be computed, and compared to the
input CMB power spectrum to estimate the effective beam window function over 
the whole sky, or over a given region of the sky. 
\end{enumerate}
Monte-Carlo (MC) simulations in which the sky realisations are changed can be performed
by repeating steps 3, 4, and 5. The impact of beam model uncertainties can be studied by
including step 2 into the MC simulations.

\begin{figure}
\begin{center}
\includegraphics[width=8cm]{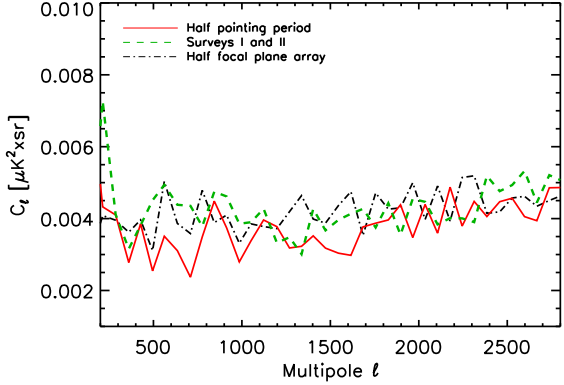}
\end{center}
\caption{\label{Fig_pknoise_SP}  Three independent noise-power-spectrum measurements in the {\tt SP} field at 353\,GHz: red continuous line, half pointing period; green dashed, surveys I and II; black dot-dashed, half focal plane array).}
\end{figure}

\paragraph*{FEBeCoP --}  
As mentioned in Sect. \ref{data}, map making reduces time-ordered data to pixelised maps. Each pixel of a map represents a convolution of the true sky with the combined effect of scanning beam and scan pattern. FEBeCoP computes this combination of beams and scans---the effective beams---as is, in the pixel space.
The FEBeCoP methodology and algorithm has been described in \citet{mitra2010}, and \citet{planck2011-1.5}.
Below, we list for completeness the essential steps made in computation of the beam window functions:
\begin{enumerate} 

\item For each pixel $i$ in the map (or CIB field) we  computed the Fourier-Legendre transform, $B_\ell$, of the pixel space effective beams $B_i(\hat{\mathbf{\Omega}})$ using the formula

\begin{equation}
b_\ell \ = \ \int_{\Delta \mathbf{\Omega}_i} d\hat{\mathbf{\Omega}} \,  P_\ell(\hat{\mathbf{\Omega}}_i \cdot \hat{\mathbf{\Omega}}) \, B_i(\hat{\mathbf{\Omega}}) \, ,
\end{equation}

where $\hat{\mathbf{\Omega}}_i$ is the direction vector of the centre of the $i$th pixel on the sky, $P_\ell$ represents Legendre polynomials of order $\ell$ and the integration is performed over the (small) solid angle $\Delta \mathbf{\Omega}_i$, outside which the beam can be taken as zero. This formula can be readily transformed to a discretised form with a careful correction for the ``pixel window function'' as
\begin{equation}
b_\ell \, W^p_\ell \ \approx \ \Omega_{\rm pix} \sum_j P_\ell(\hat{\mathbf{\Omega}}_i \cdot \hat{\mathbf{\Omega}}_j) \, B_i(\hat{\mathbf{\Omega}}_j) \, ,
\end{equation}
where the summation is over pixels that fall inside the beam solid angle $\Delta \mathbf{\Omega}_i$, $\Omega_{\rm pix}$ is the area of each (equal area) pixel and $W^p_\ell$ is the pixel window function that compensates for the systematic error that is introduced when integration over a pixel is replaced by the value of the integrand at the pixel centre times the area of the pixel.

\item We then computed $b_\ell$ at uniformly sampled directions in each field to find the {\em average window functions}. The samples were chosen as the HEALPix pixel centres at a coarser resolution ($N_{\rm side} = 128$) to ensure uniform sampling. Thus we obtained the average window functions for each frequency and field.

\item To validate the average window functions obtained using the above prescription, we performed Monte-Carlo simulations separately for each field and each frequency. We simulated $16$ realisations of the sky starting from a $\propto \ell^{-2}$ angular power spectrum, which are convolved in two ways -- (1) with FEBeCoP-generated effective beams in pixel space and (2) with analytical Gaussian beam in harmonic space for a beam size appropriate for the given frequency channel. The convolved maps were then ``masked'' using a function that is unity in the given field and smoothly (in $\sim 25$\% of field radius) goes to zero outside the field. Finally, we computed the ratio of the angular power spectra of these two maps, multiplied the ratio by the theoretical window function for the same beam size and averaged over the realisations. Though these ``transfer functions'' suffer from ringing effects often seen in Fourier transforms of a narrow function, they wiggle around the average window functions, confirming the validity of the latter.
\end{enumerate}

Fig. \ref{compar_beam} shows the FICSBell and FEBeCoP effective beams at 545\,GHz.
Also shown is the Gaussian beam with a FWHM of $4.72\arcm\pm0.21\arcm$.
This is the average FWHM of the scanning beam, determined on Mars obtained by unweighted averaging the individual
detectors FWHM. Each FWHM is that of the Gaussian beam, which would have the same solid angle
as that determined by using a full Gauss-Hermite expansion on destriped data \citep[see][for more details]{planck2011-1.5}.
We see a quite good agreement between the FICSBell, all-sky and FEBeCoP, small-field effective window functions, with a 2\% difference at $\ell\simeq 2000$, the highest $\ell$ that will be considered for CIB anisotropy analysis
(see  Sect.~\ref{ang_pow}).
We also see from the figure that the error on the input scanning beam is larger than this difference and will dominate the uncertainties at high $\ell$ and high frequency (Sect. \ref{se:systes}). Below we will use the FEBeCoP window functions because they are exactly computed for each of our fields.

\begin{figure}[here]
\begin{center}
\includegraphics[width=6.5cm]{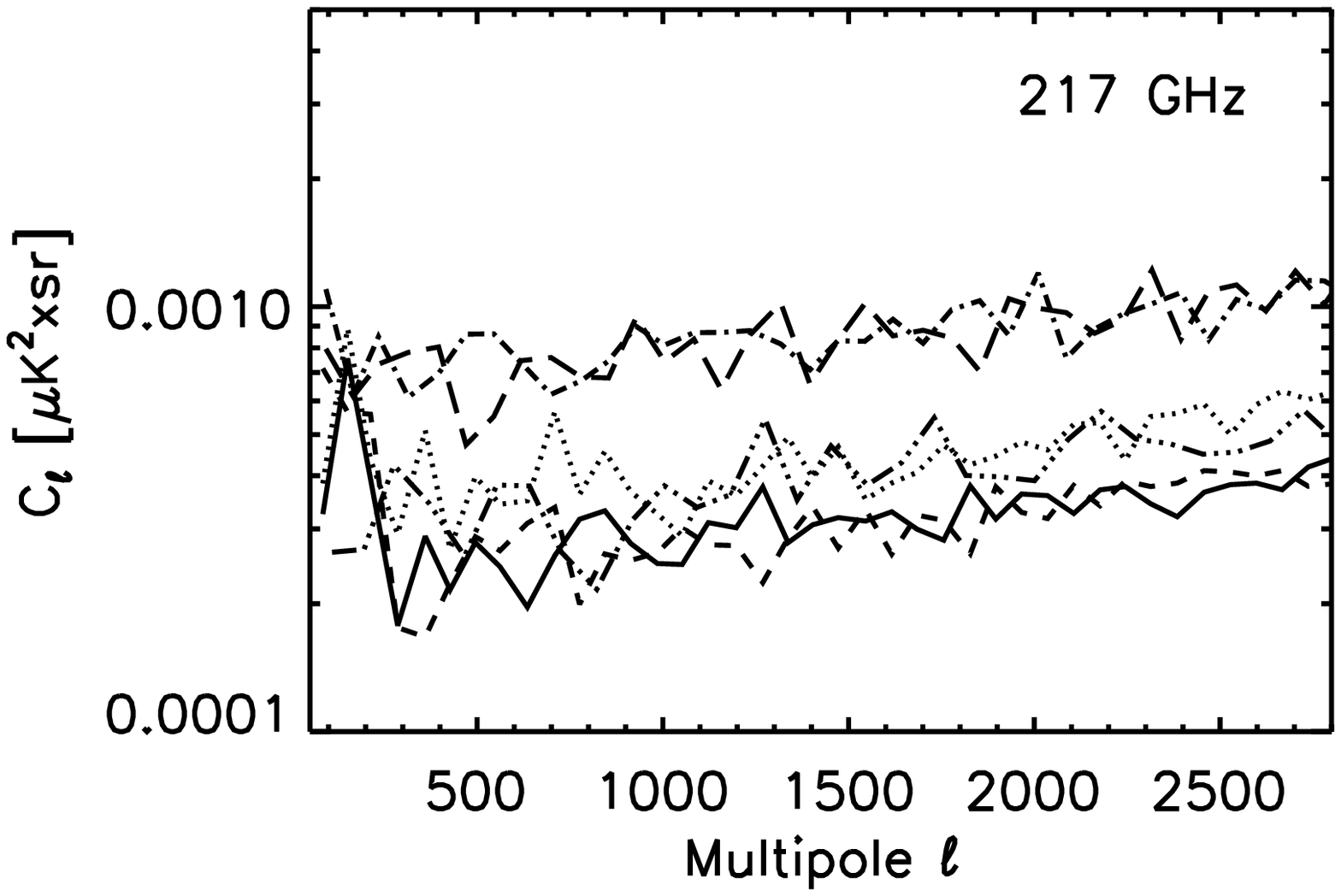}
\includegraphics[width=6.5cm]{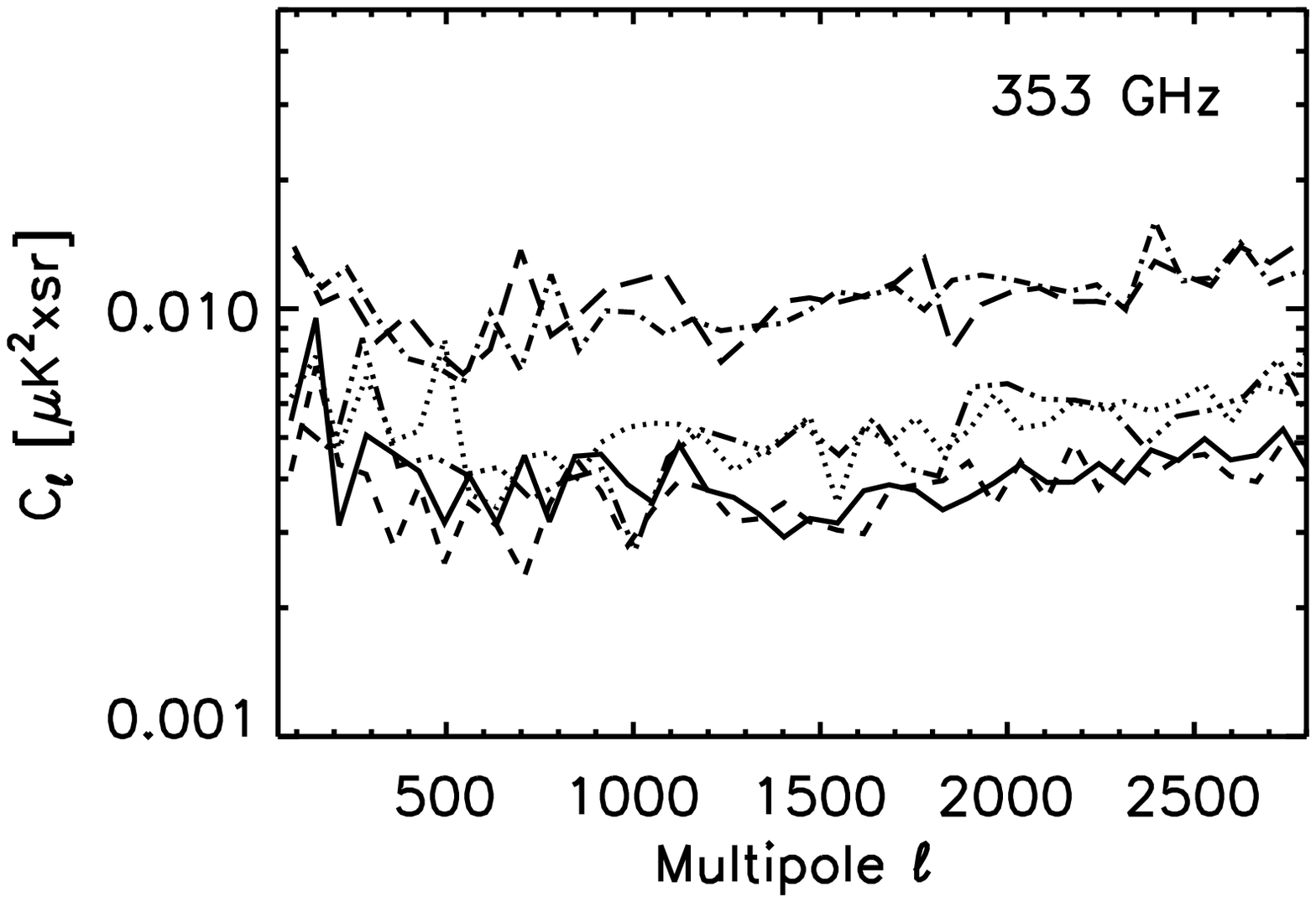}
\includegraphics[width=6.5cm]{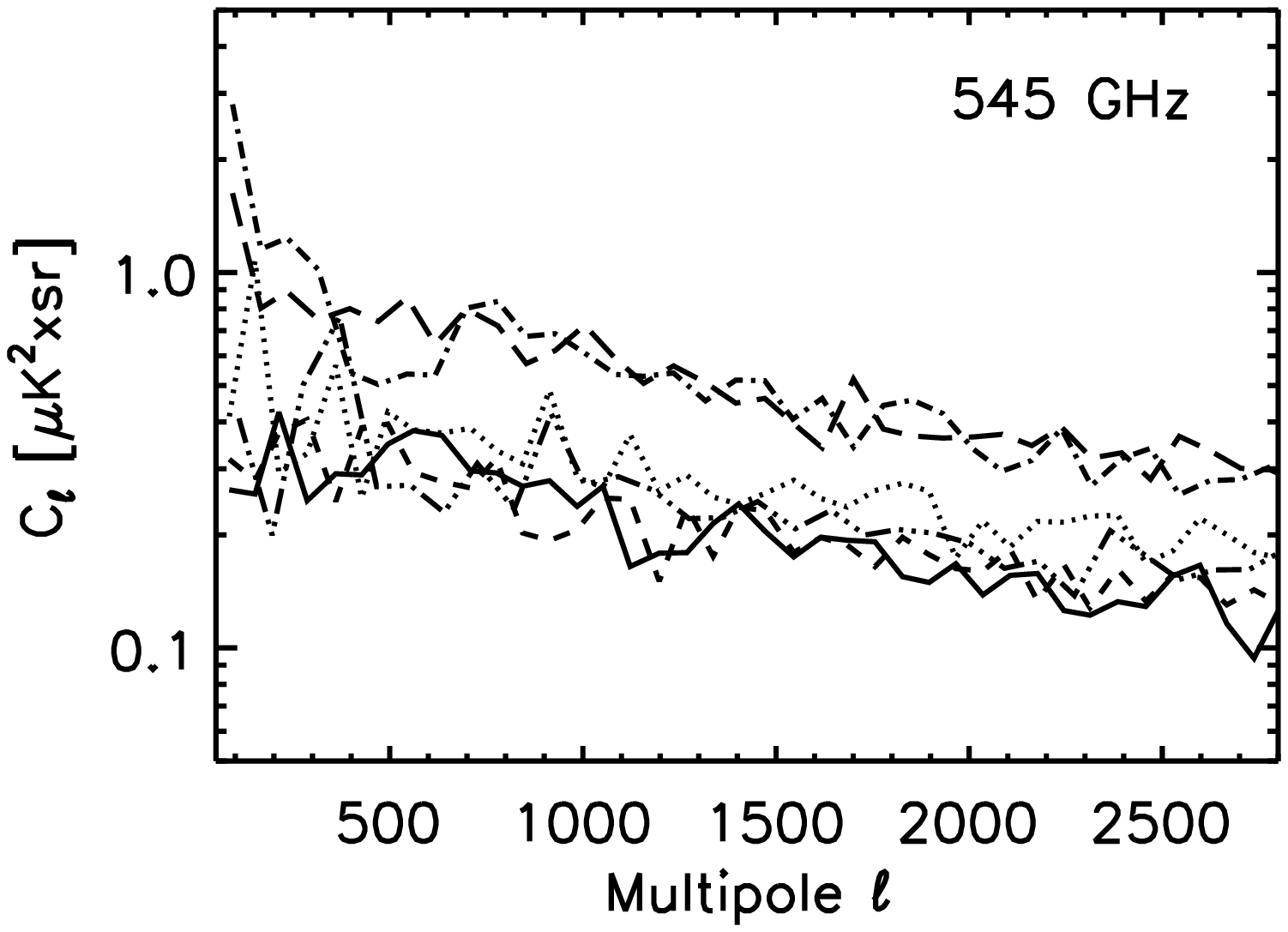}
\includegraphics[width=6.5cm]{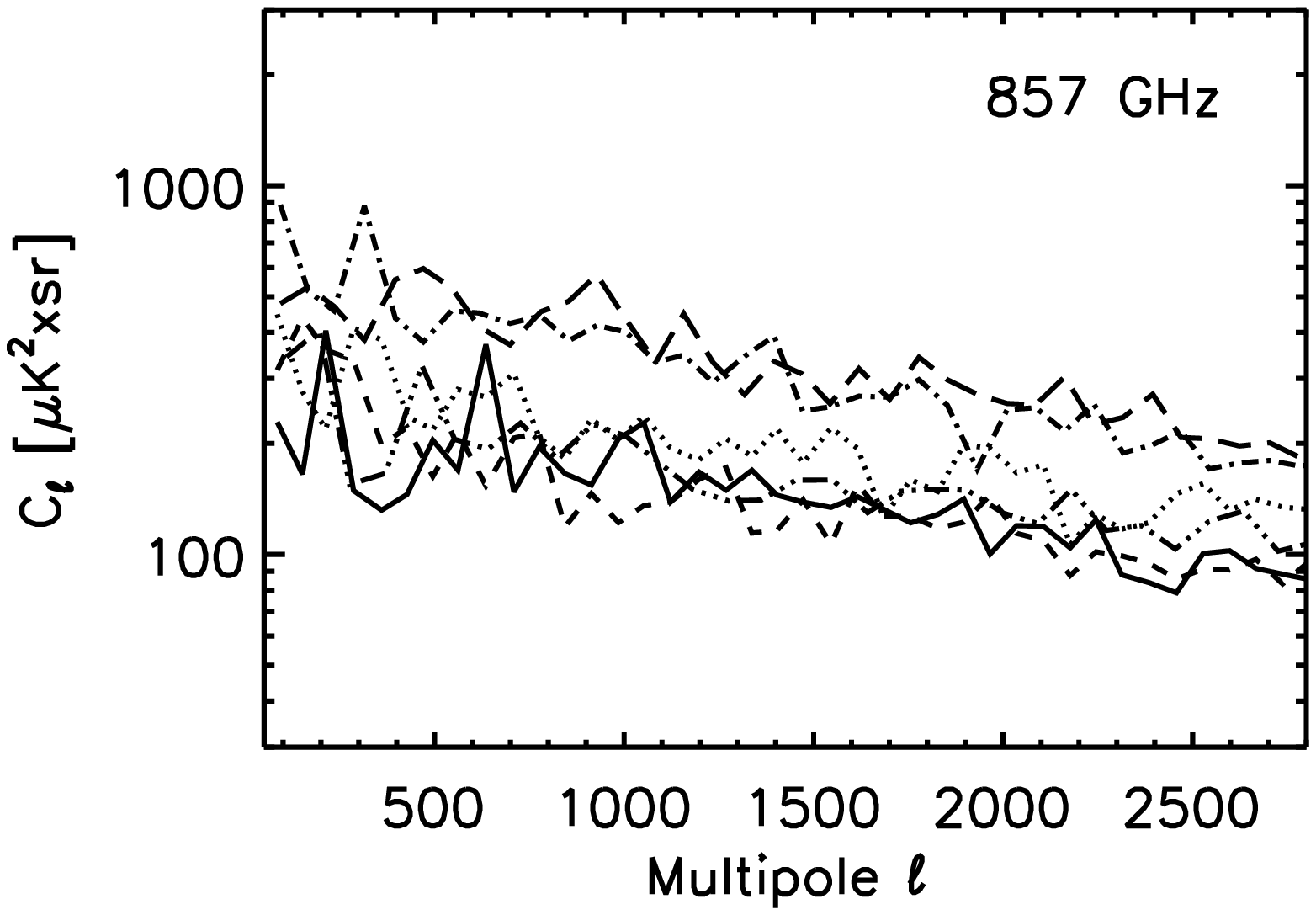}
\end{center}
\caption{\label{Fig_pknoise} Instrument noise power spectra of the six fields obtained using half-pointing period maps. From {\it top\/} to {\it bottom}: 217, 353, 545 and 857\,GHz (continuous: {\tt N1}, dotted: {\tt AG}, dashed: {\tt SP},  dash-dotted: {\tt Bootes 1}, long-dash: {\tt Bootes 2}, dash-3 dotted: {\tt LH2}).}
\end{figure}

\subsection{Instrument noise, $N_{\ell}(\nu)$ \label{se:inst_noise}}

We can use three different jack-knife difference maps to derive noise power spectra: maps made from the first and
second halves of each pointing period (a half-pointing period is of the order of 20 minutes), maps made using half of
the focal plane array, and maps using the two different surveys (surveys I and II).  In each case the noise power spectrum, $N_\ell$, is obtained by measuring the power spectrum of the difference maps.
The three methods give similar $N_\ell$, as is illustrated for one frequency and one field in
Fig.~\ref{Fig_pknoise_SP}.
We chose, however, to use the half-pointing period maps because (1) the two survey maps are only fully covered for the {\tt LH2} and {\tt SP} fields and (2) there are only three bolometers at 545 and 857\,GHz, making half-focal plane maps less accurate.
We also computed the noise power spectrum from the difference between the auto- and cross- power spectrum of the two half maps. In the range of interest, $1500\le \ell \le 2100$, where the contribution from the noise becomes important, they agree at better than 0.5, 1, 3, and 4\% at 217, 353, 545 and 857\,GHz, respectively.  Fig. \ref{Fig_pknoise} shows the noise power
spectra for all fields. They are nearly flat, the deviation from flatness is caused by the effect of deconvolution from the instrumental response at high frequency and residual low-frequency noise. Removing the ERCSC sources has no impact on the noise
determination. 

Fig. \ref{Fig_Pktot_pknoise} shows the noise power spectra compared to the HFI map power spectra for one illustrative field. We see that we have a very high signal-to-noise ratio.
At 545 and 857\,GHz, the signal is dominating even at the highest spatial frequencies.
At 217 and 353\,GHz, the residual signal ({\it i.e.\/}, CMB- and cirrus-cleaned) is comparable to the noise at high
$\ell$ ($\ell \ge 2000-2500$ depending on the field).

\subsection{\label{sect_conta} Additional corrections}

Two additional corrections linked to the CMB cleaning were made for the power spectra.
First we removed the extra instrument noise that has been introduced by CMB removal:
\begin{equation}
N^{\text{CMB}}_{\ell}(\nu)= N_{\ell}(\nu_{143})  \times w_\ell^2 \times \left( \frac{b_{\ell}(\nu)}{b_\ell(\nu_{143})} \right )^2,
\label{eq:xtranoise}
\end{equation}
with $\nu$ equal to 217 or 353\,GHz. $N_{\ell}(\nu_{143})$ is the noise power spectrum of the 143\,GHz map. It is computed as the noise in the other frequency channels, using the half-pointing period maps, following
Sect.~\ref{se:inst_noise}.

Second, owing to the lower angular resolution of the 143\,GHz channel compared to the 217 and 353\,GHz, we also had to remove the CMB contribution that is left close to the angular resolution of the 217 and 353\,GHz channels:
\begin{equation}
C^{\text{CMBres}}_{\ell}(\nu) = C^{\text{CMB}}_{\ell}(\nu) \times F_{\rm p}^2 \times b^2_\ell(\nu)\times \left [ 1 - w_\ell \right ]^2,
\label{eq:cmb_left}
\end{equation}
with $F_{\rm p}$ the pixel and reprojection transfer function (detailed in Sect.~\ref{se:poker}).\\

Finally, we had to assess the level of the astrophysical components that were removed from or added to the 217 and 353\,GHz channels, using the filtered 143\,GHz channel as a CMB template. Cirrus emission is highly correlated between 143, 217 and 353\,GHz channels.  Consequently, filtered cirrus emission was removed from the 217 and 353\,GHz. This has no impact on our CIB anisotropy analysis because this extra cirrus removal only modifies the emissivities, with no consequence on our residual maps (it should be understood for a further interpretation of the \hi-correlated dust emission, which is not the goal of this paper).
We expect the shot-noise powers to be quite decorrelated for the (143, 217) and (143, 353) sets of maps because the 143\,GHz shot noise is dominated by radio sources, whereas the 217 and 353\,GHz shot noise is dominated by dusty galaxies (see Table \ref{tab:fluc_cut}). To have an idea of the maximal effect ({\it i.e.\/}, 
perfect decorrelation between shot noise at 143, and 217, and 353\,GHz) we computed the contamination by the 143\,GHz shot noise, summing the contribution of the radio and dusty galaxies and following
\begin{equation}
\label{eq_pk_res}
C_\ell(\nu) = C^{\rm shot}(\nu_{143}) \times \left(\frac{b_\ell(\nu)}{b_\ell(\nu_{143}) }w_\ell \right )^2.
\end{equation}
The last term accounts for the filtering and "re-beaming" of the 143\,GHz map. The contamination is the highest in the 217\,GHz channel. It is a factor 1.2 and 120 smaller than the sum of the predicted radio and dusty galaxies shot-noise powers at 217 \,GHz at $\ell \simeq$ 200, and 2000, respectively, but is equivalent at $\ell \simeq$1000.
Anyway, it is smaller by factors of 20, 2.9 and 325 than the CIB anisotropies at 217\,GHz, at $\ell$= 200, 1000, and 2000, respectively. Because this is the maximal contamination and because it is quite low (and completely negligible at high $\ell$), we did not apply any correction to the CIB anisotropy power spectra. 

We still have to consider the case of CIB-correlated anisotropies at 143\,GHz. They have been marginally constrained at 150\,GHz by SPT and ACT at high $\ell$.  The power is $<5.2\times 10^{-6}\,\mu$K$^2$ and $<9.8\times 10^{-6}\,\mu$K$^2$
at $\ell=3000$ in \cite{dunkley2010} and \cite{hall2010}, respectively. This contribution is also completely negligible compared to the signal at 217\,GHz. 

In conclusion, we can ignore the CIB and cirrus components that are left in the CMB maps.

\section{Angular power spectrum estimation \label{ang_pow}}

The angular power spectrum estimator used in this work is \poker\ \citep{poker}, which is an adaptation to the flat
sky of the pseudo-spectrum technique developed for CMB analysis (see {\it e.g.} \master, \citealt{hivon2002}).
In brief, \poker\ computes the angular power spectrum of the masked data
(a.k.a., the \emph{pseudo-power spectrum}) and deconvolves it from the power spectrum of the mask to obtain an
unbiased estimate of the binned signal angular power spectrum.
We summarize the main features of \poker\ in the following section and then detail how it was used to produce the
final estimate of the CIB anisotropy power spectrum and its associated error bars.

In the following, the power spectrum associated to CIB anisotropies will be denoted $C_\ell$ and its unbiased estimator in the flat-sky approximation $P(\ell)$. As already suggested, this final estimate makes use of the power spectrum of the {\it masked\/} data. This so-called pseudo-power spectrum will be denoted $\hat{P}(\ell)$. In the flat-sky approximation, the standard angular frequencies labelled by their zenithal and azimuthal numbers, usually called $\ell$ and $m$ respectively, are replaced by an "angular" wave-vector $\vec{\ell}$; its norm $\ell$ is equal to the zenithal number \citep[see e.g., the appendix of][]{WCDH}.
Finally, we will assume that CIB anisotropies arise from a statistically isotropic process.
As is the case for the CMB, the CIB fluctuations are viewed as isotropic and homogeneous stochastic variables on the celestial sphere, leading to
\begin{equation}
	\left\langle a(\vec{\ell})a^\star(\vec{\ell}')\right\rangle=(2\pi)^2C(\ell)\delta^2(\vec{\ell}-\vec{\ell}'),
\end{equation}
with $a(\vec{\ell})$ the Fourier coefficients of CIB anisotropies.
This assumption is theoretically reasonable, moreover, we checked that $\left|a(\vec{\ell})\right|^2$ computed from our CIB maps does not depend on the direction of $\vec{\ell}$. 

\begin{figure}
\begin{center}
\includegraphics[width=\linewidth]{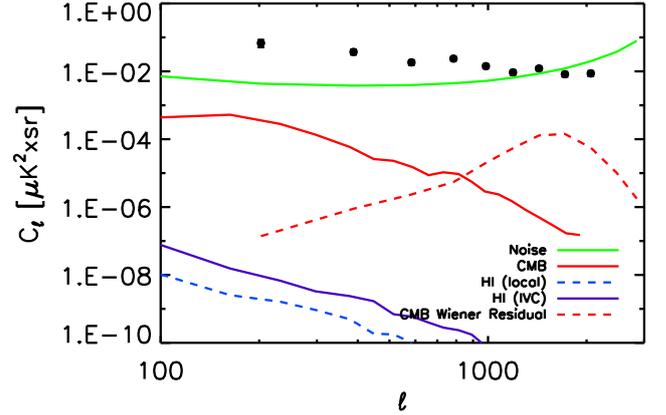}
\end{center}
\caption{\label{fig:error_budget} Contribution of residuals to the final CIB anisotropy 
estimate (illustrated here with the {\tt SP} field at 353\,GHz). The bias induced by each dust and CMB component is negligible
compared to both the CIB anisotropy signal (black dots) and the statistical noise (in green, including
cosmic variance on the noise estimate itself)}. 
\end{figure}

\subsection{\poker \label{se:poker}}

The \poker~implementation of the pseudo-spectrum approach uses the discrete Fourier transform (hereafter DFT).
For a map of scalar quantity $D_{jk}$ ($j,k$ denote pixel indices), it is defined as
\begin{eqnarray}
D_{mn} & = & \frac{1}{N_xN_y}\sum_{j,k} D_{jk}
\times e^{-2\pi i (jm/N_x+kn/N_y)}, \label{eq:dft}\\
D_{jk} & = & \sum_{m,n} D_{mn}\times e^{+2\pi i(jm/N_x+kn/N_y)},
\end{eqnarray}
where $D_{mn}$ is the set of discrete Fourier coefficients of $D_{jk}$. For a given wave-vector $\vec{\ell}$, labelled by the $m$ and $n$ indices, its corresponding norm is denoted by $\ell_{mn} = (2\pi/\Delta\theta)\sqrt{(m^\prime/N_x)^2+(n^\prime/N_y)^2}$ with
$m^\prime=m$ (respectively $n^\prime$ and n) if $m\leq N_x/2$ and $m^\prime=N_x-m$ if
$m>N_x/2$. The power spectrum of the map is defined as the square-modulus of its Fourier coefficients, {\it i.e.\/}, $P(\vec{\ell}_{mn})=\left|D_{mn}\right|^2$.

The direct DFT of the masked data relates the true Fourier coefficients to the
\emph{pseudo-}Fourier coefficients of the signal
\begin{equation}
\hat{D}_{mn} = \sum_{m'n'}W_{m,m'}^{n,n'}D_{m'n'},
\end{equation}
in which $W_{m,m'}^{n,n'}$ is a convolution kernel that depends only on the
mask DFT coefficients. Replacing $D_{mn}$ by $\hat{D}_{mn}$ in
the definition of the power spectrum of a given map leads to the power spectrum of the masked data (a.k.a. the
\emph{pseudo-power spectrum}). For a signal $T$ plus noise $N$ map, the ensemble averaged of the pseudo-spectrum tracing a statistically isotropic process, reads
\begin{equation}
\langle \hat{P}(\vec{\ell}_{mn}) \rangle =
\sum_{m'n'}\left|{W}_{m,m'}^{n,n'}\right|^2F_{m'n'}C({\ell_{m'n'}})+ \langle\hat{N}(\vec{\ell}_{mn})\rangle,
\label{eq:ps2cl}
\end{equation}
where we have introduced the total transfer function $F_{m'n'}$ accounting for
the beam, the `map-making' pixelisation effects and reprojection from curved, \healpix\ maps to flat, square
maps. The beam transfer function is given by the beam power spectrum described
in Sect.~\ref{sec:beam}. The `map-making' pixelisation effects are described by
the power spectrum of the pixel window function for full-sky maps provided by
the \healpix\ package (the initial \healpix\ maps are built with
$N_{\text{side}}=2048$ corresponding to a pixel size of $1.7$\arcm ). 
As explained in \citet{planck2011-1.5}, time-domain filtering is included as part of the
scanning beam, such that any time-domain filtering effects end up in the estimate of
the beam instead of as part of $F_{m'n'}$.
Finally, each curved map with a $1.7$\arcm\ resolution is
reprojected onto its tangent, flat space with a pixel size of 3.5\arcm. 
This induces first a repixelisation effect because the output map is less resolved than the
input one and second, a slight displacement of the pixel centres. The cumulative impact of `image deformation' and `repixelisation' is estimated via Monte-Carlo: we first generated a set of full-sky maps and computed the MC average of their pseudo-spectra. This set of maps was then re-projected onto flat maps for which MC average of their pseudo-spectra in the flat-sky approximation were computed. The ratio of the flat-sky pseudo-spectrum divided by the full-sky pseudo-spectrum gives a measurement of re-projection effect. Note that those simulations have been performed assuming different shapes for the input angular power spectra. The derived reprojection transfer functions agreed perfectly, which underlines the robustness of the approach.

An unbiased estimate of $C_\ell$ is obtained by first subtracting the noise contribution and then deconvolving the mask and beam effects encoded in the convolution kernel $\left|{W}_{m,m'}^{n,n'}\right|^2F_{m'n'}$. For the sky coverage of the considered fields, the rapid oscillations of the convolution kernels introduce strong correlations between spatial frequencies
and make its inversion numerically intractable. (Pseudo-)Power spectra are
therefore estimated in frequency bands (labelled $b$ hereafter). The binning operator is
\begin{equation}
R_b^{mn} = \left\{ \begin{array}{ll}
\displaystyle\frac{\ell_{mn}^\beta}{\Delta_{\rm b}}&\;{\rm if}\; \ell^b_{\rm low} \leq \ell_{mn} < \ell^{b+1}_{\rm low}\\
0&\;{\rm otherwise}\end{array}\right. ,
\label{eq:pmat}
\end{equation}
where $\Delta_{\rm b}$ is the number of wave vectors $\vec{\ell}_{mn}$ that fall into the bin $b$. The reciprocal
operator that relates the theoretical value of the one-dimensional binned power
spectrum $P_b$ to its value at $\ell_{mn}$ is
\begin{equation}
Q_{mn}^b = \left\{ \begin{array}{ll}
\displaystyle\frac{1}{\ell_{mn}^\beta}&\;{\rm if}\; \ell^b_{low} \leq \ell_{mn} < \ell^{b+1}_{low}\\
0&\;{\rm otherwise}\end{array}\right. 
\label{eq:qmat}
\end{equation}
For optimal results, the spectral index $\beta$ should be chosen to get $\ell^\beta C_\ell$ as flat as possible. For the CMB, $\beta \simeq 2$ is the equivalent of the standard $\ell(\ell+1)$ prefactor. For the CIB anisotropies $C_\ell$ scales roughly as $\ell^{-1}$, and we therefore adopted a binning with $\beta=1$. Nevertheless, we checked that our results were
robust against the choice of $\beta$: we simulated a power spectrum scaling as $\ell^{-1}$ but reconstructed it assuming $\beta=0,~1$ and $2$ in \poker. For each choice of $\beta$, the estimated power spectrum perfectly agreed with the input one (for a more complete discussion, see \citealt{poker}). 

The binned pseudo-power spectra is
\begin{equation}
  \hat{P}_b=\displaystyle\sum_{m,n\in b}R^{mn}_b\hat{P}(\vec{\ell}_{mn}),
\end{equation}
and the CIB power spectrum is related to its binned value, $C_b$, via
\begin{equation}
C (\ell_{m'n'}) \simeq \sum_{b'} Q^{b'}_{m'n'} C_{b'}
\end{equation}
With these binned quantities, Eq.~\ref{eq:ps2cl} can be re-written as
\begin{equation}
\left\langle\hat{P}_b\right\rangle =\sum_{b'} M_{bb'} C_{b'} + \left\langle \hat{N}_{b}\right\rangle,
\label{eq:mbb}
\end{equation}
with
\begin{equation}
M_{bb'}=\displaystyle\sum_{m,n\in b}\sum_{m',n'\in b'}R_b^{mn} \left|{W}_{m,m'}^{n,n'}\right|^2F_{m'n'}Q_{m'n'}^{b'}.
\label{eq:mbb_def}
\end{equation}
An unbiased estimate of the binned angular power spectrum of the signal is thus given by
\begin{equation}
P_b = \displaystyle\sum_{b'}M^{-1}_{bb'}\left( \hat{P}_{b'} - \langle\hat{N}_{b'}\rangle\right).
\label{eq:pb_res}
\end{equation}
It is easily checked that $\langle P_b\rangle=C_b$.

The complete recovery of the CIB anisotropy angular power spectra is therefore made in three steps:
\begin{description}
\item[A.] Given the mask, $W$, associated to our CIB map and the transfer function, $F_{m,n}$, compute and invert $M_{bb'}$ as given by Eq.~\ref{eq:mbb_def}. The different fields (with a size at most $5.1^\circ\times 5.1^\circ$)
are systematically embedded in a $10^\circ\times 10^\circ$ square map.
We use binary masks, {\it i.e.}, $W=1$ for observed pixels ({\it i.e.}, those kept in the analysis), and $W=0$ for unobserved pixels.
The estimated power spectra are binned with a bandwidth of $\Delta_{\rm b}=200$ and the first bin starting at $\ell=80$\footnote{$\ell=80$ corresponds to the inverse of the largest angular scale contained in the considered fields.}. 
\item[B.] Derive the noise bias $\langle\hat{N}_{b}\rangle$, given by first the instrument noise described in Sect.~\ref{se:inst_noise} and second the additional corrections given in Sect.~\ref{sect_conta}.
\item[C.] Compute the final estimate of $C_b$ from Eq.~\ref{eq:pb_res} and $C_b=\langle P_b\rangle$.
\end{description}

Uncertainties on $P_b$ come from sampling variance, noise variance,
astrophysical contaminants, and systematic effects. In the following section,
we present how we estimated each of these contributions from
Monte-Carlo simulations.

\subsection{Error bar estimation \label{se:pipeline}}

\subsubsection{Statistical uncertainties \label{sec:stat}}

As presented in Sect.~\ref{se:poker}, the uncertainties on $P_b$ come from signal
sampling variance, noise, and uncertainties on the subtraction of CMB and
Galactic dust. The first two are described by stochastic processes with known power
spectra, the last two come from uncertainties in the weights applied to
templates in the subtraction process at the map level.

We developed the tools necessary to
simulate maps given any input angular power spectrum for each field and frequency. The simulation pipeline
consists of simulating maps given an input power spectrum (in the case of CIB,
noise and CMB residual) and maps of template residuals (conservative Gaussian
random fractions of the templates). These maps are then combined and analysed by
the power spectrum estimator. Each realisation provides an estimated power
spectrum with the same statistical properties as our estimate on the data, and
alltogether these simulations provide the uncertainties on our
estimate. The covariance matrix of $P_b$ is 
\begin{equation}
 {\bf C}_{bb'} = \left\langle \left(P_b - \langle P_b\rangle_{\text{MC}}\right)\left(P_{b'} -
\langle P_{b'}\rangle_{\text{MC}}\right)\right\rangle_{\text{MC}},
\label{eq:cov_mat}
\end{equation}
with $\langle\cdot\rangle_{\text{MC}}$ standing for Monte-Carlo averaging. The error bar on each $P_b$ is 
\begin{equation}
\sigma_{P_b} = \sqrt{{\bf C}_{bb}},
\label{eq:sigma_pb}
\end{equation}
and the bin-bin correlation matrix is given by its standard definition
\begin{equation}
{\bf \Xi}_{bb'}=\frac{\mathbf{C}_{bb'}}{\sqrt{{\bf C}_{bb}{\bf C}_{b'b'}}}.
\label{eq:xcorr}
\end{equation}

\paragraph*{Simulation pipeline --} 

The simulated maps are $10^\circ\times 10^\circ$. They contain six components, accounting for the different ingredients supposedly present in the actual data maps:
\begin{enumerate}
\item A CIB anisotropy component obtained from a random, Gaussian realisation of the CIB anisotropy power spectrum. As a model for such a spectrum, we used a fit to $P_{\text{CIB}}$, estimated from the data additionally multiplied by the power spectrum of the beam, pixel, and reprojection transfer function;
\item a residual CMB component derived from a random, Gaussian realisation of the power spectrum given in 
Eq.~\ref{eq:cmb_left} using the known Wiener filter, beam differences
between 143\,GHz and other channels, and the WMAP  best fit CMB temperature
power spectrum;
\item the instrument noise as derived in Sect.~\ref{se:inst_noise}. Since the noise is slightly coloured, we simulate it using a fit of its measured power spectrum;
\item extra instrument noise incurred by CMB removal using Eq.~\ref{eq:xtranoise} as a model of its power spectrum.
\end{enumerate}
In addition to those four ingredients standing for signal and noise (a CMB residual being viewed as an extra-source of `noise' from the viewpoint of CIB), we added the two foreground templates that are removed, with some uncertainties:
\begin{description}
\item[5.] A CMB map with a Gaussian uncertainty distributed with 2\% and 3\% standard deviation at 217
and 353\,GHz, respectively (the CMB is negligible at higher frequencies compared to
CIB and dust). The 2\% and 3\% are justified in Sect.~\ref{se:rob};
\item[6.] an \hi~map with a 5\%, 10\% and 10\% standard deviation for its emissivity
(local, IVC and HVC components respectively), consistent with both the inter-calibration
errors (see Sect.~\ref{se:rob}) and the emissivity errors computed by the \cite{planck2011-7.12}
using Monte Carlo simulations.
\end{description}

\paragraph*{The analysis pipeline --}

The analysis pipeline works in four steps:
\begin{enumerate}
\item A first set of 1,000 MC simulations of CMB residual and noise only is performed to assess first, the pseudo-power spectrum of the instrument noise and CMB residual used to debias the simulated data pseudo-spectrum and, second, the noise variance given by
$\sigma_{N_b} = \sqrt{{\bf C}^{\text{noise}}_{bb}}$ ;
\item a second set of 1,000 MC simulations, including all the components, is performed. For a given simulated map, CMB and dust templates are removed assuming the {\it estimated\/} emissivities of Sect.~\ref{se:cirrus_remove} for dust;
\item the \poker~algorithm is then applied to these `foreground-cleaned' maps to obtain a final estimate of the CIB angular power spectrum. In this step, the bias involved in Eq.~\ref{eq:pb_res} contains the pseudo-power spectrum of the instrument noise model {\it and\/} of the CMB residual model as described in the simulation pipeline;
\item the total error bars and bin-bin correlation matrix on ${P}_b$ are obtained as the RMS of 1,000 Monte-Carlo realisations of the simulation pipeline, as described in the previous paragraph and using Eqs.~\ref{eq:sigma_pb} and \ref{eq:xcorr}.
\end{enumerate}
 
\begin{figure}
\begin{center}
\includegraphics[width=8cm, draft=false]{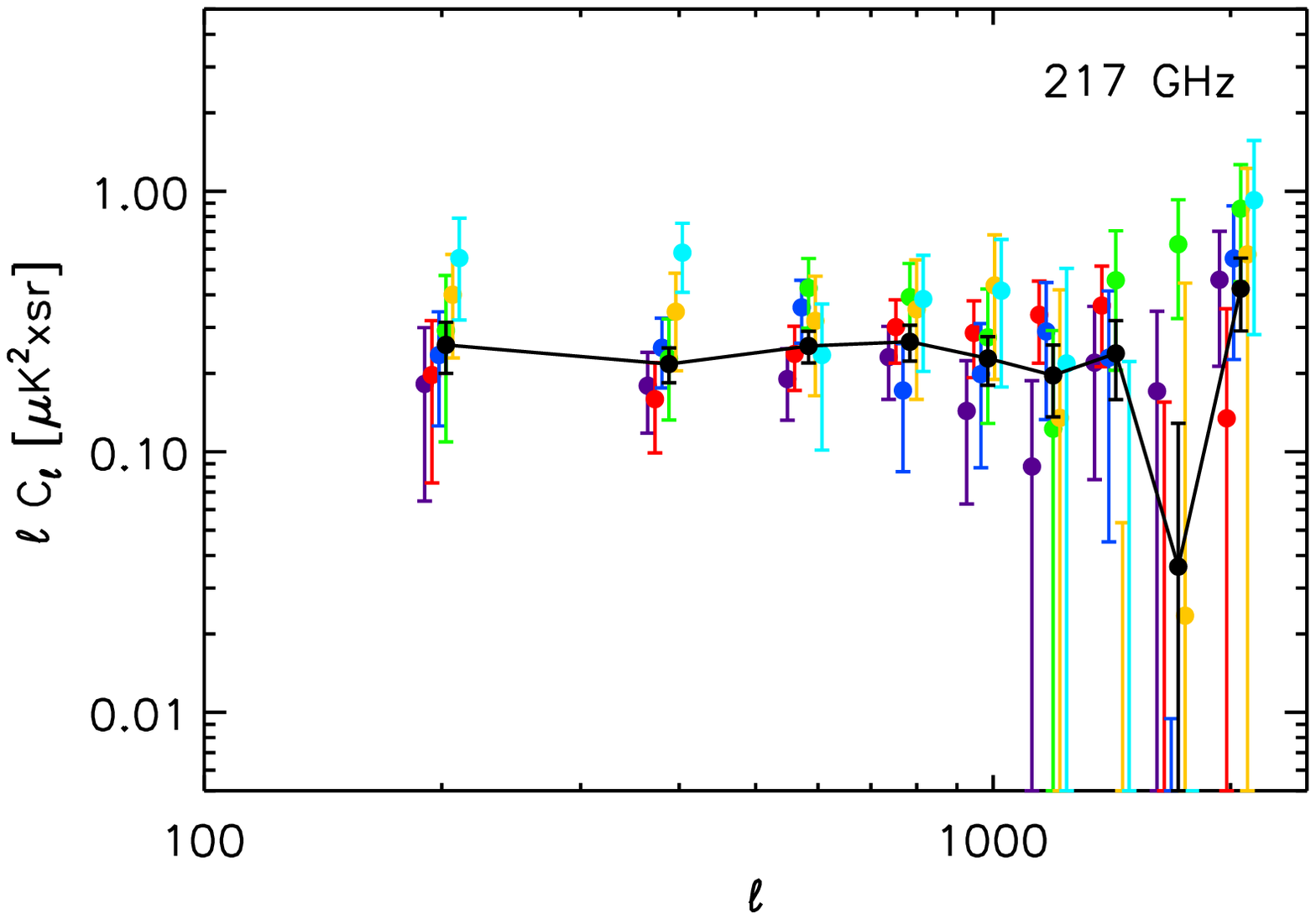}
\includegraphics[width=8cm, draft=false]{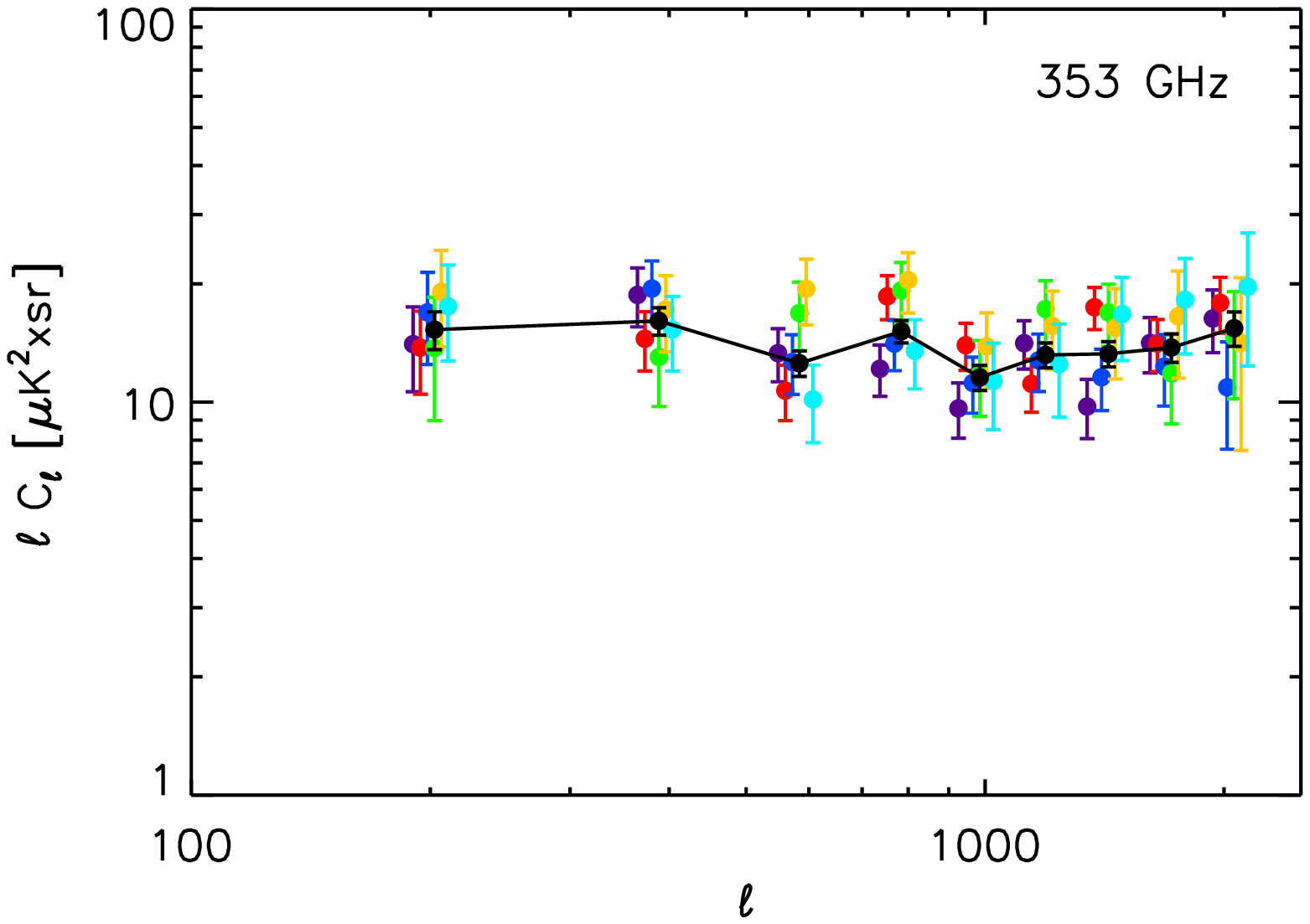}
\includegraphics[width=8cm, draft=false]{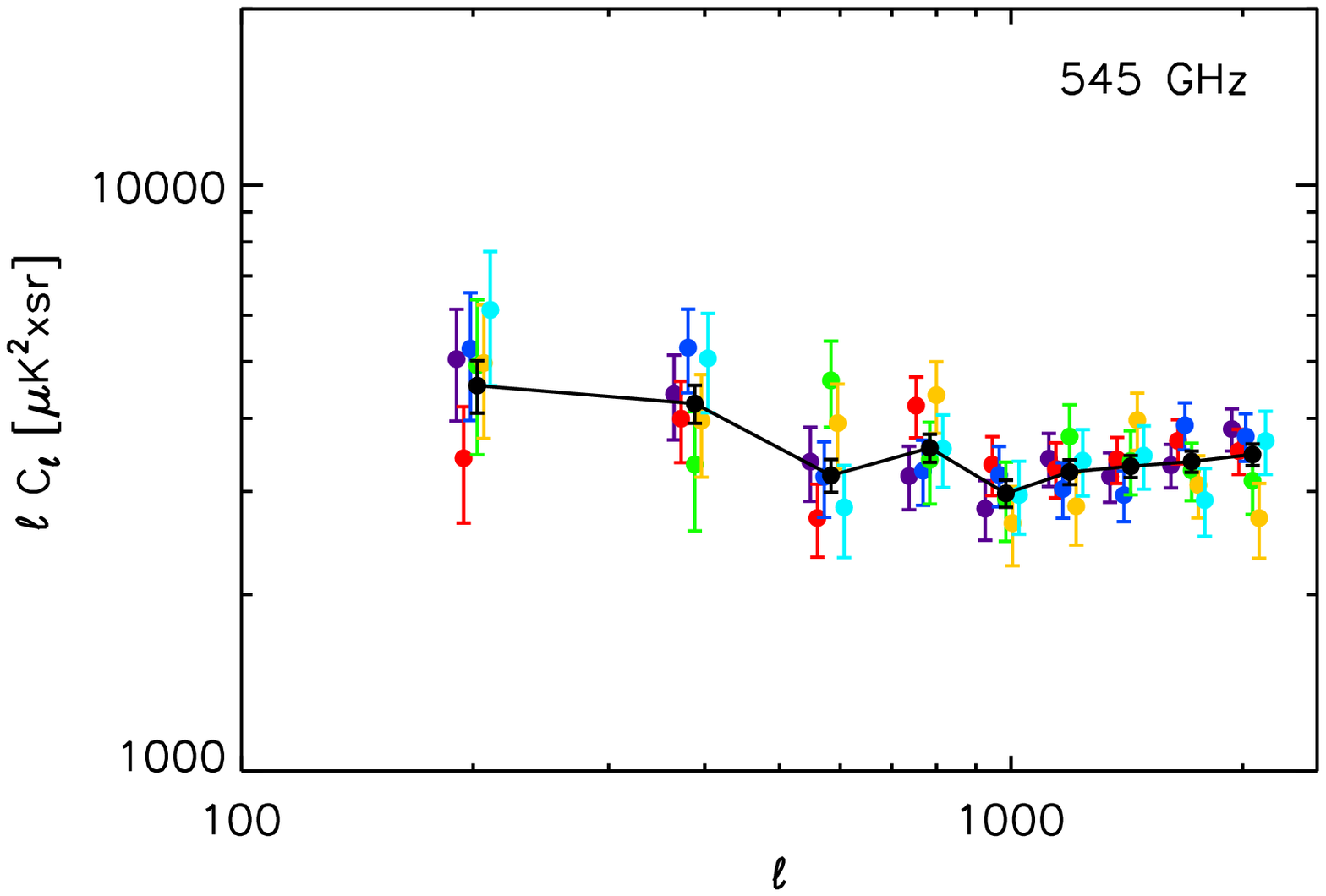}
\includegraphics[width=8cm, draft=false]{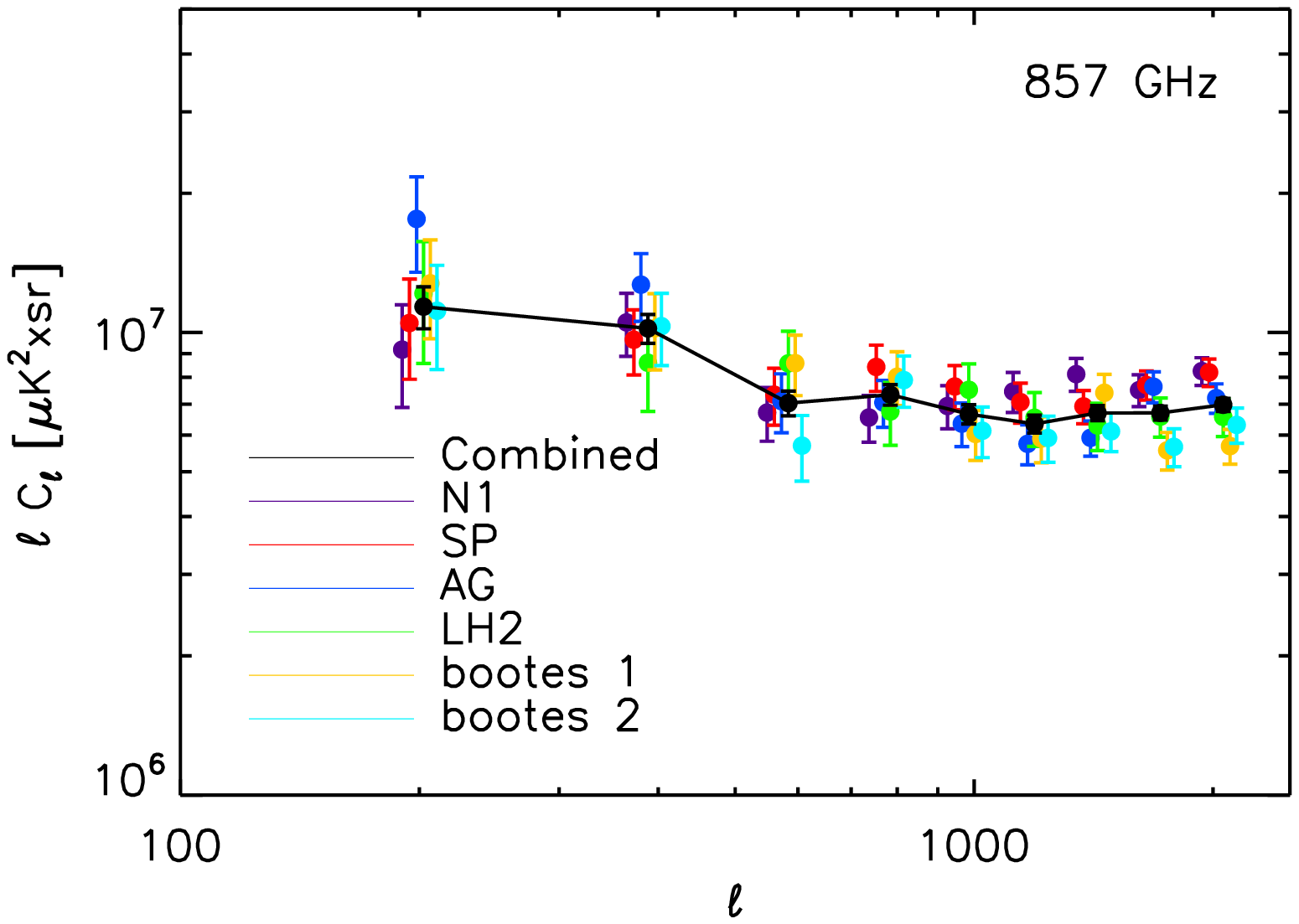}
\end{center}
\caption{\label{CIB_all_fields} CIB anisotropy power spectra of the six fields and their combined spectrum. }

\end{figure}

\begin{figure}
\begin{center}
\includegraphics[width=\linewidth, draft=false]{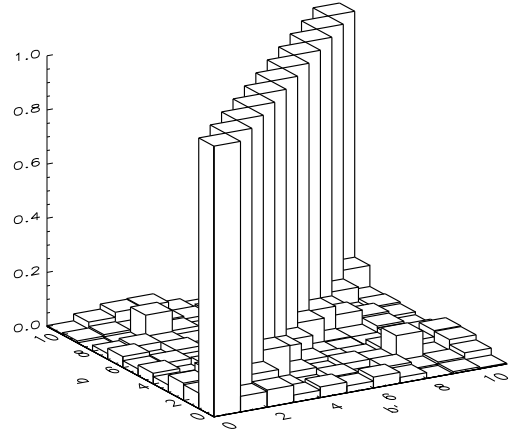}
\end{center}
\caption{\label{correl_bin} Modulus of the bin-to-bin correlation matrix derived from the simulation pipeline for the {\tt SP\/} field at 353\,GHz.  Spatial frequency bins are not correlated by more than $\simeq 10\%$.}
\end{figure}

The statistical uncertainties contain:
\begin{description}
\item[A.] sampling variance from CIB anisotropies and residual CMB as modelled in Eq.~\ref{eq:cmb_left};
\item[B.] noise variance from instrument noise and extra noise given by Eq~\ref{eq:xtranoise};
\item[C.] uncertainties on the CMB and \hi\ template subtraction.
\end{description}
In this set of simulations and analysis, we assumed the beam profile as described in Sect.~\ref{sec:beam} and ignored potential beam uncertainties (see the next section for a discussion of this systematic effect).
Below, we present the results obtained using the FEBeCoP-derived beam profiles, but the estimated power spectra using either the FICSBell-derived or the FEBeCoP-derived beam agree very well (because the difference between the two beams is small, as shown in Fig.~\ref{compar_beam}). Figure~\ref{CIB_all_fields} shows the results for all fields and frequencies.
We also display in Fig.~\ref{correl_bin} the bin-bin correlation matrix, showing that two bins are not correlated by
more than 10\%.


\subsubsection{Systematic errors \label{se:systes}}

Our estimate of each power spectrum is affected by different systematic errors that must be accounted for separately
from the statistical errors derived in the previous section. These systematic uncertainties may introduce a bias in the final estimate and/or bias our Monte-Carlo estimate of the statistical uncertainties presented above.
We review here the different sources of these systematics and evaluate their level.

\paragraph*{Mask impact--} Our power spectrum estimation is performed on a limited sky patch.
This induces power aliasing from angular scales larger than the size of the patch, an effect that increases as the signal power spectrum steepens. \poker\ is designed to account for this effect (as well as the extra aliasing induced by
holes in the map, if present) because the convolution kernel, $M_{bb'}$, contains the information on mode coupling to the larger scales.  We ran \poker\ on data maps that were embedded in larger regions that are zero-padded.
There is no general prescription as to the size of the zero-padding that one should use, but fortunately the results
are very insensitive to the particular choice and whether or not the mask is apodized.  By comparing different
choices we found uncertainties at the 2\% level, well below the statistical uncertainties.

\paragraph*{Template subtraction impact--} Imperfect template subtraction will also lead to 
`foreground' residuals that slightly bias our final estimate, $P^{\text{CIB}}_{b}$, of the CIB anisotropy angular power spectrum. This residual level is given at first order by 
\begin{equation}
\delta P^{\text{CIB}}_b\simeq(\delta\alpha)^2\displaystyle\sum_{b'}M^{-1}_{bb'}\hat{P}^{\text{temp}}_{b'},
\end{equation}
 with
$\hat{P}^{{\rm temp}}_{b_1}$ the pseudo-spectrum of the template and $\delta\alpha$ the error on the global amplitude of this template. Figure~\ref{fig:error_budget} illustrates the level of these residuals for the particular case of the {\tt SP\/} field. Although negligible compared to the statistical errors, they are accounted for in the error budget.

\paragraph*{Beam uncertainties--} Uncertainty in the beam will also bias the estimate of the power spectrum because:
\begin{enumerate}
\item The beam window function enters the
computation of the convolution kernel $M_{bb'}$. Any beam error biases our estimate of $M_{bb'}$ and thus our
final result.
\item Beam uncertainties will translate into slight misestimation of $N^{\text{CMB}}_{\ell}(\nu)$ and $C^{\text{CMBres}}_{\ell}(\nu)$, potentially biasing our final estimate.
\end{enumerate}

Moreover, any beam uncertainty will also affect our estimation of the statistical error bars. For example, the contribution of noise to the error bars scales roughly as
$\sigma_{N_b}\propto N_b/b^2_b(\nu)$
(see noise curve on Fig.~\ref{fig:error_budget}). As a consequence, any beam misestimation couples to the noise variance and leads to additional uncertainties on the final power spectrum estimate.
This additional error is given at first order by

\begin{equation}
\delta\sigma_{N_b}\approx2\sigma_{N_b}\left|\frac{\delta b_b(\nu)}{b_b(\nu)}\right|,
\end{equation}
where $\sigma_{N_b}$ is the noise error and $\delta b_b(\nu)$ is the beam uncertainty.  This may be important
at small angular scales, depending on $\delta b_b(\nu)$.
A similar argument applies equally to sample variance.

In this work we used the current, best determination of the beams.
As explained in Sect.~\ref{sec:beam}, the uncertainties on the scanning beam dominate these uncertainties.
They are $\sim 3\%$ to $\sim6\%$ of the FWHM, depending on the frequency \citep{planck2011-1.5}.

We simulated the impact of such an error using the \{simulation+analysis\} pipeline presented in Sect.~\ref{sec:stat}, assuming the appropriate frequency-dependent discrepancy between the beam used to simulate the maps and the beam used to analyse the maps. The bias induced by these uncertainties and their impact on the statistical error bars are derived by comparing the estimated power spectra and their MC-variance with and without the beam discrepancy.
The bias so induced is the dominant uncertainty, larger even than the statistical error bars at small angular scales
for measurements at 545 and 857\,GHz.
We took this bias into account in the modelling (Sect.~\ref{model_obs}). 

\subsection{Robustness \label{se:rob}}

Many additional tests have been done to test the robustness of CIB anisotropy power spectra.
First, instead of removing the CMB contribution and fitting only the \hi\ correlation coefficients,
we searched for the best fit simultaneously using the \hi\ and CMB templates
\begin{equation}
  I_\nu(x,y) = \sum_i \alpha^{i}_\nu N_{HI}^{i}(x,y) + \beta_1 I_\nu(x,y)_{CMB} + C_\nu(x,y) \,.
\end{equation}
This allowed us to take into account the photometric inter-calibration uncertainties, which are about 2\% for the CMB channels \citep{planck2011-1.5}.
We also fitted the low-frequency channels for only CMB
\begin{equation}
I_\nu(x,y) = \beta_2 I_\nu(x,y)_{CMB} + C_\nu(x,y) \,.
\end{equation}
We found that $ \beta_1$ and $ \beta_2$ agree at the $\sim$0.05\% level.
The difference between the two coefficients and unity is less than 1\% (3\%) at 217\,GHz (353\,GHz).
The determination of the $\beta_i$ is however very noisy because of the small area of our fields.
They are compatible with $\beta_{1,2}=1$ at 217\,GHz, though at 353\,GHz they fall below unity by 2--3\%.
We did not correct for these inter-calibration coefficients and took conservative errors on the residual CMB contamination in our error pipeline (2\% at 217\,GHz and 3\% at 353\,GHz, see Sect.~\ref{se:pipeline}).
Fitting for both CMB and \hi\ or just \hi\ on CMB-cleaned maps changes the dust-\hi\ emissivities by less
than 2\%. This was again taken into account in the error simulation pipeline.\\

To test the reliability of the noise power spectrum measurement, we recomputed the CIB anisotropy power spectrum using the cross-correlation between half-pointing period maps instead of the auto-correlation.
We removed from each half map the same CMB and \hi\ (with emissivities taken as those of the total map). 
On average, for the range of $\ell$ considered in this paper, the two methods agree at better than the 1\% level (1\% and 0.05\% at 217 and 857\,GHz, respectively).\\

Finally, one way to asses the robustness of our determination is to compare the CIB anisotropy power spectra for our different fields. The fields all have different noise, cirrus, and CMB contributions and they are all independent. The comparison made in Fig. \ref{CIB_all_fields} shows that they are all compatible within error bars.

\subsection{Power spectra from combined fields}

Our final estimate of the CIB anisotropy angular power spectra at different frequencies was derived by combining
each power spectrum estimated on the six different fields
\begin{equation}
	P^\text{(tot)}_b=\displaystyle\sum_{A=1}^6W^A_b\times P^A_b,
	\label{eq:pbtot}
\end{equation}
with $A$ an index running over fields and $W^A_b$ an appropriate weight, to be defined below. The same binning was adopted for each field. The bin-bin covariance of the `combined-field' estimator, $P^\text{(tot)}_b$, is a function of the bin-bin covariance of each `single-field' estimator, $P^A_b$, and the covariance between two `single-field' estimators,
$P^A_b$ and $P^B_{b'}$ follows
\begin{eqnarray}
	{\bf C}^{\text{(tot)}}_{bb'}&=&\mathrm{Cov}\left(P^\text{(tot)}_b,P^\text{(tot)}_{b'}\right) \label{eq:covpbtot} \\
	&=&\displaystyle\sum_{A=1}^6\sum_{B=1}^6W^A_bW^B_{b'}\times\mathrm{Cov}\left(P^A_b,P^B_{b'}\right). \nonumber
\end{eqnarray}
The error bars on $P^\text{(tot)}_b$ are simply given by $\sqrt{{\bf C}^{\text{(tot)}}_{bb}}$ and optimal weights are derived
by searching for those $W^A_b$ minimizing these.

Our MC simulations showed that different bins are not correlated by more than 10\%, and our fields were widely
enough separated that individual measurements are uncorrelated.  Neglecting field-to-field and bin-to-bin
covariances, the optimal weights become the usual inverse variance weighting:
\begin{equation}
W^A_b=\frac{\sigma^{-2}_{P^A_b}}{\displaystyle\sum_{B=1}^6 \sigma^{-2}_{P^B_b}} \,,
\end{equation}
where $\sigma_{P_b^A}$ is the statistical error bars derived from the MC simulations as previously described.

The final CIB anisotropy power spectra estimates were therefore computed by inserting the inverse variance weights in Eq.~\ref{eq:pbtot}. The total statistical uncertainties were obtained by inserting those weights in Eq.~\ref{eq:covpbtot}, where
$\mathrm{Cov}\left(P^A_b,P^B_{b'}\right)$ stands for the statistical uncertainties only ({\it i.e.\/} the standard quadratic summation). The total systematic errors were obtained by linearly summing the weighted systematic error on each field (any covariance between fields is neglected). We stress that irrespectively of the way weights are derived the full bin-bin covariance matrix could be computed\footnote{Assuming a vanishing bin-to-bin covariance for optimal weights computation does not prevent us from deriving the full bin-bin covariance matrix and just leads to sub-optimality in terms of error bars.}. However, the forthcoming cosmological interpretation of the derived power spectra assumes a zero bin-to-bin correlation, and we only provide here the diagonal elements, {\it i.e.\/}, error bars, of the final covariance.

\begin{figure}
\includegraphics[width=\linewidth, draft=false]{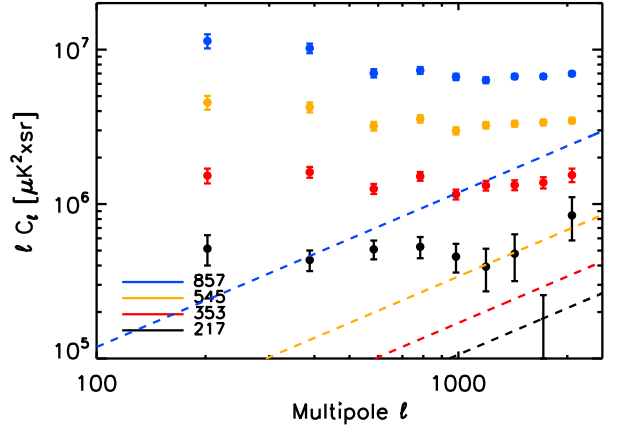}
\caption{\label{Fig_pk_CIB}  Field-combined CIB anisotropy power spectra at 217, 353, 545, and 857\,GHz.
The dashed line shows the expected sum of the dusty and radio galaxy shot-noise power
(from Table \ref{tab:fluc_cut}). The power spectra at 217, 353, and 545\,GHz were arbitrary scaled to allow for a better comparison between frequencies (they were multiplied by 2$\times$10$^6$, 10$^5$ and 10$^3$, respectively).}
\end{figure}

Our results are displayed in Fig.~\ref{Fig_pk_CIB} and the data points are given in Table \ref{tab_cib_pk}.
Though our weighting is slightly suboptimal, the final angular power spectra are measured with high signal-to-noise
compared to the single-field measurements.

\begin{table*}
\begin{center}
\begin{tabular}{c|c|c|c|c|c|c|c|c|c} 
$<\ell>$ &      202 & 388 & 583 & 784 & 985 & 1192 & 1431 & 1717 & 2060 \\  \hline
$\ell_{\rm min}$  &     80 & 280 & 480 & 680 & 880 & 1080 & 1296 & 1555 & 1866 \\  \hline
$\ell_{\rm max}$ &     280 & 480 & 680 & 880 & 1080 & 1296 & 1555.2 & 1866 & 2240\\  
\hline  \hline 
{\bf 217\,GHz}
\\
\hline
$C_{\ell}$ $\times$10$^5$  &  127.03 & 55.95 & 43.65 & 33.71 & 23.18 & 16.49 & 16.67 & 2.11 & 20.51 \\  \hline
$\Delta C_{\ell}^{\rm stat}$  $\times$10$^5$&    28.23 & 8.53 & 6.08 & 5.29 & 4.90 & 5.07 & 5.59 & 5.38 & 6.39\\  \hline
$\Delta C_{\ell}^{\rm beam}$   $\times$10$^5$& 0.10 & 0.18 & 0.33 & 0.46  & 0.50 & 0.52 &  0.76  & 0.13 & 1.90 \\   
\hline 
{\bf 353\,GHz}
\\
\hline
 $C_{\ell}$  $\times$10$^4$  &   755.17   &  414.68  &   215.18 &    193.02  &  117.35  &  110.63 &  92.76 &     80.20 &   74.84 \\  \hline
$\Delta C_{\ell}^{\rm stat}$  $\times$10$^4$ &   83.18 &   33.25 &   16.05 &  12.64 &    8.64  &   8.05  &  6.84  &   6.70  &   7.56   \\  \hline
$\Delta C_{\ell}^{\rm beam}$   $\times$10$^4$ &  0.47 &  1.04 &  1.23 &  2.01 &  1.93 &  2.65 & 3.21 &  3.99  &  5.33 \\  
\hline 
{\bf 545\,GHz}
\\
\hline
$C_{\ell}$  $\times$10$^2$  &   2246.54 &  1091.71 &   547.60  &   454.01  &  302.39  &  271.87  &  231.55  &  196.63  &  168.44 \\  \hline
$\Delta C_{\ell}^{\rm stat}$  $\times$10$^2$ &    229.92  &    81.11  &   35.42  &   24.90  &    16.12  &  13.15    &   10.0  &    8.12 &   7.12   \\  \hline
$\Delta C_{\ell}^{\rm beam}$   $\times$10$^2$ &   2.13  &  4.16  &  4.72 &  7.14&  7.50 &  9.84  &  12.07 &   14.76 &  18.18 \\  
\hline 
{\bf 857\,GHz}
\\
\hline
$C_{\ell}$  $\times$10$^{-2}$  &     561.65  &  262.93  & 120.57  &  93.57 & 67.65  &  53.24 &  46.82 &  39.02 &  33.88 \\  \hline
$\Delta C_{\ell}^{\rm stat}$  $\times$10$^{-2}$ &   58.61 & 18.82 &   7.40 &  4.79 & 3.24 &  2.36 & 1.76 & 1.35 &  1.09 \\  \hline
$\Delta C_{\ell}^{\rm beam}$   $\times$10$^{-2}$ & 0.63 &  1.18  &  1.22  &  1.72 &  1.97 &  2.26 &  2.86 & 3.44 &  4.29 \\  

\end{tabular}\\
\caption{  \label{tab_cib_pk} CIB anisotropy $C_{\ell}$, at 217, 353, 545, and 857\,GHz in $\mu K_{\rm CMB}^2 \times$sr.
The conversion to Jy${}^2\,$sr${}^{-1}$ (with the photometric convention $\nu I_{\nu}$=constant) involves multiplication
by 231483, 83135, 3391.5 and 4.99 at 217, 353, 545, and 857\,GHz, respectively. $\Delta C_{\ell}^{\rm stat}$ are the statistical errors;  $\Delta C_{\ell}^{\rm beam}$ are the systematic errors introduced by the beam uncertainty (see Sect.~\ref{se:systes}).}
\end{center}
\end{table*}

\section{Overview and comparison with previous work \label{sect_5}}
The measured power and the shot noise predicted by the model discussed previously is shown in
Fig.~\ref{Fig_pk_CIB}.
Our measurements do not allow us to detect the shot-noise component, which will dominate on
smaller scales than those probed here.  However, the predictions are quite close to the measured power at
the highest $\ell$, indicating that further analysis of the CIB in \Planck\ up to $\ell\sim 3000$ might allow a
measurement of the shot-noise component.
The figure also reveals that the shape of the power spectrum is remarkably similar at the four frequencies,
being identical within the $1\,\sigma$ statistical errors for all data points but the last two at 217\,GHz
({\it i.e.}, $\ell \simeq 1717$ and 2060, which are $1.6\,\sigma$ from the points at the other frequencies).
This suggests that over the range of frequency and $\ell$ probed here the clustering properties do not evolve much and/or the galaxy populations responsible for CIB anisotropies are the same. We will return to this point in Sect.~\ref{model_obs}.
We start in this section by analysing the frequency dependence of the CIB anisotropies and CIB mean, and comparing our
anisotropy measurements with previous measurements at the same (or nearby) frequencies.

\subsection{Comparing the CIB mean and anisotropy SEDs \label{sect_cib_ciba}}

The rms fluctuation in the CIB is related to the anisotropy power spectrum as
 \begin{equation}
 \sigma^2 = \frac{1}{2\pi} \int \ell\,d\ell\ C_\ell \,.
 \label{eq:sigma}
 \end{equation}
 
Table \ref{tab:CIB_CIBA_SED} gives approximations to this integral using the measured values of $\ell\,C_\ell$
over the range $200<\ell<2000$.
Statistical error bars on $\sigma$ are computed with Monte Carlo simulations using the statistical errors on
the power spectra. The table also gives systematic errors (the second error term) corresponding to the photometric calibration
uncertainties.
Those values can be compared to the CIB absolute level. Cosmic infrared background determinations based on FIRAS data from two groups can be used:
\begin{itemize}
\item \citet{fixsen1998} used three different methods to obtain the CIB mean. They average the three spectra to obtain one CIB mean spectrum, and then fit it by a modified black body.
They find a dust temperature $T=18.5\pm1.2\,$K, an optical depth $\tau=(1.3\pm0.4)\times 10^{-5}$ and an emissivity index $\beta=0.64\pm0.12$.
The FIRAS spectrum is quite noisy so that the uncertainties on the parameters are large and the three
parameters are highly degenerate.
\item \citet{lagache1999} made two determinations of the CIB mean spectrum, using different methods to remove the foregrounds (\hi\ and {\sc Hii})
than \citet{fixsen1998}. One CIB mean spectrum is obtained in the Lockman Hole and one on 51\% of the sky (to test isotropy). The two spectra agree very well \citep[see Fig.~6 of][]{lagache1999}.
There is a refinement of the measurement in \citet{lagache2000}, which agrees within errors with the previous
measurement. \citet{gispert2000} fit the Lockman Hole spectrum with a modified black body to derive the CIB mean values and uncertainties at certain wavelengths (their Fig.~5).
The best fit has $T=13.6\pm1.5\,$K, $\tau=(8.9\pm2.9)\times 10^{-5}$ and $\beta=1.4\pm0.2$.
These parameters are very different to those found by \citet{fixsen1998}, but because of their degeneracy, the spectrum is quite close to that of \citet{fixsen1998} in the relevant frequency range (200--1500\,GHz).
\end{itemize}

\begin{figure}
\begin{center}
\includegraphics[width=\linewidth, draft=false, width=\linewidth]{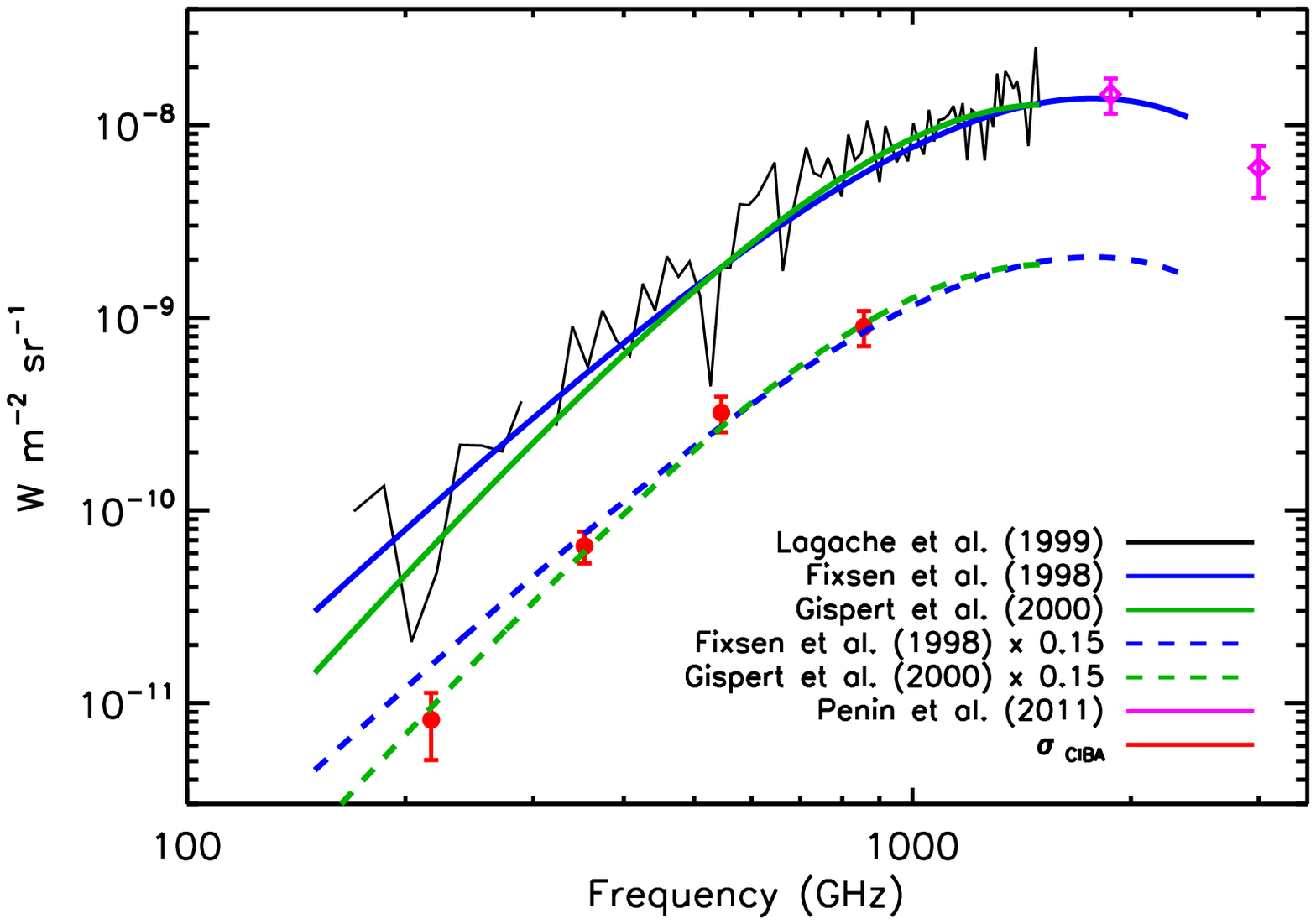}
\caption{\label{Fig_sed_ciba_cib}  Comparison of the observed CIB mean and anisotropy SED. The CIB measurements are from \citet{lagache1999} (FIRAS spectrum in black) and \citet{penin2011a} ({\it Spitzer} and IRIS, pink diamond data points). The green and blue continuous (dashed) lines are the CIB fits from \cite{gispert2000} and \citet{fixsen1998} (multiplied by 0.15). The rms fluctuations of the CIB anisotropies, measured for $200<\ell<2000$, are shown with the red dots. Their error bars include both statistical and photometric calibration systematic errors  (linearly added), as given in Table \ref{tab:cib}. This figure shows that the CIB anisotropy SED is steeper than the \citet{fixsen1998} best fit but very close to the \citet{gispert2000} best fit. We see no evidence for different CIB mean and anisotropy SED.}
\end{center}
\end{figure}

We integrated the two CIB mean fits through the HFI bandpass filters to obtain the values for the absolute signal given in Table \ref{tab:CIB_CIBA_SED}.  The two determinations agree to better than 10\%, 1\%, and 20\% at 857, 545, and 353\,GHz, respectively.
At 217\,GHz they differ by 60\%, but are compatible within errors.
It is unknown which of the determinations is the best, because the errors on the spectrum are dominated by systematic effects linked to foreground removal that are difficult to estimate.
For the uncertainties listed in Table \ref{tab:CIB_CIBA_SED} we fixed T and $\beta$ to their best-fit values and considered only the uncertainty on $\tau$ (since the errors on the three parameters are highly correlated).
Comparison between the CIB mean and anisotropy SED is shown on Fig.~\ref{Fig_sed_ciba_cib}.
We see that the CIB anisotropy SED is well described by the CIB mean spectrum of \citet{gispert2000} with the amplitude scaled by 0.15.
The CIB mean spectrum of \citet{fixsen1998} is flatter, with an 857/353 colour of 4.1 compared to 5.3 and 5.4 and 
an 857/217 colour of 12, compared to 27 and 21, for CIB anisotropies and the \citet{gispert2000} CIB mean, respectively.
A steeper rise in the CIB anisotropy SED compared to the \citet{fixsen1998} CIB mean has also been seen in \cite{hall2010}.
These authors combine SPT and CIB anisotropy data from the literature and obtain an 857/217 colour of $\simeq 25$ at $\ell=3000$.
This compares very well to the CIB anisotropy colour obtained with {\it Planck}, integrated over $200<\ell<2000$.\\

In conclusion, we do not see any evidence for a different CIB mean and anisotropy SED, which is consistent with
the galaxies that dominate the CIB mean being those responsible for CIB anisotropies.
The \citet{gispert2000} fit of the \citet{lagache1999} CIB mean spectrum ($200\le \nu \le 1500\,$GHz) describes both the CIB mean and the CIB anisotropy SED equally well.

\begin{table*}
\begin{center}
\begin{tabular}{l|c|c|c|c} 
& 217\,GHz & 353\,GHz & 545\,GHz & 857\,GHz \\ \hline

$\sigma_{\rm obs}$ &  $(3.7\pm1.3\pm0.08)\times 10^{-3}$  & $(1.9\pm0.3\pm0.04)\times 10^{-2}$   & $(5.9\pm0.8\pm0.40)\times 10^{-2}$  &  $(1.0\pm0.1\pm0.07)\times  10^{-1}$  \\  \hline

CIB$^{\rm f}_{\rm obs}$  & $(5.4\pm1.7)\times  10^{-2}$ & $(1.6\pm0.5)\times  10^{-1}$ & $(3.7\pm1.1)\times 10^{-1}$ &  $(6.5\pm2.0)\times  10^{-1}$ \\ \hline

CIB$^{\rm g}_{\rm obs}$  & $(3.4\pm1.1)\times  10^{-2}$ & $(1.3\pm0.4)\times  10^{-1}$ & $(3.7\pm1.2)\times 10^{-1}$ &  $(7.1\pm2.3)\times  10^{-1}$ \\ \hline


CIB$_{\rm mod}$ & $(3.2\pm0.6)\times  10^{-2}$  & $(1.2\pm0.1)\times 10^{-1}$ & $(3.5\pm0.2)\times  10^{-1}$&   $(6.3\pm0.3)\times  10^{-1}$\\ 

\end{tabular}
\caption{  \label{tab:CIB_CIBA_SED} RMS fluctuations in  the CIB computed from Eq.~\ref{eq:sigma} and 
CIB mean levels at 217, 353, 545 and 857\,GHz. The subscripts `obs' and `mod' stand for observational and model values, respectively.  The CIB model is taken from \citet{bethermin2010}. CIB$^{\rm f}$ and CIB$^{\rm g}$ are the \citet{fixsen1998} and \citet{gispert2000} best fits to the CIB spectra, respectively. The best fits and the model have been integrated in the HFI bandpasses. For the rms fluctuations both statistical and photometric calibration systematic errors are given. All numbers are given in MJy/sr for the photometric convention $\nu I_{\nu}$=constant.}
\label{tab:cib}
\end{center}
\end{table*}

\begin{figure}
\begin{center}
\includegraphics[width=\linewidth, draft=false, width=\linewidth]{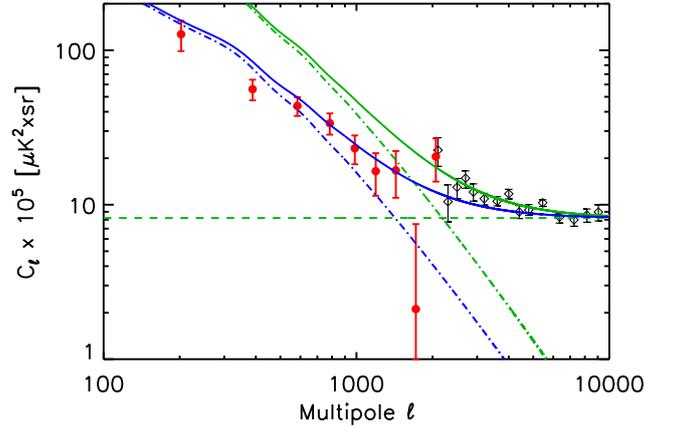} 
\caption{\label{fig_spt_hfi}  Comparison of SPT \citep[][dark open diamonds]{hall2010} and HFI measurements (red dots)
at 217\,GHz.  The green dashed line corresponds to the SPT shot noise and the green dot-dashed line to the clustering model of \citet{hall2010}, the sum of the two is the continuous green line. The clustering model over-predicts by a factor $\simeq 2.4$ the HFI power at $\ell \sim 800$. The blue dash-dotted line shows the clustering model divided by this factor. The clustering+shot noise (blue continuous line) now under-predicts the SPT data points, which may be the signature of non-linear contributions.}
\end{center}
\end{figure}

\subsection{Comparison with SPT and ACT measurements \label{sect_spt_hfi}}

SPT and ACT both measured the amplitude of CIB anisotropies, though at higher multipoles than those presented in this paper. The ACT measurement is on scales too small to be directly compared with our measurements (the amplitude is  given at $\ell=3000$ while our last data point is at $\ell=2060$).
The SPT team computed the residual bandpowers after subtracting the tSZ, kSZ, CMB, and cirrus model components,
quoting data points from $\ell=2000$ to 10{,}000.
Fig.~\ref{fig_spt_hfi} shows the comparison of the HFI measurement at 217\,GHz and the SPT measurements at 220\,GHz. 
Because the bandpass filters are not exactly the same, we applied a multiplicative correction factor (colour correction) to over-plot the SPT CIB anisotropy power spectrum on that of the HFI.
This factor, the square of the HFI/SPT colour, is computed using the CIB SED of \citet{gispert2000} convolved with the bandpass filters and is equal to 1.04. 

To interpret their data, \citet{hall2010} used a phenomenological model of CIB sources that assumed each galaxy has the same, non-evolving, modified blackbody SED and that their light was a biased tracer of the mass fluctuations, calculated in linear perturbation theory.  Moreover, the redshift distribution of the luminosity density
was set by two parameters that fix the width and peak redshift.

The green curve of Fig.~\ref{fig_spt_hfi} shows the \citet{hall2010} model, normalized to the SPT bandpowers.
We see that with this normalization, the power at large angular scales is larger by more than a factor of 2 than
the HFI data.  We also show as the blue curve the same model, except with amplitude adjusted to better agree with
the HFI data.  This downward adjustment of amplitude could arise from either a reduction in bias or in the amplitude
of the mean CIB.  This correction, of course, shifts the discrepancy to the smaller-scale SPT data.
Since we expect the linear theory assumption will be better at large scales than at small, the discrepancy between model and data at small scales may be signaling the importance of non-linear corrections.

\subsection{Comparison with BLAST and SPIRE measurements \label{compar_blast_spire}}

\cite{viero2009} presented BLAST power spectrum measurements at 1200, 857, and 600\,GHz in the GOODS-South field. They detect CIB anisotropy and shot-noise power in the range $940\le\ell \le 10{,}800$.
The measured correlations are well fitted by a power-law over scales of 5--25\arcm, with $\Delta I/I=15.1\%\pm 1.7\%$.
This level with respect to the CIB is the same as that found at the four HFI frequencies
(see Sect.~\ref{sect_cib_ciba} and Fig.~\ref{Fig_sed_ciba_cib}).
Fitting to a linear theory power spectrum, they find that the BLAST galaxies responsible for the CIB fluctuations have bias parameters, $b=3.9\pm0.6$ and $b=4.4\pm0.7$ at 857 and 600\,GHz, respectively.  They further interpret their results using the halo model and find that the simplest prescription does not fit very well.
One way to improve the fit is to increase the radius at which dark matter halos are truncated in the model
(the virial radius) and thereby distribute satellite galaxies over a larger volume.
They interpret this as being equivalent to having some star-forming galaxies at $z\ge1$ located in the outskirts of groups and clusters.

\begin{figure}
\begin{center}
\includegraphics[width=\linewidth]{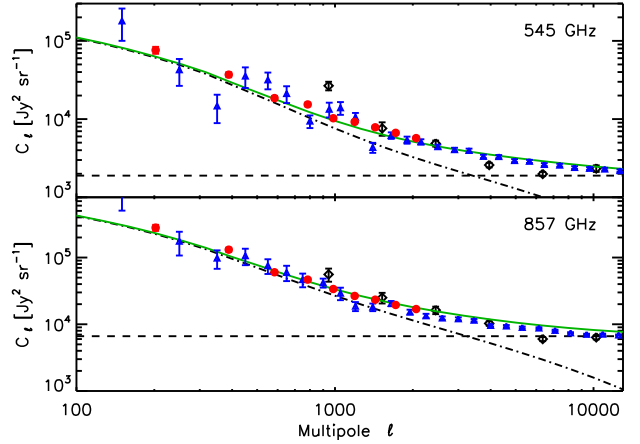} 
\caption{\label{fig_blast_hfi} Comparison of BLAST and HFI measurements at 545 and 857\,GHz.  HFI data points are the red circles; BLAST data points are the black diamonds. They were colour-corrected for the comparison (the colour was computed using the CIB SED of \citet{gispert2000}, integrated through the BLAST and HFI bandpass filters). The dashed line is the BLAST shot noise (also colour-corrected). Also shown is the BLAST best-fit clustering model (black dash-dotted line) and the total contribution (shot noise plus clustering; continuous green line). It provides a good fit to the \Planck\ data. Finally, we report in this figure a revised version of the SPIRE data points from \cite{amblard2011} (blue triangles from Fig.~\ref{fig_spire_hfi}, see text for more details).}
\end{center}
\end{figure}

We show in Fig.~\ref{fig_blast_hfi} the comparison between the BLAST and HFI measurements at 857 and 545\,GHz. Because the bandpass filters are quite different (particularly the 600 and 545\,GHz BLAST and HFI channels), we applied a colour correction as explained in Sect.~\ref{sect_spt_hfi}, 
multiplying the BLAST CIB anisotropy power spectra by 0.7 and 1.05 at 545 and 857\,GHz, respectively.
We see from Fig.~\ref{fig_blast_hfi} that the BLAST power spectra agree quite well with those from HFI, except that their largest-scale data points are systematically higher.
This may be caused by contamination by residual Galactic cirrus emission in the BLAST power spectra.
Also shown in the figure are shot-noise powers measured by BLAST.
Once the colour corrections are applied, they are 1843 and 7326\,Jy${}^2\,$sr${}^{-1}$ at 545 and 857\,GHz, respectively. Their flux cuts are comparable to ours (they removed two sources above 400\,mJy at 857\,GHz, and no sources at 600\,GHz). The measured shot noise levels are 1.6 and 1.2 times higher than the model predictions shown in Table \ref{tab:fluc_cut} at 545 and 857\,GHz, respectively.
We also plot their best-fit halo model, which has a minimum halo mass required to host a galaxy of
$\log(M_{\mathrm{min}}/\rm M_{\odot}) = 11.5_{-0.1}^{+0.4}$, and an effective bias $b_{\rm eff}\simeq 2.4$.
We see from Fig. \ref{fig_blast_hfi} that their model is a very good fit to the HFI data points.
Indeed, it provides a much better fit of the HFI data points than the BLAST data points!\\

\begin{figure}
\begin{center}
\includegraphics[width=\linewidth]{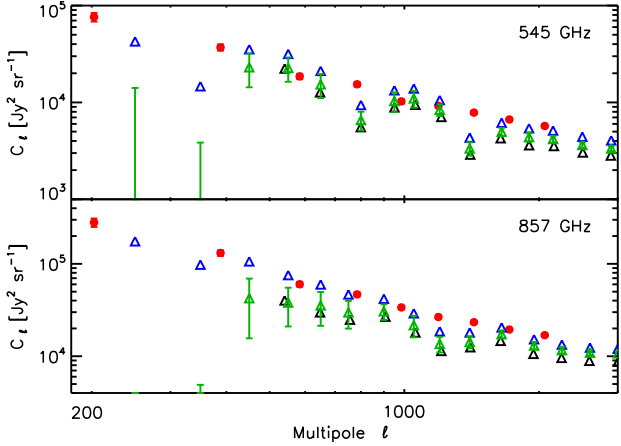} 
\caption{\label{fig_spire_hfi} Comparison of SPIRE and HFI measurements at 545 and 857\,GHz in the overlapping multipole range.  HFI data points are the red circles;  SPIRE data points from \cite{amblard2011} are the black triangles. For these data points sources down to 50~mJy have been masked, there is consequently less power compared to HFI. The green triangles  (Amblard, private communication) show the SPIRE CIB measurements identical to \cite{amblard2011}, but without a flux cut applied and thus they are directly comparable to the HFI measurement.  They should agree with HFI, but are a factor $\sim$1.7 and $\sim$1.2 below the HFI CIB data points for $400<\ell<1500$ at 857 and 545~GHZ, respectively. Indeed, they suffer from an overestimate of the cirrus contamination (by a factor 2). Moreover, preliminary cross-calibration between SPIRE and HFI is increasing the \cite{amblard2011} SPIRE power spectra by 10 and 20\% at 857 and 545 GHz, respectively (see Sect. \ref{compar_blast_spire} for more details).
When corrected from these too factors (cirrus and cross-calibration), the SPIRE (blue triangles) and HFI measurements agree well. For this figure, all SPIRE data points were colour-corrected (colours were computed using the CIB SED of \citet{gispert2000}, integrated through the SPIRE and HFI bandpass filters). Error bars include only statistical errors (for SPIRE, error bars are only shown for the green triangles for sake of clarity).}
\end{center}
\end{figure}

We now compare our results with the {\it Herschel}/SPIRE measurements \citep{amblard2011}. This comparison has to be made with caution, because the sources are masked down to a flux cut of 50 mJy in SPIRE, which is much smaller than the 540 and 710~mJy flux cut used in HFI at 545 and 857~GHz, respectively.  This large difference in the source removal will affect both the shot noise and the correlated components. We thus compare on Fig.~\ref{fig_spire_hfi} our HFI data points with a SPIRE measurement identical to the \cite{amblard2011}, but with no flux cut applied (Amblard, private communication). 
In this figure, we only show the SPIRE measurements over the multipole range of 200 to 3000 overlapping with {\it Planck}. With higher angular resolution maps, SPIRE CIB anisotropy measurements extend down to sub-arcminute angular scales or $\ell$ of $\sim2\times10^4$. For clarity we do not plot the small-scale power spectrum, but only concentrate on the consistency between HFI and SPIRE CIB anisotropy at larger angular scales.
We see from Fig. \ref{fig_spire_hfi} that SPIRE measurements (green triangles) are significantly below the HFI CIB spectra (red dots). Two elements may explain this difference: 
\begin{itemize}
\item Galactic cirrus: the cirrus signal in the SPIRE field is taken from existing measurements in the same field with IRAS 100~$\mu$m and MIPS 160~$\mu$m, and the spectrum is extrapolated from 100~$\mu$m to SPIRE wavelengths using the spectral dependence of a Galactic dust model; this procedure is less accurate than the use of \hi, and overestimates the cirrus contamination because the IRAS data contain the CIB anisotropies \citep{penin2011a}. Indeed, we checked with our LH2 field, that overlaps with the SPIRE SWIRE-Lockman field to 43\%, that in this region the cirrus contamination is negligible at the scales of interests for the comparison ($200<\ell<2000$). We accordingly added back the cirrus power spectra that were removed from the \cite{amblard2011} power spectra.
\item HFI/SPIRE cross-calibration:  SPIRE data are calibrated for point sources, with an accuracy of 15\% \citep{swinyard2010}. The point-source to diffuse-emission calibration conversion invokes an effective beam surface that is not perfectly determined yet. Planck/HFI is directly calibrated on diffuse emission. As detailed in the \cite{planck2011-1.5}, the accuracy is 7\% at high frequencies. For now, {\it Planck}/HFI therefore has a more accurate photometric calibration for diffuse emission than SPIRE. 
At this stage, it appears that assuming SPIRE beam surfaces corresponding to the integral of a Gaussian beam limited by diffraction gives a more
accurate SPIRE/HFI cross-calibration than the official beam surfaces given in \cite{swinyard2010}. This preliminary cross-calibration has been established by comparing the diffuse emission from several {\it Hershel} surveys (some HATLAS, SAG4, and Hi-Gal fields) to  {\it Planck}/HFI data. 
Compared to the beam surfaces taken in \cite{amblard2011}\footnote{\cite{amblard2011} use different values than those given in \cite{swinyard2010}, with beam surfaces of 1.77$\times$10$^{-8}$ sr and 3.99$\times$10$^{-8}$ sr at 350 and 500~$\mu$m, respectively.}, they will increase the power spectra by 10\% and 20\% at 857 and 545~GHz, respectively. 
\end{itemize}
When corrected for the cirrus overestimate and the HFI/SPIRE cross-calibration on diffuse emission, the SPIRE and HFI data points are now compatible (blue triangles and red dots on Fig.~\ref{fig_spire_hfi}, respectively). The cirrus correction is the dominant effect up to $\ell \sim$500 and $\ell \sim$1500 at 545 and 857~GHz, respectively.

\subsection{A self-consistent, cosmological, IR, galaxy evolution model \label{model_pres}}

Our interpretation of the CIB anisotropy measurements with HFI relies on a model introduced in \cite{penin2011b}.
The model builds upon the halo model formalism \citep[see][for a review]{Cooray:2002dia} and populates
dark matter halos with galaxies using a HOD,
modelling the emission of dusty galaxies using the infrared evolution model of \citet[][see Appendix \ref{A3}]{bethermin2010}.
Our main motivation for developing and using this parametric model is that it allows us to handle in a self-consistent manner the observational constraints coming from galaxy clustering and the CMB with more galaxy-evolution-centered measurements such as number counts or luminosity functions at various wavelengths and redshifts.
This is a key feature of our model.

Previous approaches, such as \citet{amblard2007} and \citet{viero2009}, have used the \citet{lagache2004} infrared-galaxy evolution model.
Compared to \citet{lagache2004} and \citet{marsden2011}, the parametric evolution of \citet{bethermin2010} better reproduces the mid-IR to millimeter statistical observations of infrared galaxies (number counts, luminosity functions, CIB, redshift distributions).  This is important because we derive from this model the mean emissivity per comoving unit volume, introduced below, which is a key quantity for interpreting CIB anisotropies.

On the scales of interest to us we can use the Limber approximation \citep{Limber54} and write the angular (cross) power spectrum of infrared emission at two frequencies, $\nu$ and $\nu'$, and at a multipole $\ell$ as \citep[e.g.,][]{knox2001}
\begin{equation}
C_\ell^{\nu\nu'} = \int{dz\ \left(\frac{d\chi}{dz}\right)\left(\frac{a}{\chi}\right)^2
  \ \bar{j}_{\nu}(z)\bar{j}_{\nu'}(z)P_{\rm gg}(k=\ell/\chi,z)} \,,
\label{eq:cl_limber}
\end{equation}
where $\chi$ is the comoving angular diameter distance to redshift $z$, $a=(1+z)^{-1}$ is the scale factor and $\bar{j}_{\nu}(z)$ is the mean emissivity per comoving unit volume at frequency $\nu$ and redshift $z$. The mean emissivity is derived using the empirical, parametric model of \citet{bethermin2010}\footnote{Note that for illustration purpose and where specified only, we will sometimes use the older phenomenological model of  \citet{lagache2004} (LDP).}:
\begin{equation}
\bar{j}_{\nu}(z)  =  \left(1+z\right)\int_0^{S_{\rm cut}} dS\ S\ \frac{d^2N}{dSdz}  \,.
\label{eq:emiss}
\end{equation}

The remaining ingredient in the model is thus $P_{\rm gg}(k,z)$.  As a foil to the HOD model for $P_{\rm gg}$ we begin
with the simple, constant bias model in which
\begin{equation}
P_{\rm gg}(k,z)=b_{\rm lin}^2 P_{\rm lin}(k,z) \,, 
\label{eq:linbias}
\end{equation}
where $b_{\rm lin}$ is a redshift- and scale-independent bias and $P_{\rm lin}(k)$ is the linear theory, dark-matter
power spectrum. We compute $P_{\rm lin}(k)$ using the fit of \citet{Eisenstein:1997ik}.
We will see that this model is not sufficient to explain the CIB anisotropies that
we measure.  This is not unexpected; at the mean distance of the sources we are probing Mpc scales
where non-linearities and scale-dependent bias are important.

By contrast the HOD model computes $P_{\rm gg}^{}(k,z)$ as the sum of the contributions of galaxies within a single
dark matter halo (1h) and galaxies belonging to two different halos (2h):
\begin{equation}
P_{\rm gg}(k) = P_{\rm 1h}(k) +P_{\rm 2h}(k) \, .
\end{equation}
The details of our assumptions for the 1h and 2h terms are given in Appendix \ref{A3}.
On large scales $P_{2h}$ reduces to a constant bias (squared) times the linear theory power spectrum, while
the 1-halo term becomes a scale-independent, shot-noise term.

Before comparing our model to \Planck\ observations, let us identify the parameters we hope to constrain with these data. The infrared galaxy evolution model of \citet{bethermin2010} satisfyingly reproduces current number count observations and luminosity function measurements at the price of introducing a luminosity function characterised by thirteen parameters (see Appendix \ref{A2}). These thirteen parameters fully define the mean emissivities, $\bar{j}_\nu(z)$, given in Eq.~\ref{eq:emiss}. The standard cosmological parameters (baryon density, tilt, etc.) mostly define the shape of the linear power spectrum in Eq.~\ref{eq:linbias} and the geometric functions like $\chi(z)$. The HOD formalism we introduce in Appendix \ref{A3} requires four more parameters.
\cite{penin2011b} investigated this full parameter space and its degeneracies and concluded, not surprisingly, that the current generation of infrared galaxy clustering measurements will not allow us to constrain all these parameters simultaneously. Furthermore, they show that most of the constraints on the luminosity function evolution come from number counts and monochromatic luminosity function measurements.  In the next section we threfore fix the luminosity function parameters to their best-fit values \citep[from][]{bethermin2010} and vary only some of the HOD parameters.

\subsection{Confronting the model with observations\label{model_obs}}

To confront the measurements with our model, we use a Levenberg-Marquardt algorithm to perform a $\chi^2$ minimization. The $\chi^2$ for our data when compared to our model is given by
\begin{equation}
\chi_\nu^2=\sum_b {(P_\nu^{\rm model}(b)-P_\nu^{\rm data}(b))^2\over \sigma_\nu(b)^2} \ \label{eq:chi2} \,.
\end{equation}
Unless specified otherwise, we use errors including statistical and systematic photometric calibration errors (2, 2, 7 and 7\% at 217, 353, 545 and 857~GHz, respectively, as defined in Sect.~\ref{data}), added linearly, and we assume diagonal, uncorrelated error bars as justified above. To reproduce the binning performed while measuring the power spectra, $P_\nu^{\rm model}(b)$ is computed taking the average of $\ell C_\ell^{model}$ at the minimum, maximum and mean $\ell$ of each bin.
We assume a Gaussian likelihood and assume  that setting the fixed parameters (e.g., the luminosity function parameters) at their best-fit value is equivalent to marginalising over them. It is important to note that the model power spectrum, $P_\nu^{\rm model}(b)$ includes a shot-noise term (SN) defined in Sect.~\ref{sect_sn}. Depending on the precise configuration we study, this shot-noise level is either fitted as an extra parameter or fixed to the predicted value. 
We also note that the models are derived for the \Planck\ bandpass filters and are colour-corrected to be in Jy$^2\,{\rm sr}^{-1}$($\nu I_{\nu}$=constant) as the data points, using the \cite{gispert2000} CIB SED (colour corrections are (1.08)$^2$, (1.08)$^2$, (1.06)$^2$ and (1.00)$^2$ at 217, 353, 545 and 857~GHz, respectively).\\

We remark here that this approach might be limited in two ways. First, we cannot exclude the possibility that our solution is a local extremum and not the global minimum of  Eq.~\ref{eq:chi2}.  While for all the results quoted above we checked the convergence with varying starting points, this particular point would have to be validated using for example a simulated annealing technique. We also did not explore the validity of the Gaussian likelihood approximation.
Some of these limitations could be resolved by implementing, for example, a Monte Carlo solver, and we defer this approach to future work.

\begin{figure}
\begin{center}
\includegraphics[angle=0,width=\linewidth, draft=false]{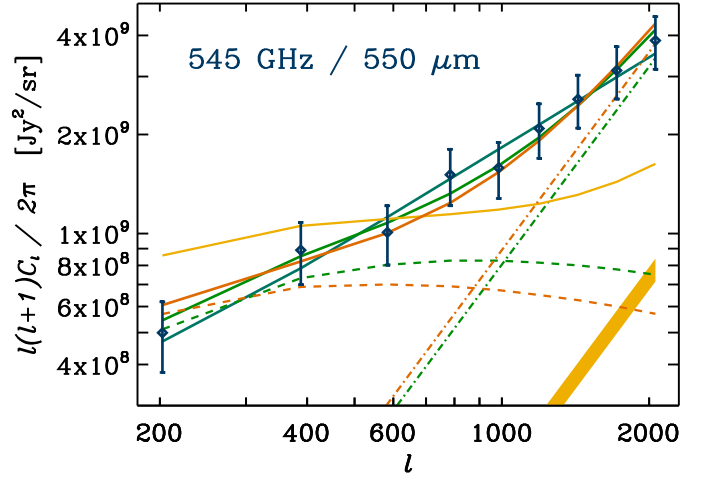}
\caption{\label{fig:lin_bias} In this plot we illustrate the constant bias model at 545\,GHz. The orange (solid, dashed and dot-dashed) lines correspond to the best-fit linear model, its clustering component, and the shot-noise level, respectively. While this fit was performed using our fiducial emissivity defined in Eq.~\ref{eq:emiss}, for illustrative purpose we plot in green the analogous fit using the LDP emissivity. In both cases the required shot-noise level (dot-dashed lines) is well above the 68\% C.L. predicted by \citet{bethermin2010} and given in Table~\ref{tab:CIB_CIBA_SED} (yellow contour). Conversely, the solid yellow line represents the best-fit curve when the shot-noise level is fixed to the expected value (Table~\ref{tab:fluc_cut}) and only the constant bias is varied. The fit is obviously unsatisfactory. These results lead us to  consider the linear bias model as unphysical, despite the good fit it provides ($\chi^2/dof\simeq 0.36$ (2.52/7)). For illustration purpose, we also show our best-fit power-law model defined in Eq.~\ref{eq:powl} (blue solid line).}
\end{center}
\end{figure}

\subsubsection{Linear bias model and power-law constraints \label{sec:linfit}}

In this section we will illustrate the discussion looking at the 545\,GHz data only, but the same conclusions are reached with the other three frequencies. The relevant results are illustrated in Fig.~\ref{fig:lin_bias}. The data points correspond to our measurements with statistical error bars only. 

Fitting simultaneously for a shot-noise level and a constant bias, $b_{\rm lin}$, defined in Eq.~\ref{eq:linbias}, we obtain a good fit ($\chi^2/dof \simeq$ 0.36) as visible from the solid orange curve. The best fit bias is  $b_{\rm lin}=2.45 \pm 0.18$ which is consistent with previous results in the literature \citep{lagache2007,viero2009}. The clustering contribution is plotted as the dashed orange line. However,  the shot-noise level required to obtain this good fit (dot-dashed orange curve) is unrealistically high: $(5.6 \pm 0.7)\times 10^3$ Jy$^2$/sr\footnote{The shot-noise levels required at the other frequencies to fit the data with the linear bias model are 27$\pm$6.7, 589$\pm$47, and 15915$\pm$1987 Jy$^2$sr$^{-1}$ at 217, 353, and 857\,GHz, respectively}.
When compared to the expected level from our model whose 68$\%$ C.L. amplitude is displayed as the yellow shaded area, our required level is $\simeq 5$ times higher in power. Given that this model reproduces well all known number counts, the monochromatic luminosity function, CIB measurements (see e.g., the last line in Table \ref{tab:cib}), and {\it Herschel}/SPIRE shot-noise measurements (when no sources are removed, see Sect. \ref{compar_blast_spire}),
 this excess level is excluded (see also the discussion in Sect.~\ref{sect_sn}). We thus conclude that the linear bias model is not a physically realistic fit to our data. The reason for this will be made clear in the next section, as it will appear clearly that the shot-noise level we fit here absorbs a strong non-linear component.

To further illustrate this point, we show the solid yellow curve that represents the best-fit model when we fix the shot noise to the level predicted by our model and only vary $b_{\rm lin}$. The fit is obviously unsatisfactory. We also note that the specific emissivity density that comes from the underlying model is unlikely to be the culprit.
The solid, dashed, and dot-dashed green curves are the analogous fit when we use the emissivity coming from LDP. The conclusions remain unchanged.

\begin{table}
\begin{center}
\begin{tabular}{c|c|c|c} 
Frequency & $A$ &  $n$  & Reduced $\chi^2$\\
(GHz) & Jy$^2$ sr$^{-1}$& & ($\chi^2/dof$)\\
\hline
217 & 51    $\pm$ 5   & -1.04 $\pm$ 0.13 & 0.98 (6.92/7)\\
353 &  1117   $\pm$ 46 & -1.03 $\pm$ 0.06 & 0.86 (6.07/7)\\
545 & (114   $\pm$ 7) $\times 10^2$& -1.09 $\pm$ 0.10 & 0.21 (1.51/7)\\
857 &   (35 $\pm$ 2) $\times 10^3$ & -1.18 $\pm$ 0.10 & 0.25 (1.75/7)\\
\end{tabular}\\
\caption{Power-law model best-fit parameters for each frequency as well as the reduced $\chi^2$. The errors corresponds to the 1$\sigma$ Gaussian error including statistical and photometric calibration systematic contributions. \label{tab:powl}}
\end{center}
\end{table}

Galaxy correlation functions can be reasonably well fitted, over a limited range of scales, by power-laws.
A power-law correlation function would project into a power-law angular correlation function, so we
also consider a power-law fit to our data,
\begin{equation}
\label{eq:powl}
  C_\ell = A \left({\ell \over 1000}\right)^{n}\ .
\end{equation}
This simple two-parameter model (A,n) is a reasonable fit at all frequencies, giving a reduced $\chi^2/dof\simeq 0.21$ (1.51/7) at 545\,GHz.
The best-fit values are given in Table \ref{tab:powl} and the best-fit model at 545\,GHz is displayed as the blue solid line in Fig.~\ref{fig:lin_bias}.
Although the power-law model provides a good fit to the data, it does not provide physical insight into
the properties of the galaxies it describes.

\begin{figure*}
\begin{center}
\includegraphics[angle=+0,width=\linewidth, draft=false]{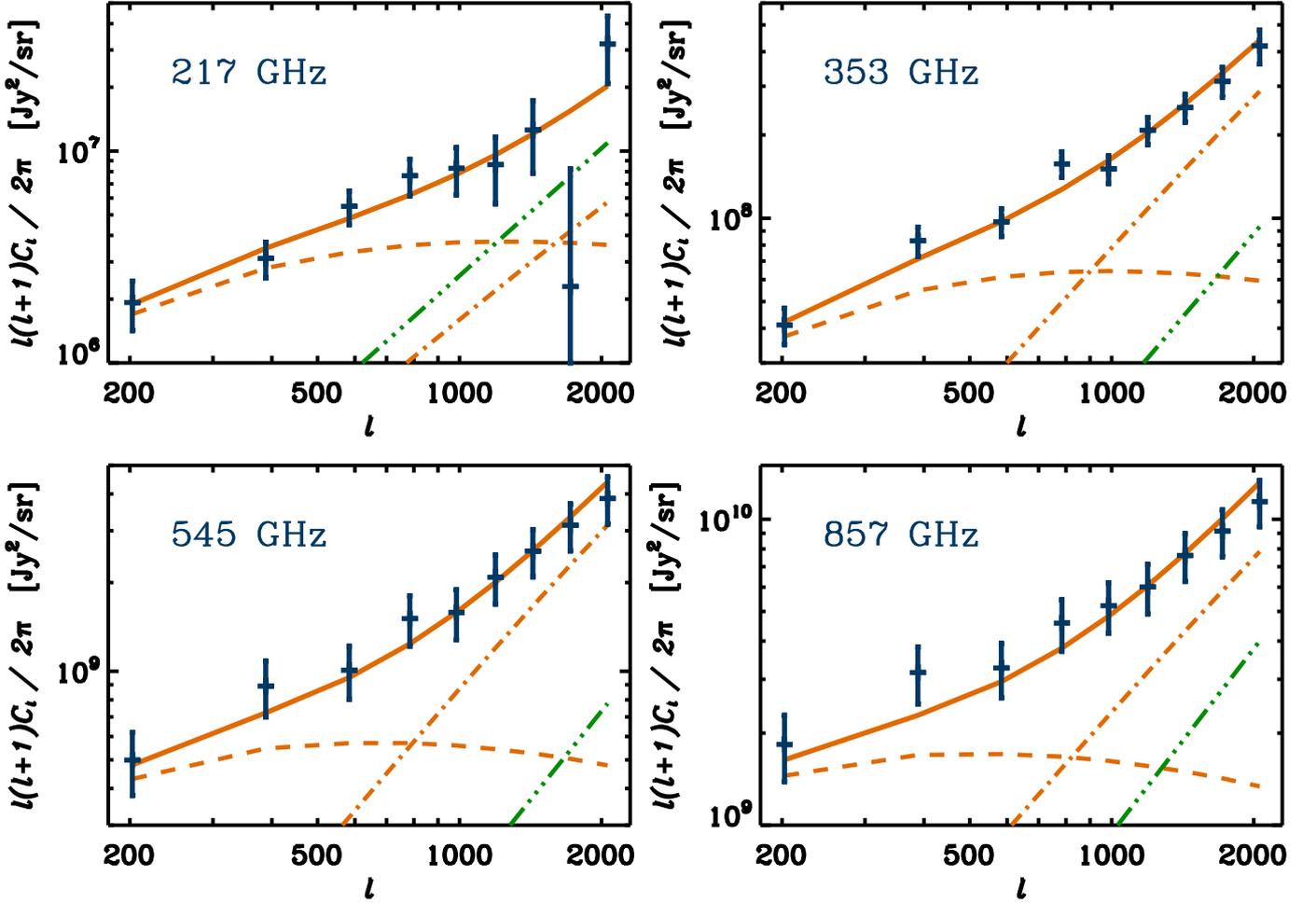}
\caption{\label{fig:l_cl_fit_4panel} Angular power spectrum of CIB anisotropies at 217, 353, 545, and 857~GHz.
Each panel corresponds to one frequency. For each frequency, the blue points correspond to the angular auto power-spectra, and the associated error bars include statistical and photometric calibration systematic contributions. The best-fit model per frequency (including shot noise) corresponds to the solid orange line. The dashed (dot-dashed) orange lines correspond to the 2h (1h) contributions. The green triple dot-dashed curve corresponds to the Poisson noise level, fixed to its expected value. To obtain these fits, three parameters per frequency were varied: $\log_{10}^{}M_{\rm min}^{}$, $\alpha_{\rm sat}$ and $j_{\rm eff}$.  The fits are obviously qualitatively very good.}
\end{center}
\end{figure*}

\begin{table*}
\begin{center}
\label{tab:chi2}
\begin{tabular}{c|c|c|c|c} 
Frequency (GHz) & \lmmin [$h^{-1}$M$_{\odot}$] &  $\alpha_{\rm sat}$  & $j_{\rm eff}$ [Jy/Mpc/sr] & Reduced $\chi^2$ ($\chi^2/dof$)\\
\hline
217 & 11.95 $\pm$ 2.10 & 1.30 $\pm$ 1.16 & 7.51 $\pm 0.75 \times10^1$ & 2.68 (16.1/6) \\
353 & 12.49 $\pm$ 0.42 & 1.39 $\pm$ 0.42 & 2.00 $\pm 0.29 \times10^2$ & 2.42 (14.5/6) \\
545 & 12.35 $\pm$ 1.01 & 1.17 $\pm$ 0.65 & 3.11 $\pm 3.85 \times10^2$ & 0.50 (3.04/6) \\
857 & 12.20 $\pm$ 0.51 & 1.02 $\pm$ 0.87 & 3.14 $\pm 17.0 \times10^2$ & 0.73 (4.40/6) \\
\hline
217 & 11.82 $\pm$ 1.92 & 1.17 $\pm$ 2.38  & N/A  & 1.14 (7.96/7) \\
353 & 12.50 $\pm$ 0.09 & 1.35 $\pm$ 0.20  & N/A  & 0.80 (5.64/7) \\
545 & 12.35 $\pm$ 0.94 & 1.17 $\pm$ 0.45  & N/A  & 0.35 (2.46/7) \\
857 & 12.21 $\pm$ 1.23 & 0.96 $\pm$ 0.73  & N/A  & 0.60 (4.22/7) \\
\end{tabular}\\
\caption{Best-fit values for each frequency, as well as the reduced $\chi^2$. The errors correspond to the 1$\sigma$ Gaussian errors, including statistical and photometric calibration systematic contributions. Systematic errors introduced by the beam uncertainty (see Sect.~\ref{se:systes}) are not included here, but contribute less than an extra 10\% to the error budget. The upper half of the array allows for a freely varying $j_{\rm eff}^{}$ per frequency, while in the bottom half $j_{\rm eff}$ is fixed to the extrapolation coming from our model.}\label{tab:best_fit_chi2} \end{center}
\end{table*}

\subsubsection{HOD model constraints}

We now consider the HOD model introduced in Sect.~\ref{model_pres}. In the most general configuration, our parametrisation of this model allows for four different parameters: $M_{\rm min}$, $M_{\rm sat}$, $\alpha_{\rm sat}$ and $\sigma_{\log M}$ (see Appendix \ref{A3}). The full exploration of this parameter space turns out to be a difficult task beyond our scope in this paper.  Therefore, we will restrict ourselves to only two parameters, $M_{\rm min}$ and $\alpha_{\rm sat}$, which describe the mass above which we expect a halo to host a CIB-contributing
galaxy and the slope of the high-mass end of the HOD.  By varying these parameters we control the mean,
galaxy-weighted halo mass and the satellite fraction in the model, which in turn control the amplitudes of the 1h
and 2h terms.

We impose $M_{\rm sat}=3.3\,M_{\rm min}$ \citep{penin2011b} and choose $\sigma_{\log M} = 0.65$, motivated by the clustering observed in optical surveys \citep[see][for discussion and references]{Tinker:2009mx}. We did not find $\sigma_{\log M}$ to be a critical parameters for our fit, but letting it vary drives the fit to physically unrealistic region of parameter space. For each frequency, we fixed the Poisson noise level to the one expected from our model (see Sect. \ref{sect_sn}).  To add to the robustness of our interpretation, we took one more conservative step.

As stated above, our interpretation of these data requires the knowledge of the emissivity. While our fiducial model is well tested and reproduces all relevant current observations \citep{bethermin2010}, these do not extend beyond a redshift of about 3.5. As such, extrapolations to higher redshifts are unconstrained by previous observations except for the integral constraints provided by the CIB measurement discussed in Sect.~\ref{sect_cib_ciba}. To let our conclusions be as model-independent as possible and also to isolate and constrain the high-$z$ contribution, we made the extra assumption that the emissivity, $j$, is constant for $z>3.5$ and we fitted for it simultaneously while solving for \lmmin and $\alpha_{\rm sat}$. More precisely,  we rewrite Eq.~\ref{eq:cl_limber} as 
\begin{eqnarray}
C_\ell^{\nu\nu'}  & = &  \int_{0}^{3.5}{dz\frac{d\chi}{dz}\frac{a^2}{\chi^2}\bar{j}_{\nu}(z)\bar{j}_{\nu'}(z)P_{\rm gg}(k=\ell/\chi,z)}\nonumber \\
& +& \left(j_{\rm eff}^{\nu\nu'}\right)^2\int_{3.5}^{7}{dz\frac{d\chi}{dz}\frac{a^2}{\chi^2}P_{\rm gg}(k=\ell/\chi,z)} \,,
\end{eqnarray}
where we have introduced the effective redshift-independent emissivity, $j_{\rm eff}^{}$, that we will also solve for. We adopt the arbitrary but reasonable $z=7$ cut-off of \citet{bethermin2010}.

We treated the four frequencies as independent and performed a single minimization per frequency. The results are illustrated in Fig.~\ref{fig:l_cl_fit_4panel}. The solid orange line represents the best-fit model per frequency. Our three-parameter model obviously fits each frequency very well. The orange dashed line represents the 2-halo (linear) term, while the orange dot-dashed line represents the 1-halo (non-linear) term.
The green curve corresponds to the assumed Poisson noise level.
Clearly, the angular scales we probe require a modelling of both the linear and the non-linear contribution to the power spectrum for all frequencies, and the 1-halo term is similar in slope to the shot-noise term, which leads to a model degeneracy.

Quantitative results are given in Table \ref{tab:best_fit_chi2}, where we quote a reduced $\chi^2$ as a goodness-of-fit measure. The errors quoted in Table \ref{tab:best_fit_chi2}  correspond to the Gaussian errors computed from the Fisher matrix at the best-fit values.  Each $\ell$ bin at any given frequency is considered independent from the others at all frequencies.  For reference, we also give the results of a fit where we fixed $j$ to the value given by the \cite{bethermin2010} model and fitted for only \lmmin and $\alpha_{\rm sat}$.  While the best-fit values are consistent between the two models, it is clear from the table that allowing $j_{\rm eff}$ to vary degrades strongly the constraints on \lmmin and $\alpha_{\rm sat}$. In fact, a strong degeneracy between $j_{\rm eff}$ and \lmmin is observed, as might be expected.

As is the case with optical data, our data do not appear to require a departure from $\alpha_{\rm sat}$=1. We considered different ratios of $M_{\rm sat}/M_{\rm min}$, {\it i.e.\/}, 2, 5, 10 or 20.  None of them provided a better fit, and most of them required similar values of $\alpha_{\rm sat}$ \citep[see][for a recent summary]{Tinker:2009mx}. 

Taking our emissivity model at face value, the best-fit angular power spectrum at $\ell=2000$ receives the following redshift contribution at 217 (353/545/857)~GHz:   5\% (4/34/71) between $0<z<1$, 7\% (7/23/22) between $1<z<2$, and 88\% (89/43/7) for $2<z$.

It is obvious from our Fig.~\ref{fig:l_cl_fit_4panel} that the non-linear contribution is degenerate with the Poisson noise level. This explains the problem faced by the linear model discussed in Sect.~\ref{sec:linfit}.
Our data by themselves are not sufficient to explore this degeneracy and we thus rely on our model. For similar reasons, we do not discuss details of the implementation of the 1h term (e.g., the truncation radius in $u(k,M)$) unlike \citet{viero2009}. We expect that a future joint analysis with higher angular resolution measurements from {\it Herschel\/} and SPT/ACT will allow us to alleviate this degeneracy. 

As described in Sect.~\ref{sect_cib_ciba}, our fiducial model predicts a mean CIB consistent with observations. While it is not possible to translate our constraints on $j_{\rm eff}^2$ (a weighted integral of $j^2$ over redshift) into a prediction for the integral of $j$ with the different weighting required to compute the CIB mean, rough estimates suggest that our values are consistent with the FIRAS measurements.

\begin{figure}[t]
\begin{center}
\includegraphics[width=\linewidth, draft=false]{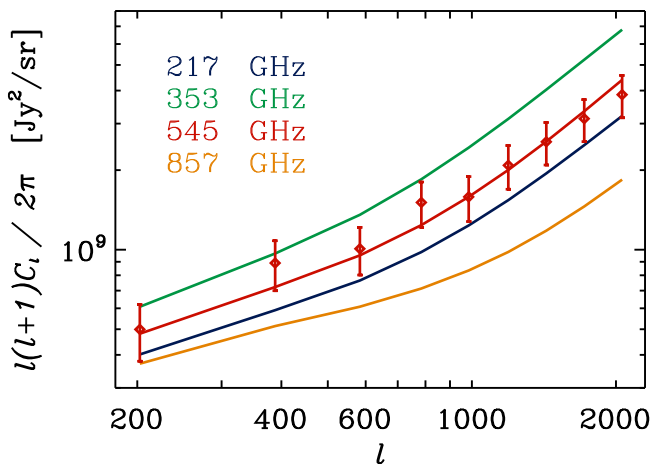}
\caption{\label{fig:l_cl_model_545} Predicted 545\,GHz power spectra derived from each frequency's best-fit model.
For the best-fit model at 217, 353, 545 and 857\,GHz, we plot the predicted clustering plus shot-noise power spectra at 545~GHz. Also shown are the HFI data points at~545 GHz (red diamonds). This plot suggests that the fits across frequencies are fairly different, which hints at an evolution in the population of galaxies we probed. We note, however, that the uncertainties associated with each prediction are not fully characterised by our method.}
\end{center}
\end{figure}

The relatively good consistency between the best-fit values of \lmmin and $\alpha_{\rm sat}$ observed across frequencies raises an interesting question: does a single model fit all our data? Or to put it another way, are the differences between each frequency HOD subsantial? Different frequencies, loosely speaking, correspond to different redshifts for the dominant galaxy population. As such, consistency in \lmmin and $\alpha_{\rm sat}$ could imply that the CIB fluctuations arise from a single subset of galaxies whose redshift evolution we capture well with our emissivity model, mass function, and HOD prescription, which is constant in redshift. To illustrate this hypothesis, Fig.~\ref{fig:l_cl_model_545} shows for each frequency's best-fit model its prediction at 545\,GHz. Even though this plot does not convey the uncertainty associated with each prediction, it clearly illustrates that each HOD leads to substantially different predictions and thus that the frequency dependence of the HOD may be significant. We postpone to future work more quantitative statements on the implications for the clustering of galaxies at high redshift.

\section{Conclusion \label{sect_cl}}

We presented the first measurement of CIB anisotropies with \Planck, detecting power from 10\arcmin\ to 2$^\circ$.
Owing to the exceptional quality of the data, and using a complete analysis of the different steps that lead to the CIB anisotropy power spectra, we were able to measure the clustering of dusty, star-forming galaxies at 217, 353, 545, and 857\,GHz with unprecedented precision.

We worked on six independent fields, chosen to have high angular-resolution \hi\ data and low foreground contamination.  The CIB maps were cleaned using templates: \hi\ for Galactic cirrus; and the \Planck\ 143\,GHz maps
for CMB.  Having \hi\ data is necessary to cleanly separate CIB and cirrus fluctuations.
Because the CIB anisotropies and Galactic dust have similar SEDs, blind component separation methods
do not adequately distinguish CIB anisotropies from cirrus emission.
The 143\,GHz \Planck\ channel, cleaned from sources and filtered, provides a good template for the CMB because it has
low instrument noise and an angular resolution close to the higher frequency channels from which we measure the CIB.
It also has the advantage of being an``internal" template, meaning its noise, data reduction process, photometric calibration, and beam are well known.

We obtained CIB anisotropy maps that reveal structures produced by the cumulative emission of high-redshift, dusty, star-forming galaxies. The maps are highly correlated at high frequencies. They decorrelate at lower frequency,
as expected from models of the redshift distribution of sources producing the CIB anisotropies \citep[e.g.][]{fernandez-conde2008,penin2011b}.
In these models, at 217\,GHz the contribution of $z\ge2$ galaxies is becoming dominant, while at higher frequencies
the dominant sources are at lower $z$.
We computed the power spectra of the maps and their associated errors using a dedicated pipeline, based on the \poker\ algorithm \citep{poker}.
After a careful examination of many systematic effects, use of the best determination of the beam window function
and instrument noise determined from jack-knife methods, we ended up with measurements of the angular power spectrum of the CIB anisotropy, $C_{\ell}$, at 217, 353, 545, and 857\,GHz, with high signal-to-noise ratio over the range $200<\ell<2000$ (see Table \ref{tab_cib_pk} and Fig.~\ref{CIB_all_fields}). 

The SED of CIB anisotropies is not different from the CIB mean SED, even at 217\,GHz. This is expected from the model of \citet{bethermin2010} and reflects the fact that the CIB mean and anisotropies are produced by the same population of sources.
Our measurement compares very well with previous measurements at higher $\ell$, at 220\,GHz by SPT \citep{hall2010} and at 600 and 857\,GHz by BLAST \citep{viero2009}. On the contrary, {\it Herschel}/SPIRE measurements from \cite{amblard2011}, but with no sources removal to be comparable to HFI (Amblard, private communication), are significantly lower than our measurements at high frequencies owing to an overestimate of the cirrus contribution. Preliminary cross-calibration between SPIRE and HFI is also increasing the \cite{amblard2011} power spectra by 10 and 20\% at 857 and 545 GHz, respectively.

From the \Planck\ data alone we can exclude a model where galaxies trace the linear theory matter power spectrum with a scale-independent bias: that model requires an unrealistic high level of shot noise to match the small-scale power we observe.  Consequently, we developed an alternative model that couples the dusty galaxy, parametric evolution model of \citet{bethermin2010} with a halo model approach. Characterised by only two parameters, this model  provides an excellent fit to our measured anisotropy angular power spectrum for each frequency treated independently. Whereas in principle these two parameters could offer us unique insights into the clustering and the nature of dusty galaxies at high redshift, the current uncertainties in the underlying model prevent us from drawing detailed inferences. Our results suggest that a different HOD is required at each frequency, which is consistent with the fact that we expect each frequency to be dominated by contributions from different redshifts. We find that half of the contribution to the power spectrum at $\ell$=2000 comes from redshifts lower than 0.8 and 1.5 at 857 and 545\,GHz, respectively. Those numbers are quite robust against exact evolution of dusty galaxies comoving emissivity at high-redshift ($z\ge$3.5). This is not the case at lower frequencies and our best-fit model predicts that about 90\%  of the anisotropies power at $\ell$=2000 come from redshifts $z>$2 at 353~GHz and 217\,GHz.

Further modelling and interpretation of the CIB anisotropy will be aided by the use of cross-power spectra between bands, and by the combination of the \Planck\ and {\it Herschel\/} data at 857 and 545/600\,GHz and \Planck\ and SPT/ACT data at 220\,GHz.
This combination will measure the CIB anisotropy power spectrum over a wide range of scales, covering the three regimes where we expect the 2-halo, 1-halo and shot-noise contributions to dominate.
More progress could be made by measuring the CIB anisotropies over more sky and at lower frequencies (at least 143\,GHz) with \Planck.
Going to lower frequency extends our reach in redshift, and is also important for CMB analysis and measurement of the SZ power spectrum.
Additional information will be obtained by cross-correlating the CIB maps with external tracers of the density field, like the galaxy and quasar distributions in large area catalogues (such as those from the SDSS, VIKING/VISTA and KIDS/VST surveys).
This will additionally constrain 1) the populations contributing most to the CIB; and
2) the relative bias between the external tracer and the distribution of far-infrared emission.
A particularly interesting cross-correlation may be between the CMB lensing convergence and the CIB maps \citep{song2003}.
The lensing and CIB anisotropies are expected to have a high degree of overlap, and lead to a signal readily detectable
by \Planck.  This signal will give a direct and independent measure of the bias.

\begin{acknowledgements}
This paper has made use of modelling tools that were made available by Matthieu B\'ethermin and Aur\'elie P\'enin.
The Planck Collaboration acknowledges the support of: ESA; CNES and CNRS/INSU-IN2P3-INP (France); ASI, CNR, and INAF (Italy); NASA and DoE (USA); STFC and UKSA (UK); CSIC, MICINN and JA (Spain); Tekes, AoF and CSC (Finland); DLR and MPG (Germany); CSA (Canada); DTU Space (Denmark); SER/SSO (Switzerland); RCN (Norway); SFI (Ireland); FCT/MCTES (Portugal); and DEISA (EU). A description of the Planck Collaboration and a list of its members, indicating which technical or scientific activities they have been involved in, can be found at http://www.rssd.esa.int/Planck.
\end{acknowledgements}

\bibliographystyle{aa}

\bibliography{main_bib,Planck_bib}

\appendix

\section{\label{A1}From HFI maps to CIB power spectra: flow charts of the different steps}
\begin{figure}
\begin{center}
\includegraphics[width=\linewidth, draft=false]{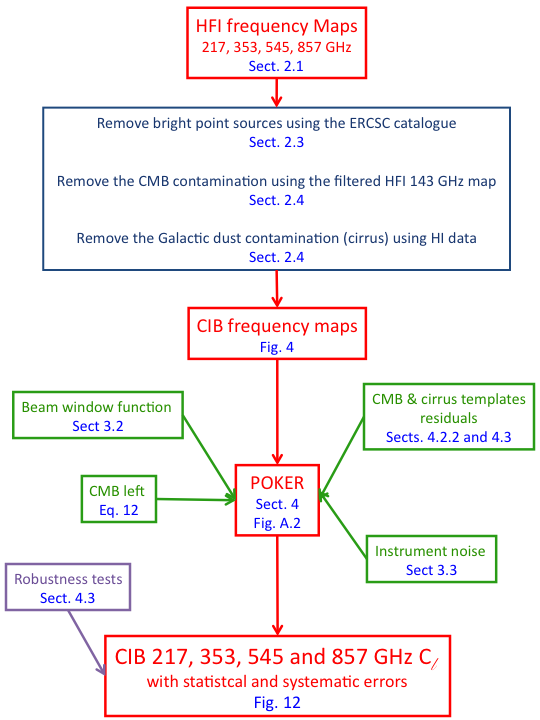}
\caption{\label{fig:flowchart_all} Cartoon illustrating the different steps from HFI frequency maps to CIB power spectra for one field.}
\end{center}
\end{figure}

\begin{figure*}
\begin{center}
\includegraphics[width=\linewidth, draft=false]{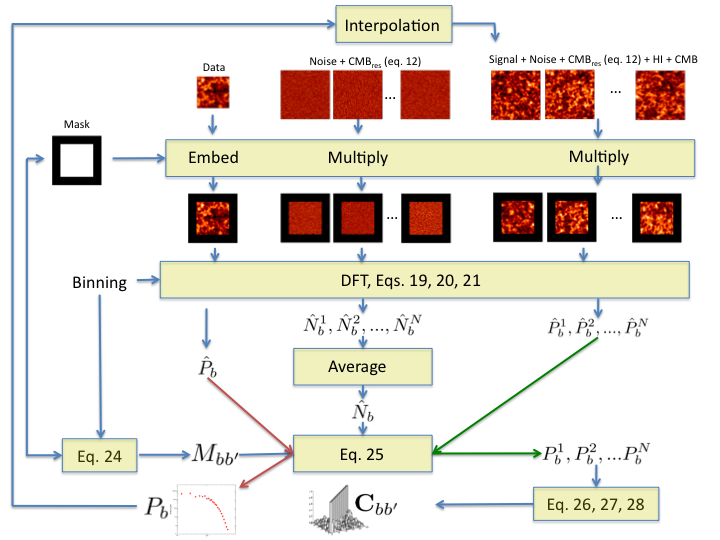}
\caption{\label{fig:flowchart_poker} Cartoon illustrating the angular power spectrum measurement and errors estimate using the \poker\ algorithm \citep{poker}.}
\end{center}
\end{figure*}

We summarize with the two flow charts presented in Fig. \ref{fig:flowchart_all} and \ref{fig:flowchart_poker} the procedures of data 
preparation and cleaning, and power spectra measurements and errors evaluation.

\section{\label{A2}The parametric backward evolution model of dusty star-forming galaxies}
In this appendix we give some details about the dusty star-forming galaxies evolution model we are using in our modelling of CIB anisotropies. The model is fully described in \cite{bethermin2010}. Two ingredients come into play: the spectral energy distribution (SED) of galaxies and the luminosity function (LF) evolution. \\
The SEDs are from the templates library of \cite{lagache2004}, which consists of two galaxy populations: a star-forming galaxy population with SEDs that vary with IR bolometric luminosities, and a normal-galaxy population with a template SED that is fixed and is colder than the star-forming galaxy templates and is scaled with IR bolometric luminosities. The normal and star-forming galaxies are dominant at low- and high-luminosity, respectively. The fraction of each galaxy population as a function of the bolometric luminosity was given using a smooth function\\
\begin{equation}
\label{eq:mixpop}
\frac{\Phi_{\rm starburst}}{\Phi} = \frac{1+th \bigl [ {\rm log}_{10}(L_{\rm IR}/L_{\rm pop}) /\sigma_{\rm pop} \bigl ]}{2},
\end{equation}
where $th$ is the hyperbolic tangent function, $L_{\rm pop}$ the luminosity at which the number of normal and star-forming galaxies are equal, and $\sigma_{\rm pop}$ characterises the width of the transition between the two populations. At $L_{\rm IR}=L_{\rm pop}$, the starbursts fraction is 50\%.\\

They assume that the luminosity (LF) is a classical double exponential function: \\
\begin{equation}
\label{eq:lf}
\Phi(L_{\rm IR}) = \Phi^\star \, \times \bigl  (\frac{L_{\rm IR}}{L^\star})^{1-\alpha} \,  \times \,exp \bigl [ -\frac{1}{2\sigma^2} {\rm log}_{10}^2 (1+\frac{L_{\rm IR}}{L^\star}) \bigl ],
\end{equation}
where $\Phi(L_{\rm IR})$ is the number of sources per logarithm of luminosity and per comoving volume unit for an infrared bolometric luminosity $L_{\rm IR}$, $\Phi_{\star}$ is the normalization constant characterising the density of sources, $L_{\star}$ is the characteristic luminosity at the break, and $1-\alpha$ and $1-\alpha-1/\sigma^2/ln^2(10)$ are the slope of the asymptotic power-law behaviour at low and high luminosity respectively.\\
A continuous LF redshift-evolution in luminosity and density is assumed following $L^\star \propto (1+z)^{r_L}$  and $\Phi^\star \propto (1+z)^{r_\Phi}$, where $r_L$ and $r_\phi$ are parameters driving the evolution in luminosity and density, respectively. It is impossible to reproduce the evolution of the LF with constant $r_L$ and $r_\phi$. They consequently authorize their value to change at two specific redshifts. The position of the first redshift break is a free parameter and converges to the same final value ($z\sim0.9$) for initial values $0<z<2$. To avoid a divergence at high redshift, the second break is fixed at z=2. The position of the breaks are the same for both $r_{\rm L}$ and $r_\phi$.\\

The model has 13 free parameters that are summarized in Table \ref{tab:params}. The parameters were determined by fitting the model to published measurements of galaxy number counts and monochromatic LF measured at given redshifts:
\begin{itemize}
\item Number counts: {\it Spitzer} counts at 24, 70 and 160~$\mu$m, {\it Herschel} counts at 250, 350 and 500~$\mu$m, and AzTEC counts at 1.1~mm.
\item Monochromatic LFs: IRAS local LF at 60~$\mu$m, {\it Spitzer} LF at 24~$\mu$m at z=0, at 15~$\mu$m at z=0.6, at 12~$\mu$m at z=1, and at 8~$\mu$m at z=2.
\item FIRAS CIB spectrum between 200~$\mu$m and 2~mm.
\end{itemize}
Measured redshift distributions were not used because the cosmic variance and the selection effects were currently poorly quantified.
The best-fit parameters as well as their uncertainties and degeneracies were obtained using a Monte Carlo Markov chain (MCMC) Metropolis-Hastings algorithm. The model adjusted on deep counts and monochromatic LFs at key wavelengths also reproduces recent very discriminating observations well, such as the \cite{jauzac2011} measured redshift distribution of the CIB. \\

We used the so-called {\it mean model}, which is obtained using the mean value of the parameters as given in Table~\ref{tab:params} without the lensing contribution (dndsnudz\_arr\_nolensing\_meanmodel\_final file on the http://www.ias.u-psud.fr/irgalaxies/ web page).

\section{\label{A3}The halo model}

In this appendix we give the details of our halo modelling.
Neglecting scale-dependent halo bias, the distinction between central and satellite galaxies and halo exclusion,
and assuming a Poisson distribution of galaxies, the 1h and 2h terms in $P_{\rm gg}$ have
simple analytic expressions \citep{Cooray:2002dia}:
\begin{eqnarray}
P_{\rm 1h}(k) & = & \int dM\ {dN\over dM}{\langle N_{\rm gal}(N_{\rm gal}-1) \rangle \over \bar n_{\rm gal}^2} u^2(k,M)\,; \\
P_{\rm 2h}(k) & = & P_{\rm lin}(k)\left[\int dM\ {dN\over dM}b(M){\langle N_{\rm gal} \rangle \over \bar n_{\rm gal}}
                       u(k,M)\right]^2 \;.
\end{eqnarray}
Here $M$ is the halo mass, $dN/dM$ is the halo mass function, $u(k,M)$ is the Fourier transform of the (normalized) halo density profile, $b(M)$ the halo bias and $\langle N_{\rm gal} \rangle$ is the mean number of galaxies in a halo of mass $M$.

The mean number density of galaxies, $\bar n_{\rm gal}$, can be written
\begin{equation}
  \bar{n}_{\rm gal} = \int dM\ \frac{dN}{dM}\langle N_{\rm gal}\rangle \,.
\end{equation}
Note that on large scales $u(k\to 0,M)\simeq 1$ so that we can define the ``effective'' bias as 
\begin{equation}
  b_{\rm eff}^{}(z)  =  \int dM\ {dN\over dM}b(M){\langle N_{\rm gal} \rangle \over \bar n_{\rm gal}} \,,
\label{eq:beff} 
\end{equation}
and the 1-halo term becomes scale-independent ({\it i.e.\/}, a shot-noise term).
 
We used the fitting function of \citet{Eisenstein:1997ik} to compute
$P_{\rm lin}$, an NFW profile \citep{Navarro:1996gj} truncated at the
virial radius to compute $u(k,M)$, and we relied on the mass function fit
of \citet{Tinker:2008ff} with its associated halo bias prescription
\citep{Tinker:2009jp}.
All these relations were calibrated through the use of N-body
simulations.
Our definition of halo mass is the mass interior to a radius within which the mean density is
$200$ times the mean density of the Universe.

\begin{table*}
\begin{tabular}{llr}
\hline
\hline
Parameter & Description & Value\\
\hline
$\alpha$ & Faint-end slope of the infrared bolometric LF & 1.223 $\pm$  0.044 \\
$\sigma$ & Parameter driving the bright-end slope of the LF & 0.406 $\pm$  0.019 \\
$L_\star$(z=0) ($\times 10^{10}~L_\odot$) & Local characteristic luminosity of the LF & 2.377 $\pm$  0.363 \\
$\phi_\star$ (z=0) ($\times 10^{-3}$ gal/dex/Mpc$^3$) & Local characteristic density of the LF & 3.234 $\pm$  0.266 \\
$r_{L_\star,lz}$ & Evolution of the characteristic  luminosity between 0 and $z_{break,1}$ & 2.931 $\pm$  0.119 \\
$r_{phi_\star,lz}$ & Evolution of the characteristic  density between 0 and $z_{break,1}$  &0.774 $\pm$  0.196 \\
$z_{break,1}$ & Redshift of the first break & 0.879 $\pm$  0.052 \\
$r_{L_\star,mz}$ & Evolution of the characteristic  luminosity between $z_{break,1}$ and $z_{break,2}$ &  4.737 $\pm$  0.301 \\
$r_{phi_\star,mz}$& Evolution of the characteristic  density of between $z_{break,1}$ and $z_{break,2}$ & -6.246 $\pm$  0.458 \\
$z_{break,2}$ &  Redshift of the second break  & 2.000 \, \,   (fixed) \\
$r_{L_\star,hz}$ &   Evolution of the characteristic luminosity for z$>z_{break,2}$ & 0.145 $\pm$  0.460 \\
$r_{phi_\star,hz}$ & Evolution of the characteristic density for z$>z_{break,2}$ &-0.919 $\pm$  0.651 \\
$L_{pop}$ ($\times 10^{10}~L_\odot$) & Luminosity of the transition between normal and starburst templates & 23.677 $\pm$  2.704 \\
$\sigma_{pop}$ & Width of the transition between normal and starburst templates & 0.572 $\pm$  0.056 \\
\hline
\end{tabular}
\caption{\label{tab:params} Dusty star-forming galaxies evolution model parameters, fitted to selected infrared observations (table extracted from \citealt{bethermin2010}). The errors are derived from a MCMC analysis. }
\end{table*}

The HOD describes the way galaxies populate the dark matter halos.
While we do not distinguish between central and satellite galaxies in the above, the functional form we adopt
for the mean occupation is modelled on the form frequently used in optical observations
\citep[e.g.,][]{Zheng:2004id}
\begin{equation}
\langle N_{\rm gal}\rangle  =N_{\rm cen} + N_{\rm sat}
\end{equation}
with
\begin{equation}
N_{\rm cen} = \frac{1}{2}\left[1+\mbox{erf}\left(\frac{\log M-\log M_{\rm min}}{\sigma_{\log M}}\right)\right] \,,
\label{eq:hod_cent}
\end{equation}
and
\begin{equation}
N_{\rm sat}  = \frac{1}{2}\left[1+\mbox{erf}
  \left(\frac{\log M-\log 2M_{\rm min}}{\sigma_{\log M}}\right)\right]\left(\frac{M}{M_{\rm sat}}\right)^{\alpha_{\rm sat}}
\label{eq:hod_sat} \,.
\end{equation}
These definitions ensure that $M_{\rm min}$ is the halo mass at which a halo has a probability of 50\% of
having a central galaxy.  We introduce $\sigma_{\log M}$ to allow scatter in this relation between halo mass
and observable, which is important on large scales.  Following \citet{Zheng:2004id}, we assume a Poisson distribution for $N_{sat}$ and write
\begin{equation}
\langle N_{\rm gal}(N_{\rm gal}-1) \rangle = 2\langle N_{\rm sat}^{}\rangle + \langle N_{\rm sat}^{}\rangle^2\ .
\end{equation}
Within this parametrisation, halos with $M\ll M_{\rm min}$ will not host any galaxies, whereas those
with $M\gg M_{\rm min}$ are almost certain to contain one.
The satellite occupation has a similar cut-off, but the mass is chosen to be twice $M_{\rm min}$, so that halos with a low probability of having a central galaxy are unlikely to contain a satellite galaxy \citep[see][for a further discussion of this form]{Tinker:2009mx}.
 
We note that this parametrisation was introduced to reproduce the observed clustering of luminosity-threshold samples of optical galaxies at $0<z<2$.  We are therefore making substantial assumptions when applying this same parametric form to dusty star-forming galaxies at higher redshift. As we shall see, however, our constraints on even this form of the HOD are weak enough to argue against introducing additional degrees  of freedom in the model. 

\end{document}